\documentclass[aps,prd,reprint,superscriptaddress,showpacs,nofootinbib,longbibliography]{revtex4-1}
\usepackage{graphicx}
\usepackage{amsfonts}
\usepackage{amsmath}
\usepackage{mathrsfs}
\usepackage{epsfig}
\usepackage{slashed}
\usepackage{dcolumn}%

\usepackage{ulem}
\usepackage{color}
\usepackage{caption}
\usepackage{subcaption}
\usepackage{hyperref}

\def\nat{Nature\ }
\def\aap{Astron.\ Astrophys.\ }
\def\apj{Astrophys.\ J.\ }
\def\apjl{Astrophys.\ J.\ Lett.\ }

\def\aj{Astron.\ J.\ }
\def\mnras{Mon.\ Not.\ Roy.\ Astron.\ Soc.\ }
\def\physrep{Phys.\ Rept.\ }
\def\prd{Phys.\ Rev.\ D\ }

\def\jcap{J.\ Cosmol.\ Astropart.\ Phys.\ }

\def\pasp{Publications\ of\ the\ Astronomical\ Society\ of\ the\ Pacific}

\newcommand\diff{\,\mathrm{d}}
\def\kpc{\,\mathrm{kpc}}
\def\Tesla{\,\mathrm{Tesla}}
\def\km{\,\mathrm{km}}
\def\TeV{\,\mathrm{TeV}}
\def\GeV{\,\mathrm{GeV}}
\def\MeV{\,\mathrm{MeV}}
\def\GV{\,\mathrm{GV}}

\def\cm{\,\mathrm{cm}}
\def\m{\,\mathrm{m}}
\def\s{\,\mathrm{s}}
\def\p{\,\mathrm{p}}
\def\e{\,\mathrm{e}}
\def\He{\,\mathrm{He}}
\def\A{\,\mathrm{A}}
\def\sr{\,\mathrm{sr}}
\newcolumntype{p}{D{,}{\pm}{-1}}
\def\pbar{\,\bar{\text{p}}}
\def\pbarp{\,\bar{\text{p}}/\text{p}}
\def\psr{\,\mathrm{psr}}
\def\DM{\,\mathrm{DM}}
\def\sec{\,\mathrm{sec}}

\def\phinuc{\phi_{\mathrm{nuc}}}
\def\phipbar{\phi_{\pbar}}
\def\phipos{\phi_{e^{+}}}
\def\cpos{c_{e^{+}}}
\def\pos{e^{+}}
\def\lep{e^{-}+e^{+}}
\def\Mdm{m_{\chi}}

\def\sigv{\langle \sigma v \rangle}
\def\etae{\eta_{e}}
\def\etamu{\eta_{\mu}}
\def\etatau{\eta_{\tau}}
\def\eebar{e^{-}e^{+}}
\def\mumubar{\mu \bar{\mu}}
\def\tautaubar{\tau \bar{\tau}}

\begin{document}


\title{Bayesian Analysis of the break in  DAMPE Lepton Spectra}

\author{Jia-Shu Niu}
\email{jsniu@itp.ac.cn}
\affiliation{CAS Key Laboratory of Theoretical Physics, Institute of Theoretical Physics, Chinese Academy of Sciences, Beijing, 100190, China}
\affiliation{School of Physical Sciences, University of Chinese Academy of Sciences, No.~19A Yuquan Road, Beijing 100049, China}

\author{Tianjun Li}%
\email{tli@itp.ac.cn}
\affiliation{CAS Key Laboratory of Theoretical Physics, Institute of Theoretical Physics, Chinese Academy of Sciences, Beijing, 100190, China}
\affiliation{School of Physical Sciences, University of Chinese Academy of Sciences, No.~19A Yuquan Road, Beijing 100049, China}
 
\author{Ran Ding}%
\affiliation{Center for High-Energy Physics, Peking University, Beijing, 100871, P. R. China}
 
\author{Bin Zhu}%
\affiliation{Department of Physics, Yantai University, Yantai 264005, P. R. China}
 
\author{Hui-Fang Xue}
\affiliation{Astronomy Department, Beijing Normal University,    Beijing 100875, P.R.China}
\author{Yang Wang}
\affiliation{School of Mathematical Sciences, Shanxi University, Shanxi 030006, P.R. China.}

\date{\today}

\begin{abstract}

Recently, DAMPE has released its first results on the high-energy cosmic-ray electrons and positrons (CREs) from about $25 \GeV$ to $4.6 \TeV$, which directly detect a break at $\sim 1 \TeV$. This result gives us an excellent opportunity to study the source of the CREs excess. In this work, we used the data for proton and helium flux (from AMS-02 and CREAM), $\pbarp$ ratio (from AMS-02), positron flux (from AMS-02) and CREs flux (from DAMPE without the peak signal point at $\sim 1.4 \TeV$) to do global fitting simultaneously, which can account the influence from the propagation model, the nuclei and electron primary source injection and the secondary lepton production precisely. For extra source to interpret the excess in lepton spectrum, we consider two separate scenarios (pulsar and dark matter annihilation via leptonic channels) to construct the bump ($\gtrsim 100 \GeV$) and the break at $\sim 1 \TeV$. The result shows: (i) in pulsar scenario, the spectral index of the injection should be $\nu_{psr} \sim 0.65$ and the cut-off should be $R_{c} \sim 650 \GV$; (ii) in dark matter scenario, the dark matter particle's mass is $\Mdm \sim 1208 \GeV$ and the cross section is $\sigv \sim 1.48 \times 10^{-23} \cm^{3} \s^{-1}$.  Moreover, in the dark matter scenario, the $\tautaubar$ annihilation channel is highly suppressed, and a DM model is built to satisfy the fitting results. 
\end{abstract}


                              
\maketitle


\section{Introduction}

Recently, DAMPE (DArk Matter Particle Explorer) \citep{Chang2014,Chang2017} Satellite, which has been launched on December 17, 2015,  has released its first data on high-energy cosmic-ray electrons and positrons (CREs) \citep{DAMPE2017}. DAMPE has measured the CREs (i.e.,  $e^{-}+e^{+}$) spectrum in the range of $25 \GeV - 4.6 \TeV$  with unprecedented energy resolution (better than $1.2\%$  $\gtrsim 100 \GeV$). The results shows a bumps at about $100 \GeV - 1 \TeV$ which is consistent with previous results \citep{Adriani2009,Adriani2010lepton,AMS02_lepton_sum,Ackermann2012,CALET2017,Fermi2017}. More interesting, a break at $\sim 1 \TeV$ and a peak signal at $\sim 1.4 \TeV$ have been detected. All of these features cannot be described  by a single power law and provide us an opportunity to study the source of high-energy CREs.

The peak signal at $\sim 1.4 \TeV$ has been studied by many works which employed nearby pulsars wind, supernova remnants (SNRs) and dark matter (DM) sub-structures \citep{Malyshev2009,Kuhlen2009,Brun2009,Gendelev2010,Profumo2012,Panov2013,Fang2017,Yuan2017_dampe,Athron2017,Fan2017,Duan2017,Gu2017,Liu2017,Cao2017,Profumo2017}. At the same time, considering the statistical confidence level of this signal is about $3 \sigma$ which needs more counts in future, we exclude the peak signal and do a global fitting on the left points in DAMPE CREs spectrum in this work. As a result, if we refer to the DAMPE CREs flux in this work, the peak point is excluded except special emphasis.

In cosmic ray (CR) theory, the CR electrons are expected to be accelerated during the acceleration of CR nuclei at the sources, e.g. SNRs. But the CR positrons are produced as secondary particles from CR nuclei interaction with the interstellar medium (ISM) \citep{Adriani2009,AMS2013,Barwick1997,AMS01}. From the results of the flux of positrons and electrons \citep{AMS02_fraction01,AMS02_fraction02,AMS02_lepton,AMS02_lepton_sum}, we can infer that there should be some extra sources producing electron-positron pairs. 
This can be interpreted both by the astrophysical sources' injection \citep{Shen1970,Zhang2001,Yuksel2009,Hooper2009,Profumo2012,Blasi2009,Hu2009,Fujita2009} and DM annihilation or decay \citep{Bergstrom2008,Barger2009,Cirelli2009,Zhang2009,Bergstrom2009,Yin2013,Dev2014}.

As a result, the CREs data contains the primary electrons, the secondary electrons, the secondary positrons and the extra source of electron-positron pairs. If we want to study the properties of the extra source, we should deduct the primary electrons and secondary electrons/positrons first. The primary electrons are always assumed to have a power-law form injection and the secondary electrons/positrons are determined dominatingly by the CR proton and helium particles interact with ISM. Consequently, we should do global fitting to these data simultaneously which can avoid the bias of choosing the lepton background parameters..

Considering the situations of  high-dimentional parameter space of propagation model and precise data sets, we employ a Markov Chain Monte Carlo (MCMC \citep{Lewis2002}) method (embeded by {\sc dragon}) to do global fitting and sample the parameter space of all the related parameters to reproduce the CREs spectrum \citep{Liu2010,Lin2015,Yuan2017,Niu2017}.

Moreover, because of the significant difference in the slopes of proton and helium, of about $\sim 0.1$ \citep{ATIC2006,CREAM2010,PAMELA2011,AMS02_proton,AMS02_helium}, has been observed, we use separate primary source spectra settings for proton and helium. Note also that we consider propagation of nuclei only up to $Z = 2$ and neglect possible contributions from the fragmentation of $Z > 2$ nuclei, which should be a good approximation since their fluxes are much lower than the p and He fluxes \citep{Korsmeier2016}. In this condition, all the secondary particles (antiprotons and leptons) are produced from the interactions between proton, helium and ISM, which give us a self-consistent way to combine the nuclei and lepton data together.

This paper is organized as follows. We first introduce the setups of our work in Sec. II. The global fitting method and the chosen data sets and parameters is given in Sec. III. After present the fitting results and add some discussions in Sec. IV, we summarize our results in Sec. VI.

\section{Setups}
\label{sec:setups}

In this section, we just listed some of the most important setups in this work which is different from our previous work \citep{Niu2017}. More detailed description can be found in Ref. \citep{Niu2017}.

\subsection{Propagation model}

In this work, we use the diffusion-reacceleration model which is widely used and can give a consistent fitting results to the AMS-02 nuclei data (see for e.g., \citep{Niu2017,Yuan2017}). A uniform diffusion coefficient ($D_{xx} = D_0\beta \left( R/R_0 \right)^{\delta}$) is used in the whole propagation region.

At the same time, because high-energy CREs loss energy due to the process like inverse Compton scattering and synchrotron radiation, we parameterize the interstellar magnetic field in cylinder coordinates $(r,z)$ as 
\begin{equation}
\label{eq:magnetic_field}
 B(r,z)=B_{0} 
 \exp\left(-\frac{r-r_{\odot}}{r_{B}}\right)
 \exp\left(-\frac{|z|}{z_{B}} \right),
 \end{equation}
to calculate the energy loss rate. In Eq. \ref{eq:magnetic_field}, $B_{0}=5\times 10^{-10} \Tesla$, $r_{B}=10 \kpc$, $z_{B}=2 \kpc$~\citep{Strong2000}, and  $r_{\odot}\approx 8.5 \kpc$ is the distance from the Sun to the galactic center.

\subsection{Primary Sources}

In this work, considering the fine structure of spectral hardening for primary nuclei at $\sim 300 \GeV$ (which was observed by ATIC-2 \citep{ATIC2006}, CREAM \citep{CREAM2010}, PAMELA \citep{PAMELA2011}, and AMS-02 \citep{AMS02_proton,AMS02_helium}) and the observed significant difference in the slopes of proton and helium (of about $\sim 0.1$ \citep{Adriani2011,AMS02_proton,AMS02_helium}), we use separate primary source spectra settings for proton and helium and each of them has 2 breaks at rigidity $R_{\A1}$ and $R_{\A2}$. The corresponding slopes are $\nu_{\A1}$ ($R \le R_{\A1}$), $\nu_{\A2}$ ($R_{\A1} < R \le R_{\A2}$) and $\nu_{\A3}$ ($R > R_{\A3}$).
For cosmic-ray electrons primary source, we followed the same configuration as proton and helium. But due to the DAMPE lepton data range ($20 \GeV - 4 \TeV$), we use 1 break $R_{e}$ for electron primary source, and the corresponding slopes are $\nu_{e1}$ ($R \le R_{e}$) and $\nu_{e2}$ (($R > R_{e}$)).

\subsection{Secondary sources} 
 
The secondary  cosmic-ray particles are produced in collisions of primary cosmic-ray particles with ISM. 
The secondary antiprotons are generated dominantly from inelastic pp-collisions and pHe-collisions. At the same time, the secondary electrons and positrons are the final product of decay of charged pions and kaons which in turn mainly created in collisions of primary particles with gas. As a result, the corresponding source term of secondary particles can be expressed as
\begin{equation}
  \label{eq:secondary_source}
q_{\sec}=
\frac{c}{4 \pi} \sum_{i=\text{H,He}}  n_{i} \sum_{j}
\int \diff p' \beta n_{j}(p') \frac{\diff \sigma_{i,j}(p,p')}{\diff p} 
\end{equation}
where $n_{i}$ is the number density of interstellar hydrogen (helium), $\diff \sigma_{i,j}(p,p')/\diff p$ is the differential production cross section, $n_{j}(p')$ is the CR species density and $p'$ is the total momentum of a particle. 

To partially take into account the uncertainties when calculating the secondary fluxes, we employ a parameter $c_{\pbar}$ and $\cpos$ to re-scale the calculated secondary flux to fit the data \citep{Tan1983,Duperray2003,Kappl2014,diMauro2014,Lin2015}. Note that the above mentioned uncertainties may not be simply represented with a constant factor, but most probably they are energy dependent \citep{Delahaye2009,Mori2009}. Here we expect that a constant factor is a simple  assumption.

\subsection{Extra sources}

In this work, 2 kind of extra lepton sources are considered. The pulsar scenario account the extra lepton source to the pulsar ensemble in our galaxy, which is able to generate high energy positron-electron pairs from their magnetosphere. The injection spectrum of the CREs in such configuration can be parameterized as a power law with an exponential cutoff:
\begin{equation}
  q_e^{\psr}(p) = N_{\psr}(R/\mathrm{10\GeV})^{-\nu_{\psr}} \exp{(-R/R_\mathrm{c})},
  \label{pulsar_injection}
\end{equation}
where $N_{\psr}$ is the normalization factor, $\nu_{\psr}$ is the spectral index, $R_\mathrm{c}$ is the cutoff rigidity. The spatial distribution of this pulsar ensemble which provide continuous and stable CREs injection obeys the form as Eq. (5) in Ref. \citep{Niu2017}, with slightly different parameters $a=2.35$ and $b=5.56$ \cite{Lin2015}.

The DM scenario ascribe the extra lepton source to the annihilation of Majorana DM particles distributed in our galaxy halo, whose source term always has the form:
\begin{equation}
\label{eq:dm_source}
Q(\boldsymbol{r},p)=\frac{\rho(\boldsymbol{r})^2}{2 m_{\chi}^2}\langle \sigma v \rangle 
\sum_{f} \eta_{f} \frac{dN^{(f)}}{dp} ,
\end{equation}
where $\rho(\boldsymbol{r})$ present the DM density distribution, $\langle \sigma v \rangle$ is the velocity-averaged DM annihilation cross section multiplied by DM relative velocity,  and $dN^{(f)}/dp$ is the injection energy spectrum  of CREs from DM annihilating into standard model (SM) final states through all possible channels $f$ with $\eta_{f}$ (the corresponding branching fractions). In this work, we considered DM annihilation via leptonic channels, the corresponding branching fractions for $\eebar$, $\mumubar$, and $\tautaubar$ are $\etae$, $\etamu$, and $\etatau$ respectively ($\etae + \etamu + \etatau = 1$). We use the results from PPPC 4 DM ID \citep{Cirelli2011}, which includes the electroweak corrections \citep{Ciafaloni2011}, to calculate the electron (positron) spectrum from DM annihilation by different channels. At the same time, we use Einastro profile~\citep{Navarro2004,Merritt2006,Einasto2009,Navarro2010} to describe the DM spatial distribution in our galaxy, which has the form:
\begin{equation}
\rho(r)=\rho_\odot \exp
\left[
-\left( \frac{2}{\alpha}\right)
\left(\frac{r^{\alpha}-r_\odot^{\alpha}}{r_{s}^{\alpha}} \right)
\right] ,
\end{equation}
with $\alpha\approx 0.17$, $r_{s}\approx 20 \kpc$ and $\rho_\odot \approx 0.39 \GeV \cm ^{-3}$ is the local DM energy density \cite{Catena2010,Weber2010,Salucci2010,Pato2010,Iocco2011}.

\subsection{Solar modulation}

We adopt the force-field approximation \citep{Gleeson1968} to describe the effects of solar wind and helioshperic magnetic field in the solar system, which contains only one parameter the so-called solar-modulation $\phi$. Considering the charge-sign dependence solar modulation represented in the previous fitting \citep{Niu2017}, we use $\phinuc$ for nuclei (proton and helium) data and $\phipbar$ for $\pbar$ data to do the solar modulation. At the same time, we use $\phipos$ to modulate the positron flux. Because the DAMPE lepton data $\gtrsim 20 \GeV$, we did not consider the modulation effects on electrons (or leptons).

\subsection{Numerical tools}

The public code  {\sc dragon} \footnote{https://github.com/cosmicrays/DRAGON} \citep{Evoli2008} was used to solve the diffusion equation numerically, because its good performance on clusters. Some custom modifications are performed in the original code, such as the possibility to use specie-dependent injection spectra, which is not allowed by default in {\sc dragon}.

In view of some discrepancies when fitting with the new data which use the default abundance in {\sc dragon} \citep{Johannesson2016}, we use a factor $c_{\He}$ to rescale the helium-4 abundance (which has a default value of $7.199 \times 10^4$) which help us to get a global best fitting.

The radial and $z$ grid steps are chosen as $\Delta r = 1 \kpc$, and $\Delta z = 0.5 \kpc$. The grid in kinetic energy per nucleon is logarithmic between $0.1 \GeV$ and $220  \TeV$ with a step factor of $1.2$. The free escape boundary conditions are used by imposing $\psi$ equal to zero outside the region sampled by the grid.

\section{Fitting Procedure}
\label{sec:fitting_pro}

\subsection{Bayesian Inference}
\label{sec:Bayes}

As our previous works \citep{Niu2017}, we take the prior PDF as a uniform distribution and the likelihood function as a Gaussian form. The algorithms such as the one by \citet{Goodman2010} instead of classical Metropolis-Hastings is used in this work for its excellent performance on clusters. The algorithm by \citet{Goodman2010} was slightly altered and implemented as the Python module {\tt emcee}\footnote{http://dan.iel.fm/emcee/} by \citet{Mackey2013}, which makes it easy to use by the advantages of Python. Moreover, {\tt emcee} could distribute the sampling on the multiple nodes of modern cluster or cloud computing environments, and then increase the sampling efficiency observably.

\subsection{Data Sets and Parameters}

In our work, the proton flux (from AMS-02 and CREAM \citep{AMS02_proton,CREAM2010}), helium flux (from AMS-02 and CREAM \citep{AMS02_helium,CREAM2010}) and $\pbarp$ ratio ( from AMS-02 \citep{AMS02_pbar_proton}) are added in the global fitting data set to determine not only the propagation parameters but also the primary source of nuclei injections which further produce the secondary leptons. The CREAM data was used as the supplement of the AMS-02 data because it is more compatible with the AMS-02 data when $R \gtrsim 1 \TeV$.  The errors used in our global fitting are the quadratic summation over statistical and systematic errors.

On the other hand, the AMS-02 positrons flux \citep{AMS02_lepton} is added to set calibration to the absolute positron flux in DAMPE CREs flux \citep{DAMPE2017}. Although the electron energy range covered by AMS-02 is under $\TeV$ and there are systematics between the AMS-02 and DAMPE CREs data, fittings to the AMS-02 leptonic data provide a self-consistent picture for the extra source models. As the extra sources accounting for the AMS-02 results may provide contribution to the $\TeV$ scale, the AMS-02 data could also constrain the properties of the predicted $\lep$ spectrum above $\sim \TeV$. Considering the degeneracy between the different lepton data, we use the positron flux from AMS-02 and CREs flux from DAMPE together to constraint the extra source properties. The systematics are dealt with by employing a re-scale factor $\cpos$ on positron flux.

Altogether, the data set in our global fitting is 
 \begin{align*}
D = &\{D^{\text{AMS-02}}_{\p}, D^{\text{AMS-02}}_{\He},  D^{\text{AMS-02}}_{\pbarp}, D^{\text{CREAM}}_{\p}, \\
&D^{\text{CREAM}}_{\He}, D^{\text{AMS-02}}_{\pos}, D^{\text{DAMPE}}_{\lep} \}~.
\end{align*}

The parameter sets for pulsar scenario is 
\begin{align*}
\boldsymbol{\theta}_{\psr} =  &\{ D_{0}, \delta, z_{h}, v_{A}, | N_{\p}, R_{\p1}, R_{\p2}, \nu_{\p1}, \nu_{\p2}, \nu_{\p3}, \\
&  R_{\He1},  R_{\He2}, \nu_{\He1}, \nu_{\He2}, \nu_{\He3}, |  c_{\pbar}, c_{\He},  \phinuc, \phipbar, | \\
& N_{\e}, R_{\e1}, \nu_{\e1}, \nu_{\e2}, |   \\
& N_{\psr}, \nu_{\psr}, R_{c}, | \\
& c_{\pos}, \phipos \}~,
\end{align*}
for DM scenario is 
\begin{align*}
\boldsymbol{\theta}_{\DM} =  &\{ D_{0}, \delta, z_{h}, v_{A}, | N_{\p}, R_{\p1}, R_{\p2}, \nu_{\p1}, \nu_{\p2}, \nu_{\p3}, \\
&  R_{\He1},  R_{\He2}, \nu_{\He1}, \nu_{\He2}, \nu_{\He3}, |  c_{\pbar}, c_{\He},  \phinuc, \phipbar, | \\
& N_{\e}, R_{\e1}, \nu_{\e1}, \nu_{\e2}, |   \\
& \Mdm, \sigv, \eta_{e}, \eta_{\mu}, \eta_{\tau},  | \\
& c_{\pos}, \phipos \}~.
\end{align*}

Note that, most of these 2 scenarios' parameters in the set $\boldsymbol{\theta}_{\psr}$ and $\boldsymbol{\theta}_{\DM}$ is the same with each other except those who account the extra sources of lepton.

\section{Fitting Results and Discussion}

\begin{figure*}[!htbp]
  \centering
  \includegraphics[width=0.46\textwidth]{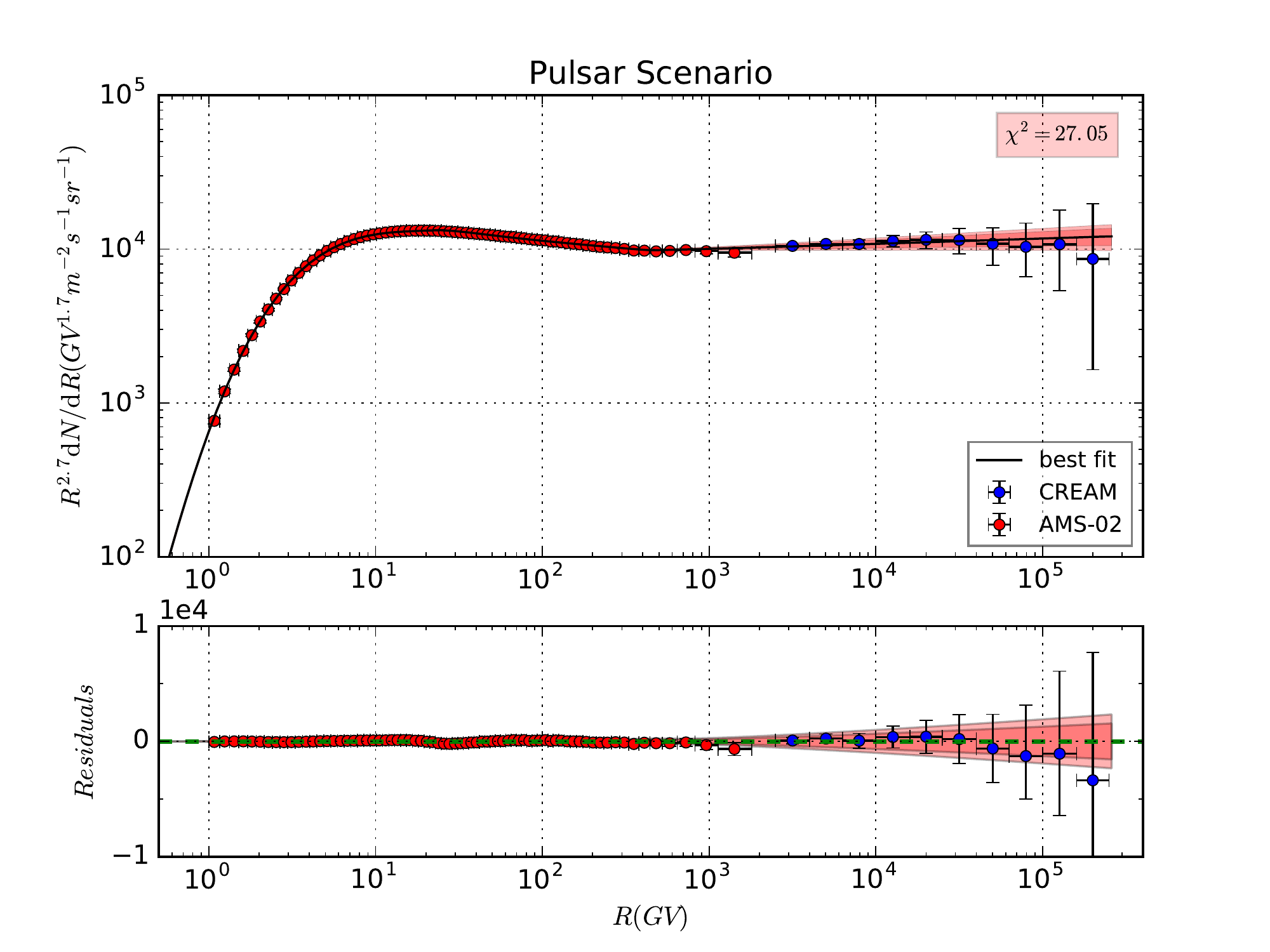}
  \includegraphics[width=0.46\textwidth]{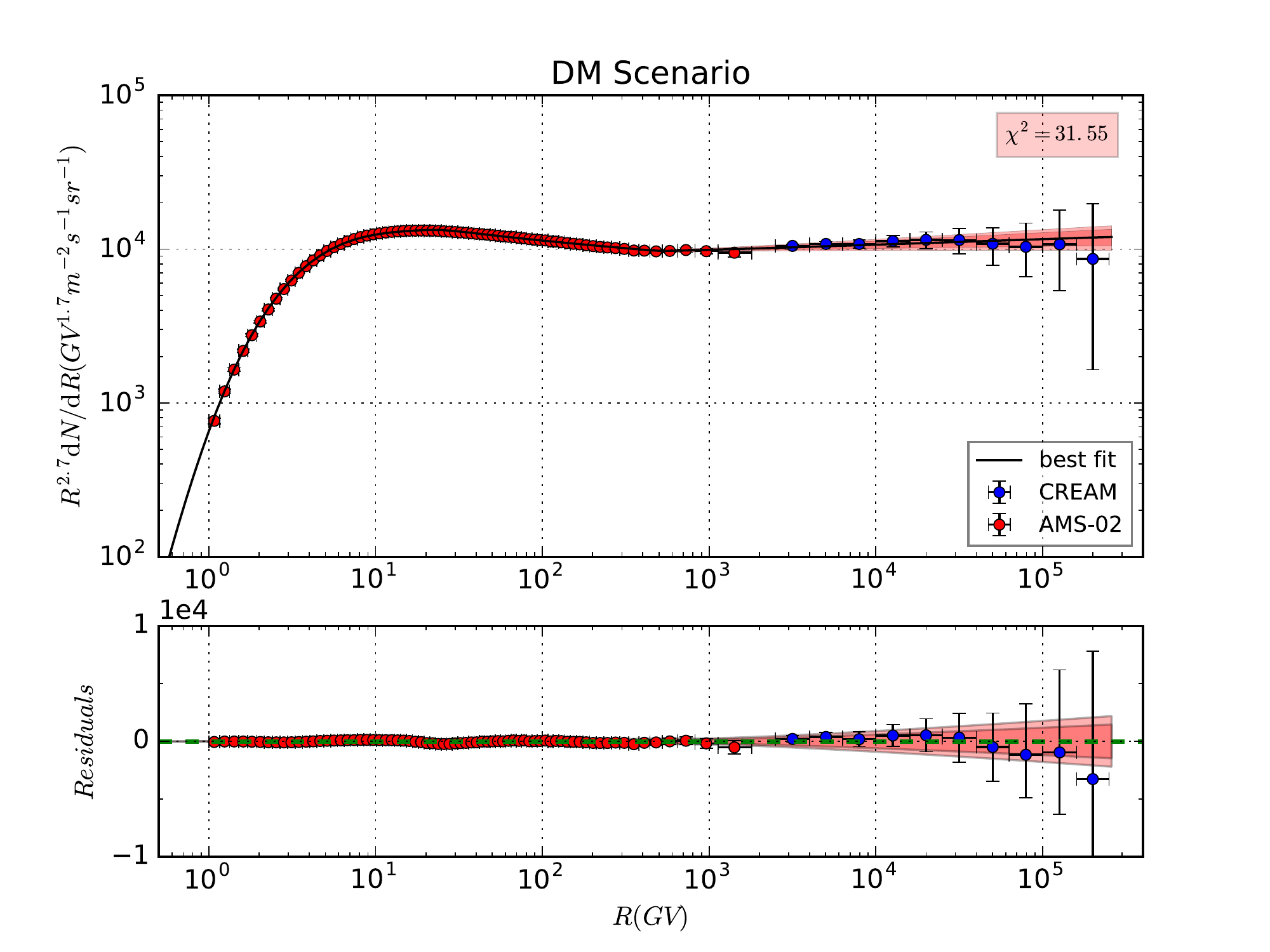}
  \includegraphics[width=0.46\textwidth]{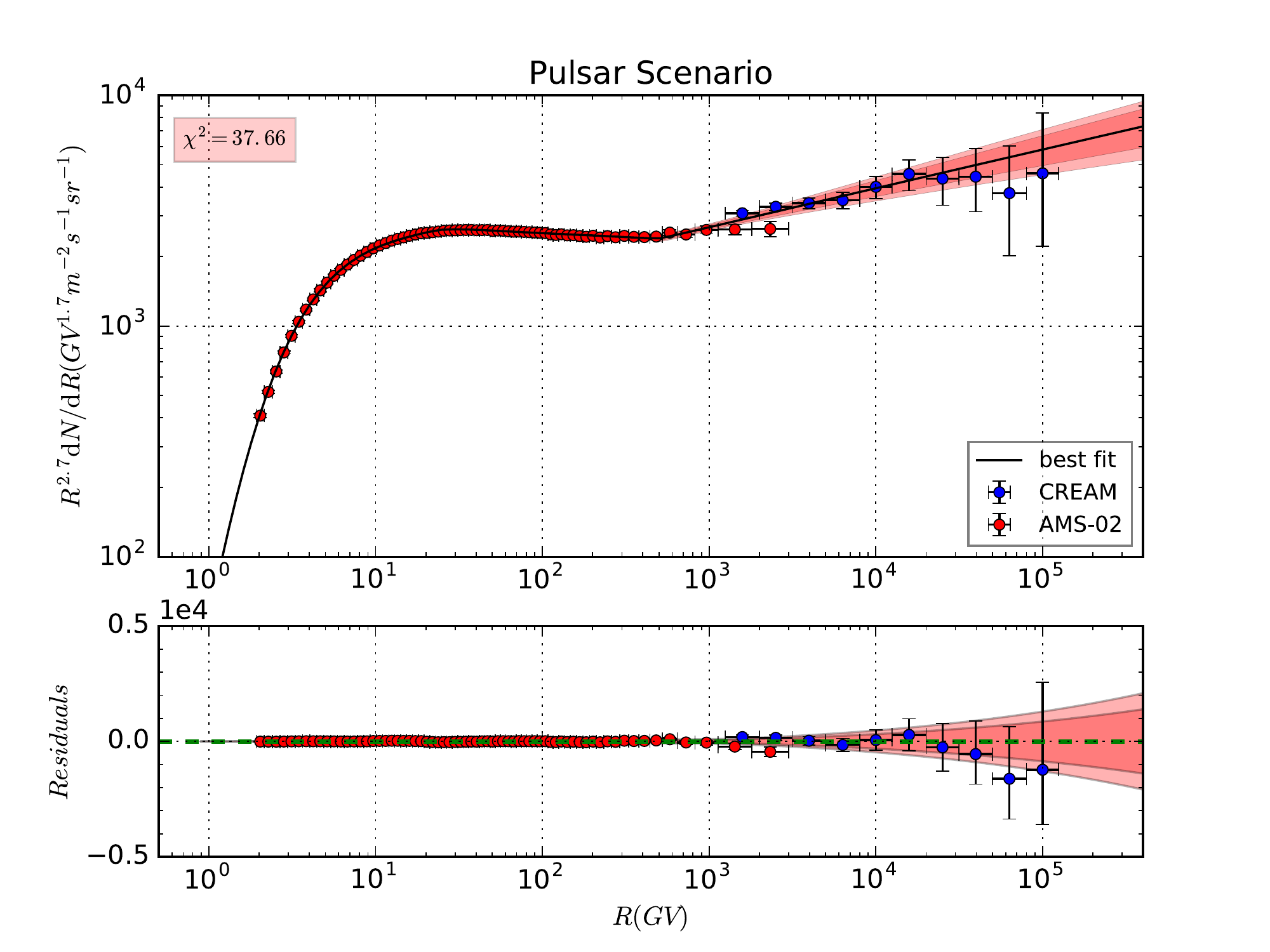}
  \includegraphics[width=0.46\textwidth]{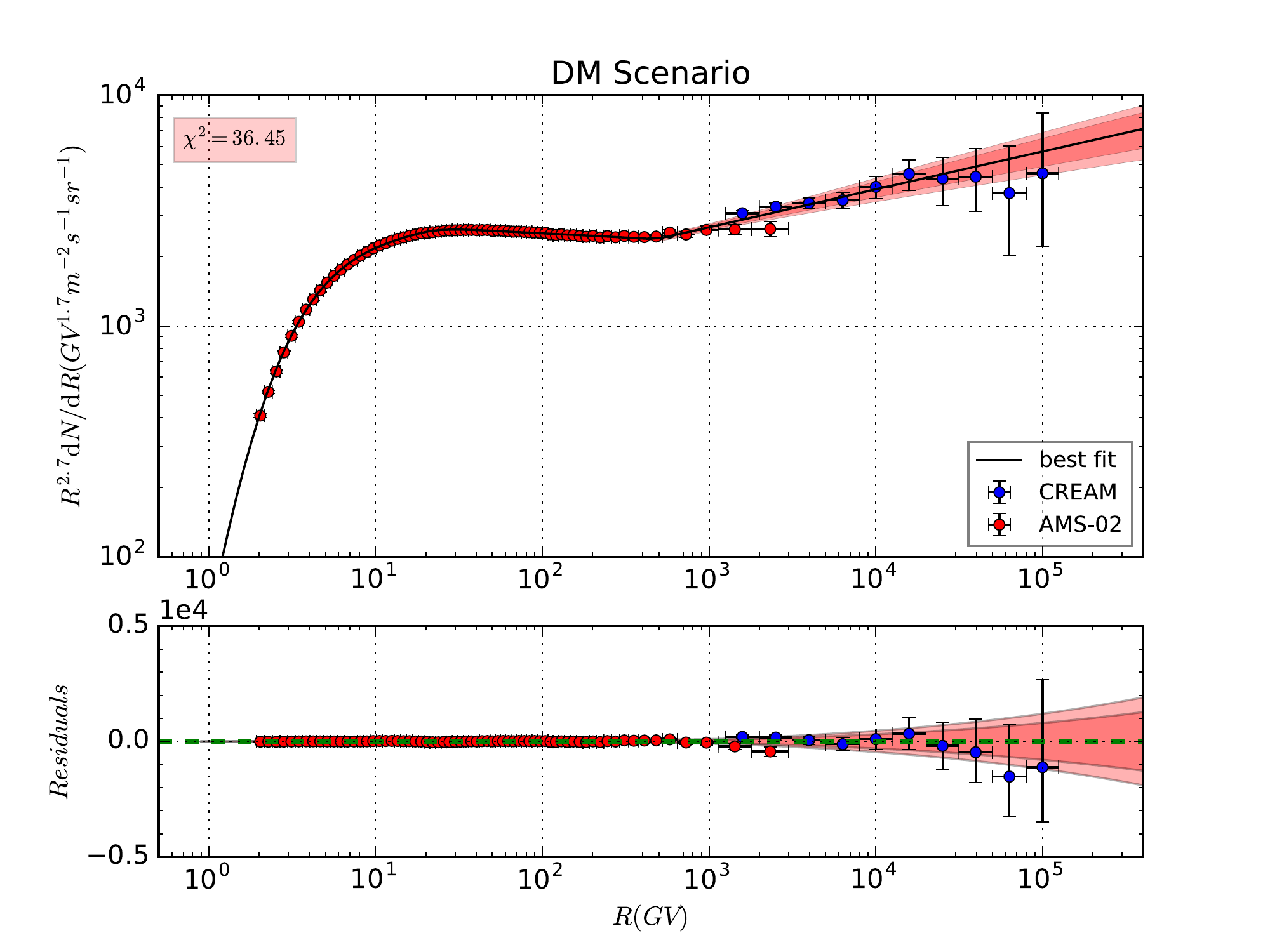}
  \includegraphics[width=0.46\textwidth]{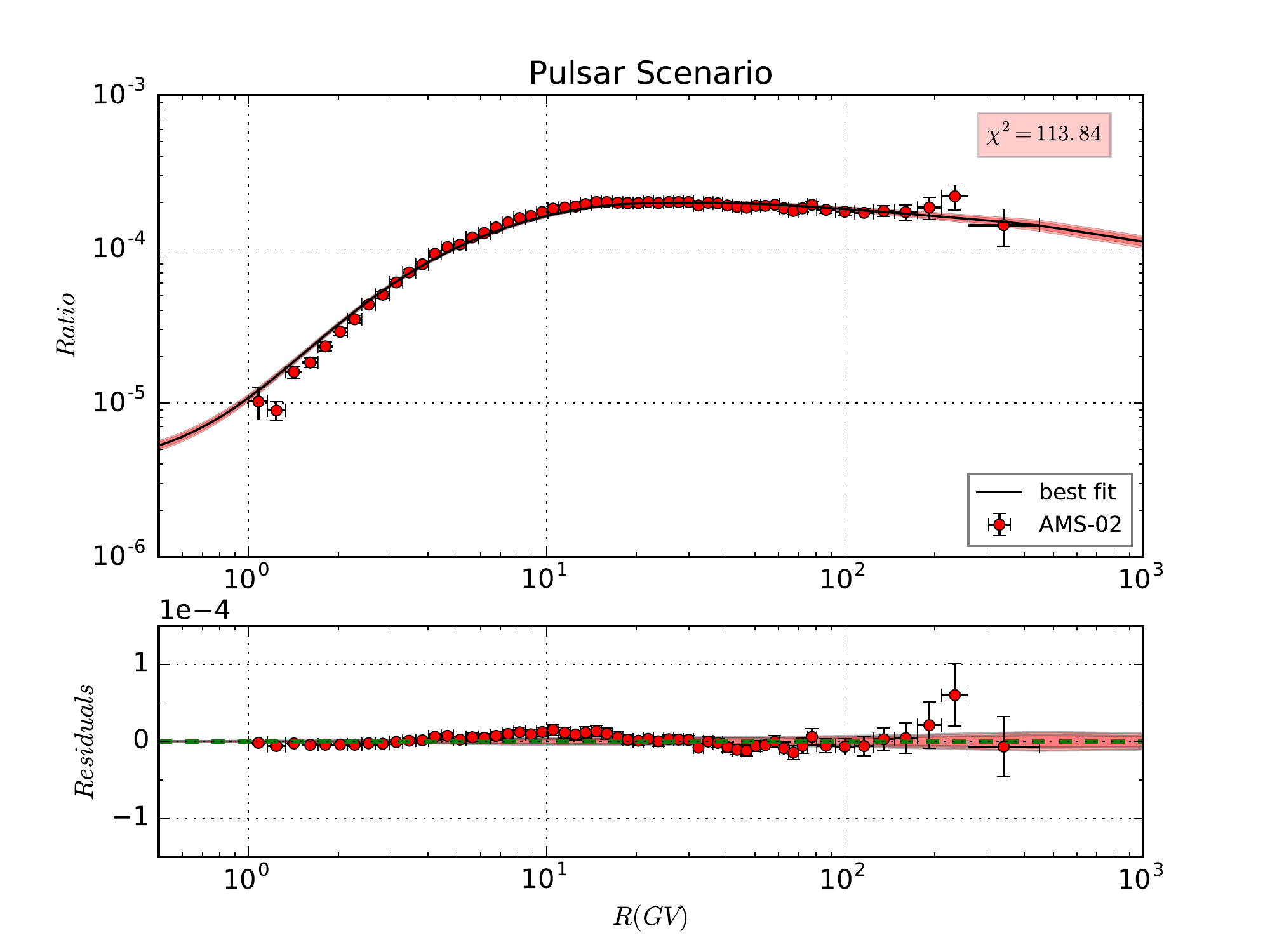}
  \includegraphics[width=0.46\textwidth]{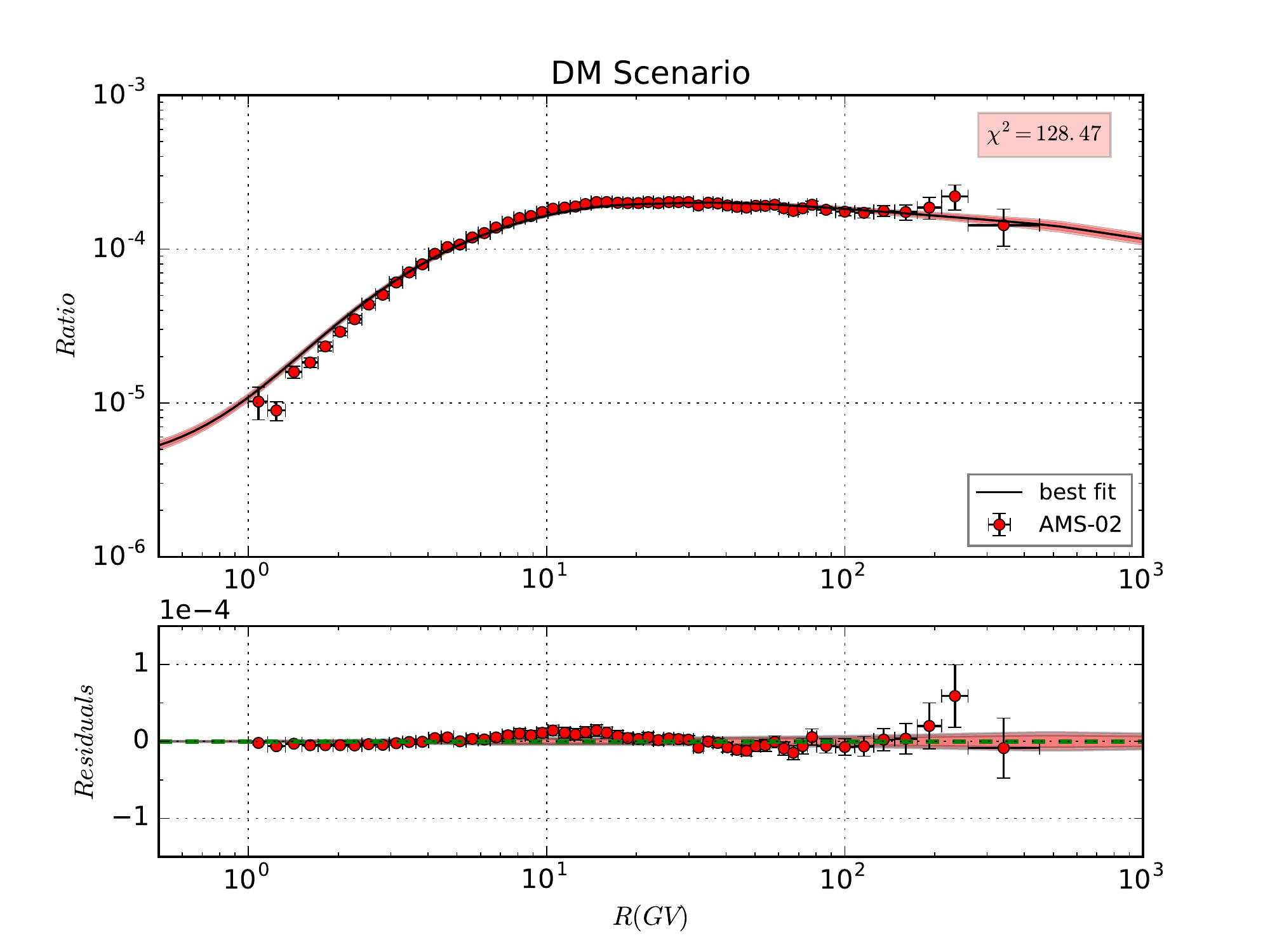}
  \caption{The global fitting results and the corresponding residuals to the proton flux, helium flux and $\pbarp$ ratio for 2 scenarios. The $2\sigma$ (deep red) and $3\sigma$ (light red) bound are also showed in the figures.}
\label{fig:nuclei_results}
\end{figure*}

\begin{figure*}[!htbp]
  \centering
  \includegraphics[width=0.46\textwidth]{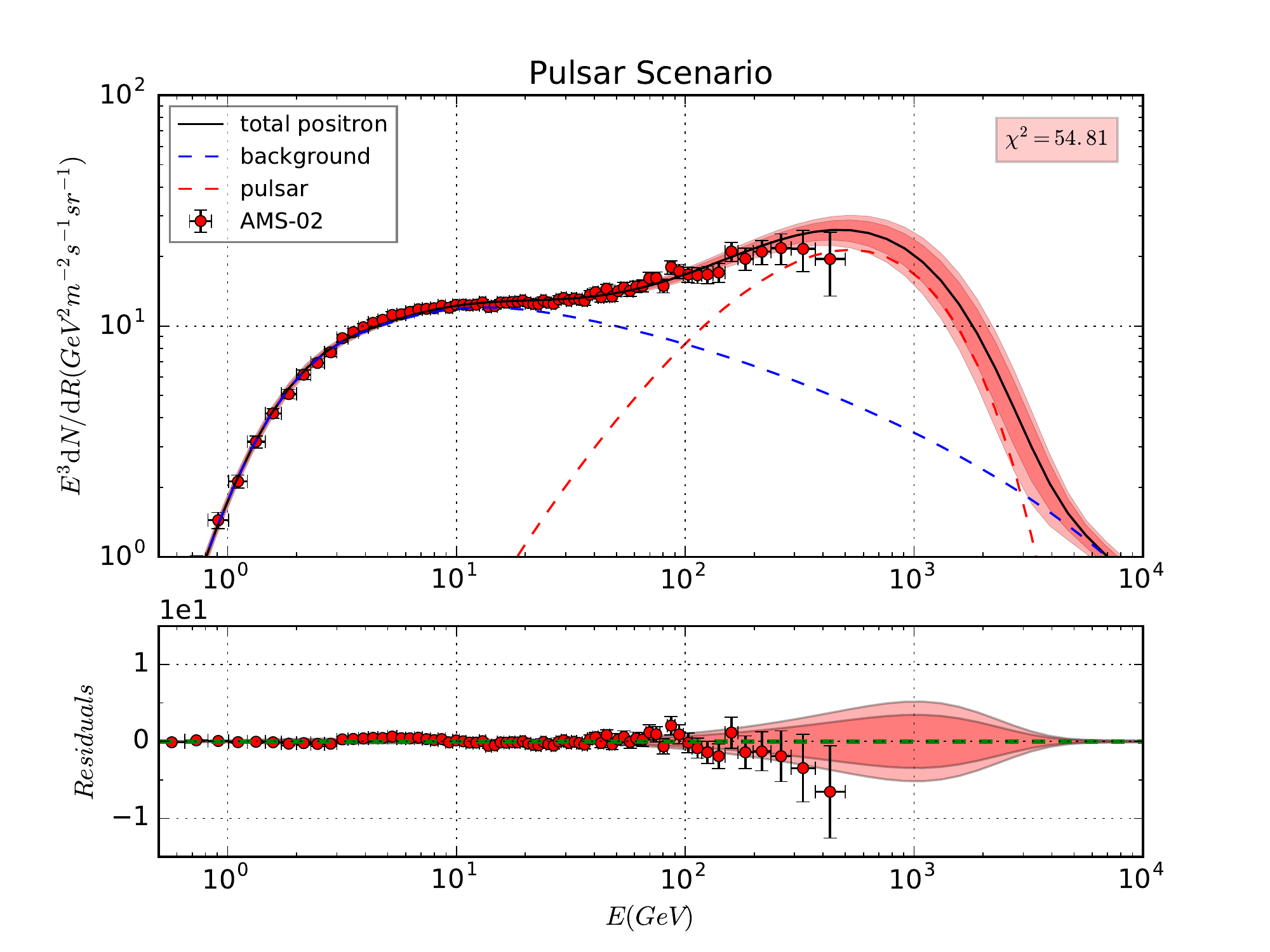}
  \includegraphics[width=0.46\textwidth]{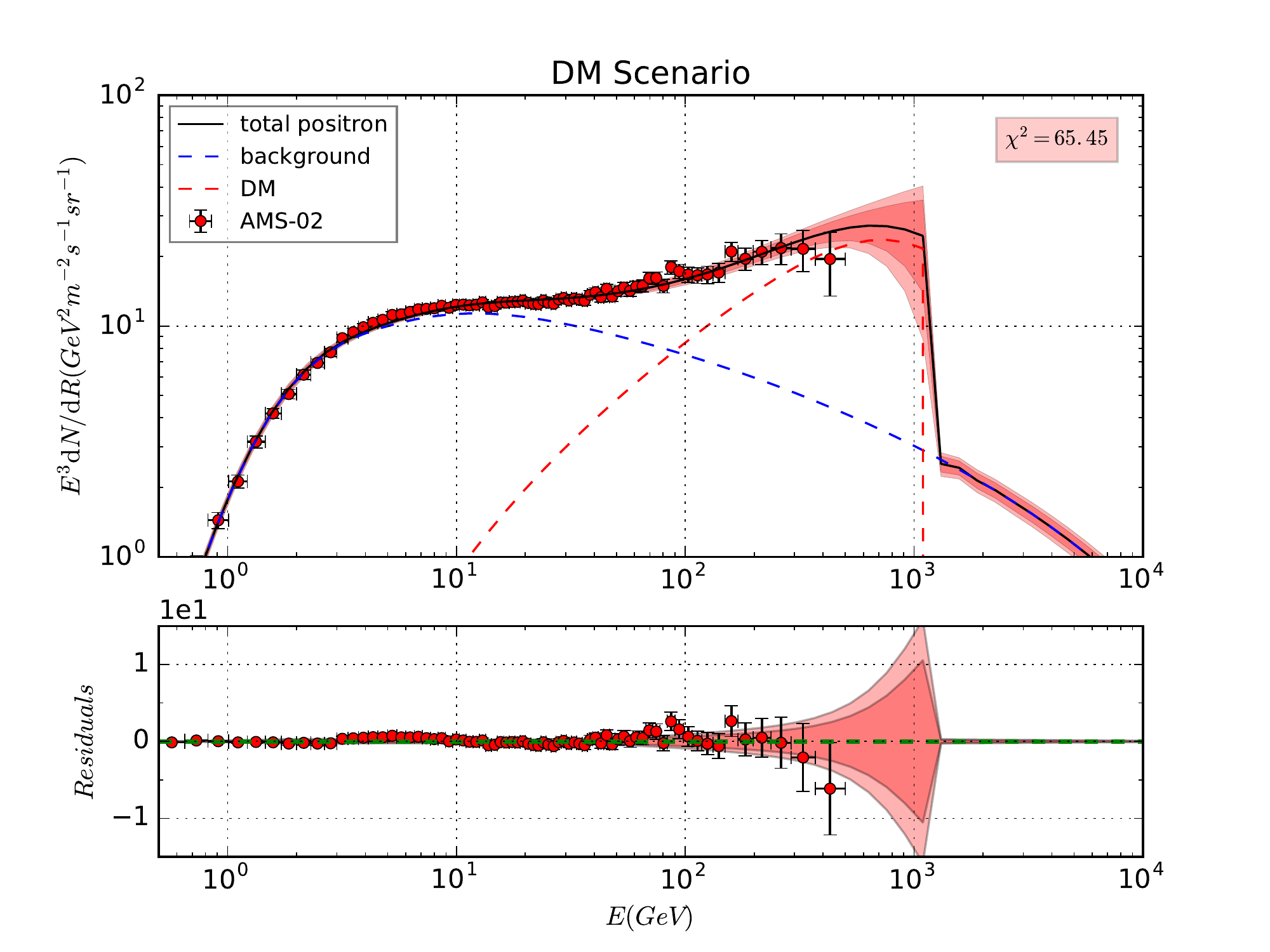}
  \includegraphics[width=0.46\textwidth]{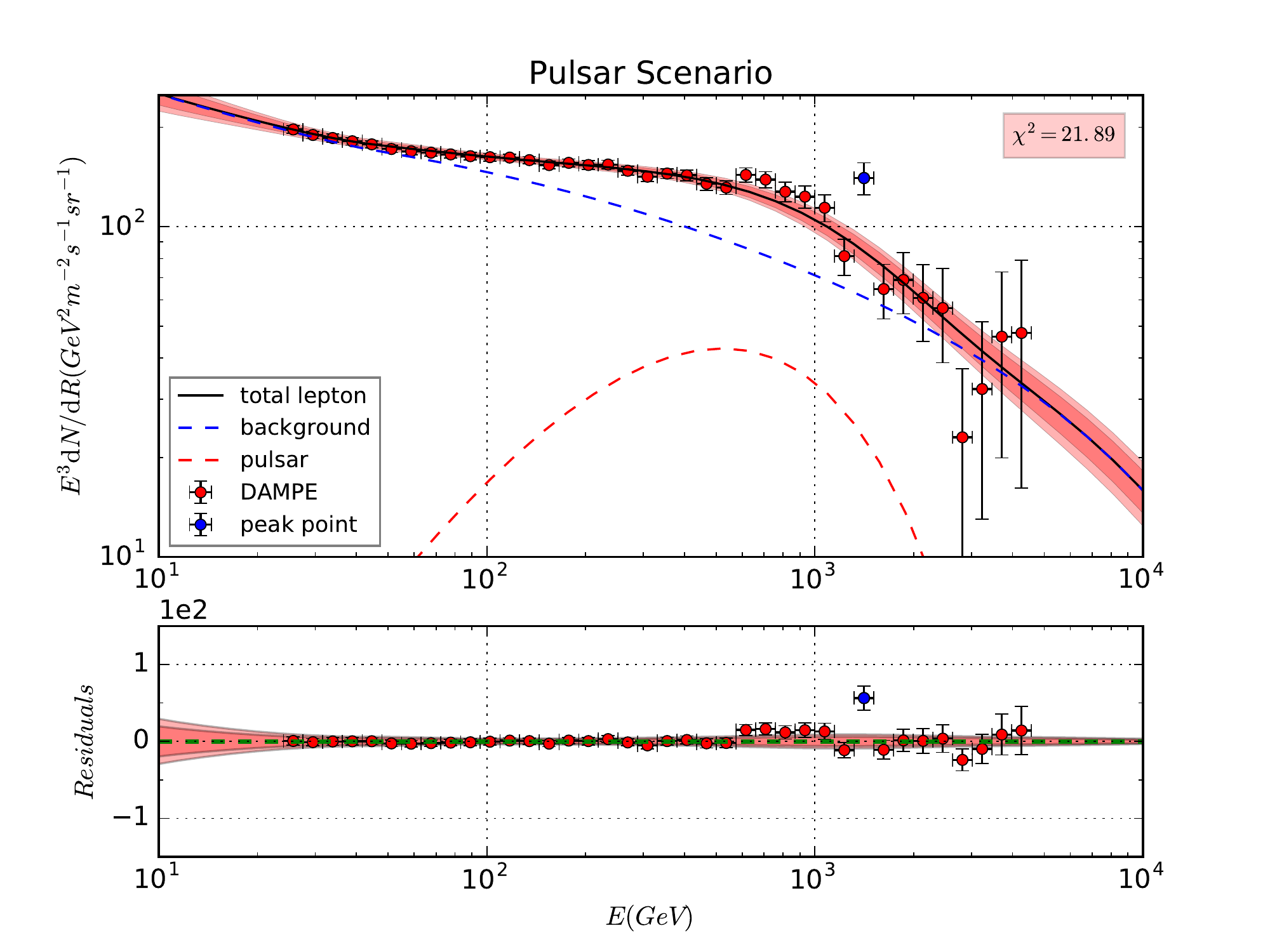}
  \includegraphics[width=0.46\textwidth]{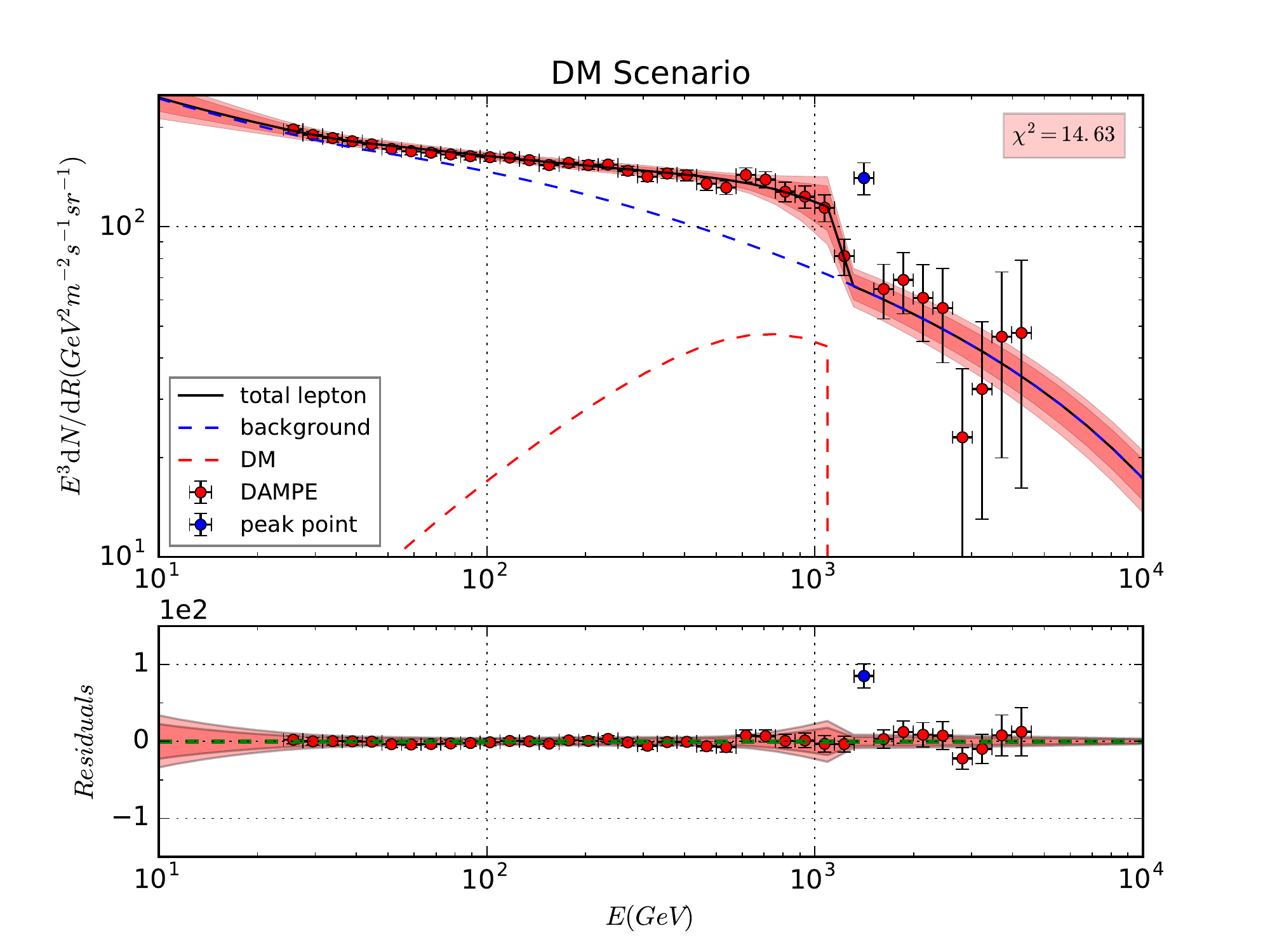}
  \caption{The global fitting results and the corresponding residuals to the AMS-02 positron flux and DAMPE lepton flux. The $2\sigma$ (deep red) and $3\sigma$ (light red) bound are also showed in the figures. The first column shows the fitting results of pulsar and the second shows the fitting results of DM. For DAMPE CREs flux only, we got $\chi^{2} = 21.89$ for pulsar scenario and $\chi^{2} = 14.63$ for DM scenario. }
  \label{fig:lepton_results}
\end{figure*}

The MCMC algorithm was used to determine the parameters in the 2 scenarios. When the Markov Chains have reached their equilibrium state, we take the samples of the parameters as their posterior PDFs. The best-fitting results and the corresponding residuals of the proton flux, helium flux and $\pbarp$ ratio for 2 scenarios are showed in Fig. \ref{fig:nuclei_results}, and the corresponding results of the positron and CREs flux are showed in Fig. \ref{fig:lepton_results}.  
The best-fit values, statistical mean values, standard deviations and allowed intervals at $95 \%$ CL for parameters in set $\boldsymbol{\theta}_{\psr}$ and $\boldsymbol{\theta}_{\DM}$  are shown in Table \ref{tab:params_psr} and Table \ref{tab:params_dm}, respectively. For best fit results of the global fitting, we got $\chi^{2}/d.o.f = 255.24/298 $ for pulsar scenario and $\chi^{2}/d.o.f = 276.56/296 $ for DM scenario. \footnote{Considering the correlations between different parameters, we could not get a reasonable reduced $\chi^{2}$ for each part of the data set independently. As a result, we showed the $\chi^{2}$ for each part of the data set in Figs. \ref{fig:nuclei_results}, \ref{fig:lepton_results}.  }

\begin{table*}[htb]
\begin{center}
\begin{tabular}{lllll}
  \hline\hline
ID  &Prior & Best-fit &Posterior mean and   &Posterior 95\%    \\
    &range&value  &Standard deviation & range  \\
\hline
$D_{0}\ (10^{28}\cm^{2}\s^{-1})$
    &[1, 20]  &14.37  &14.38$\pm$0.16     &[13.95, 14.74]   \\

$\delta$
  &[0.1, 1.0] &0.318  &0.317$\pm$0.003     &[0.311, 0.326]     \\

$z_h\ (\kpc)$
  &[0.5, 30.0]  &25.08  &25.13$\pm$0.22     &[24.55, 25.69]     \\

$v_{A}\ (\km/\s)$
  &[0, 80]  &41.34  &41.34$\pm$0.38     &[40.37, 42.32]    \\

\hline
$N_{p}\ \footnote{Post-propagated normalization flux of protons at 100 GeV in unit $10^{-2}\m^{-2}\s^{-1}\sr^{-1}\GeV^{-1}$}$
  &[1, 8] &4.46   &4.46$\pm$0.01    &[4.44, 4.49]  \\

$R_{\p1}\ (\GV)$
  &[1, 30]  &25.88  &25.78$\pm$0.20      &[25.43, 26.41]     \\
  
$R_{\p2}\ (\GV)$
  &[60, 1000]  &428.98  &429.05$\pm$7.44      &[409.86, 447.63]     \\
  
$\nu_{\p1}$
  &[1.0, 4.0] &2.196  &2.198$\pm$0.006     &[2.180, 2.209]    \\

$\nu_{\p2}$
  &[1.0, 4.0] &2.465  &2.464$\pm$0.005     &[2.453, 2.474]    \\

$\nu_{\p3}$
  &[1.0, 4.0] &2.348  &2.349$\pm$0.008     &[2.332, 2.368]    \\
  
$R_{\He1}\ (\GV)$
  &[1, 30]  &12.07  &12.09$\pm$0.15      &[11.67, 12.50]    \\
  
$R_{\He2}\ (\GV)$
  &[60, 1000]  &244.83  &246.41$\pm$8.14      &[220.09, 265.47]    \\

$\nu_{\He1}$
  &[1.0, 4.0] &2.186  &2.188$\pm$0.007     &[2.170, 2.199]     \\

$\nu_{\He2}$
  &[1.0, 4.0] &2.422  &2.422$\pm$0.005     &[2.411, 2.431]     \\

$\nu_{\He3}$
  &[1.0, 4.0] &2.219  &2.219$\pm$0.012     &[2.197, 2.241]     \\

\hline
$\phinuc\ (\GV)$
    &[0, 1.5] &0.73   &0.73$\pm$0.01    &[0.71, 0.76]   \\
$\phipbar\ (\GV)$
    &[0, 1.5] &0.28   &0.28$\pm$0.01    &[0.26, 0.30]   \\

$c_{\He}$
  &[0.1, 10.0] &3.93  &3.89$\pm$0.11     &[3.66, 4.22]   \\
$c_{\pbar}$
  &[0.1, 10.0] &1.37  &1.37$\pm$0.02     &[1.34, 1.41]   \\

\hline

$\log (N_{\e})\ \footnote{Post-propagated normalization flux of electrons at 25 GeV in unit $\m^{-2}\s^{-1}\sr^{-1}\GeV^{-1}$}$
  &[-4, 0] &-1.936   &-1.936$\pm$0.006    &[-1.950, -1.926]  \\

$\log (R_{\e} / \GV)$
  &[0, 3]  &1.64  &1.64$\pm$0.03      &[1.55, 1.75]     \\
  
$\nu_{\e1}$
  &[1.0, 4.0] &2.56  &2.57$\pm$0.02     &[2.50, 2.61]    \\

$\nu_{\e2}$
  &[1.0, 4.0] &2.39  &2.39$\pm$0.01     &[2.36, 2.42]    \\
  
$\log (N_{\psr})\ \footnote{Post-propagated normalization flux of electrons at 300 GeV in unit $\m^{-2}\s^{-1}\sr^{-1}\GeV^{-1}$}$
  &[-8, -4]  &-6.15  &-6.15$\pm$0.02      &[-6.19, -6.11]    \\
  
$\nu_{\psr}$
  &[0, 3.0] &0.65  &0.65$\pm$0.01     &[0.61, 0.69]     \\

$\log (R_{c} / \GV))$
  &[2, 5]  &2.81  &2.80$\pm$0.02      &[2.78, 2.86]    \\

\hline
$\phipos\ (\GV)$
    &[0, 1.5] &1.37   &1.37$\pm$0.01    &[1.36, 1.39]   \\

$\cpos$
  &[0.1, 10.0] &5.09  &5.08$\pm$0.05     &[5.03, 5.15]   \\

  \hline\hline
\end{tabular}
\end{center}
\caption{
Constraints on the parameters in set $\boldsymbol{\theta}_{\psr}$. The prior interval, best-fit value, statistic mean, standard deviation and the allowed range at $95\%$ CL are listed for parameters. With $\chi^{2}/d.o.f = 255.24/298 $ for best fit result.}
\label{tab:params_psr}
\end{table*}

\begin{table*}[htb]
\begin{center}
\begin{tabular}{lllll}
  \hline\hline
ID  &Prior & Best-fit &Posterior mean and   &Posterior 95\%    \\
    &range&value  &Standard deviation & range  \\
\hline
$D_{0}\ (10^{28}\cm^{2}\s^{-1})$
    &[1, 20]  &15.72  &15.76$\pm$0.14     &[15.47, 15.96]   \\

$\delta$
  &[0.1, 1.0] &0.307  &0.307$\pm$0.004     &[0.302, 0.313]     \\

$z_h\ (\kpc)$
  &[0.5, 30.0]  &28.59  &28.39$\pm$0.22     &[28.07, 28.78]     \\

$v_{A}\ (\km/\s)$
  &[0, 80]  &42.46  &42.60$\pm$0.48     &[41.69, 43.32]    \\

\hline
$N_{p}\ \footnote{Post-propagated normalization flux of protons at 100 GeV in unit $10^{-2}\m^{-2}\s^{-1}\sr^{-1}\GeV^{-1}$}$
  &[1, 8] &4.50   &4.48$\pm$0.02    &[4.45, 4.51]  \\

$R_{\p1}\ (\GV)$
  &[1, 30]  &23.18  &23.19$\pm$0.20      &[22.92, 23.60]     \\
  
$R_{\p2}\ (\GV)$
  &[60, 1000]  &497.28  &492.08$\pm$8.41      &[480.08, 507.07]     \\
  
$\nu_{\p1}$
  &[1.0, 4.0] &2.222  &2.226$\pm$0.009     &[2.212, 2.239]    \\

$\nu_{\p2}$
  &[1.0, 4.0] &2.477  &2.477$\pm$0.006     &[2.468, 2.486]    \\

$\nu_{\p3}$
  &[1.0, 4.0] &2.357  &2.352$\pm$0.009     &[2.338, 2.368]    \\
  
$R_{\He1}\ (\GV)$
  &[1, 30]  &11.06  &11.23$\pm$0.17      &[10.97, 11.57]    \\
  
$R_{\He2}\ (\GV)$
  &[60, 1000]  &237.29  &232.95$\pm$8.88      &[219.91, 248.52]    \\

$\nu_{\He1}$
  &[1.0, 4.0] &2.206  &2.207$\pm$0.008     &[2.196, 2.221]     \\

$\nu_{\He2}$
  &[1.0, 4.0] &2.435  &2.435$\pm$0.005     &[2.426, 2.443]     \\

$\nu_{\He3}$
  &[1.0, 4.0] &2.232  &2.232$\pm$0.013     &[2.213, 2.257]     \\

\hline
$\phinuc\ (\GV)$
    &[0, 1.5] &0.77   &0.78$\pm$0.01    &[0.76, 0.80]   \\
$\phipbar\ (\GV)$
    &[0, 1.5] &0.25   &0.26$\pm$0.01    &[0.24, 0.27]   \\

$c_{\He}$
  &[0.1, 10.0] &3.68  &3.56$\pm$0.11     &[3.38, 3.74]   \\
$c_{\pbar}$
  &[0.1, 10.0] &1.47  &1.47$\pm$0.02     &[1.44, 1.50]   \\

\hline

$\log (N_{\e})\ \footnote{Post-propagated normalization flux of electrons at 25 GeV in unit $\m^{-2}\s^{-1}\sr^{-1}\GeV^{-1}$}$
  &[-4, 0] &-1.940   &-1.943$\pm$0.007    &[-1.958, -1.928]  \\

$\log (R_{\e} /\GV)$
  &[0, 3]  &1.62  &1.63$\pm$0.04      &[1.57, 1.74]     \\
  
$\nu_{\e1}$
  &[1.0, 4.0] &2.55  &2.54$\pm$0.03     &[2.46, 2.60]    \\

$\nu_{\e2}$
  &[1.0, 4.0] &2.37  &2.37$\pm$0.01     &[2.34, 2.40]    \\
  
$\log (\Mdm / \GeV)$
  &[1, 6]  &3.082  &3.085$\pm$0.006      &[3.076, 3.096]    \\
  
$\log (\sigv)$\ \footnote{In unit $\cm^{3} \s^{-1}$}
  &[-28, -18] &-22.83  &-22.80$\pm$0.06     &[-22.93, -22.70]     \\

$\etae$
  &[0, 1]  &0.484  &0.479$\pm$0.007      &[0.466, 0.488]    \\
  
$\etamu$
  &[0, 1]  &0.508  &0.508$\pm$0.008      &[0.493, 0.518]    \\
  
$\etatau$
  &[0, 1]  &0.008  &0.013$\pm$0.010      &[0.001, 0.032]    \\

\hline
$\phipos\ (\GV)$
    &[0, 1.5] &1.32   &1.31$\pm$0.01    &[1.296, 1.332]   \\

$\cpos$
  &[0.1, 10.0] &5.02  &5.03$\pm$0.03     &[4.97, 5.08]   \\

  \hline\hline
\end{tabular}
\end{center}
\caption{
The same as Table. \ref{tab:params_psr}, but for the ones in set $\boldsymbol{\theta}_{\DM}$. With $\chi^{2}/d.o.f = 276.56 / 296 $ for best fit result.}
\label{tab:params_dm}
\end{table*}

In Fig. \ref{fig:nuclei_results}, we can see that the nuclei data is perfectly reproduced, which would provide a good precondition for the subsequent fitting on the lepton data. The proton and helium particles $\gtrsim \TeV$ would produce the secondary particles (including anti-protons and positrons) in lower energy range. Although the CREAM proton and helium data in $\gtrsim \TeV$ has a relative large uncertainties, the spectral hardening at $\sim 300 \GeV$ is accounted and then its influence on secondary products is included.

The best-fitting results and the corresponding residuals of the lepton and positron spectra are showed in Fig. \ref{fig:lepton_results}. The corresponding best-fit values, statistical mean values, standard deviations and allowed intervals at $95 \%$ CL for these parameters are shown in Table \ref{tab:params_psr} and Table \ref{tab:params_dm}.

In Fig. \ref{fig:lepton_results}, the lepton data can be fitted within fitting uncertainties. Although we got smaller reduced $\chi^{2}$ from global fitting on pulsar scenario, if we consider the DAMPE CREs flux alone, the best fit results shows $\chi^{2} = 21.89$ for pulsar scenario and $\chi^{2} = 14.63$ for DM scenario.

\subsection{Propagation parameters}

\begin{figure}
\centering
\includegraphics[width=0.5\textwidth]{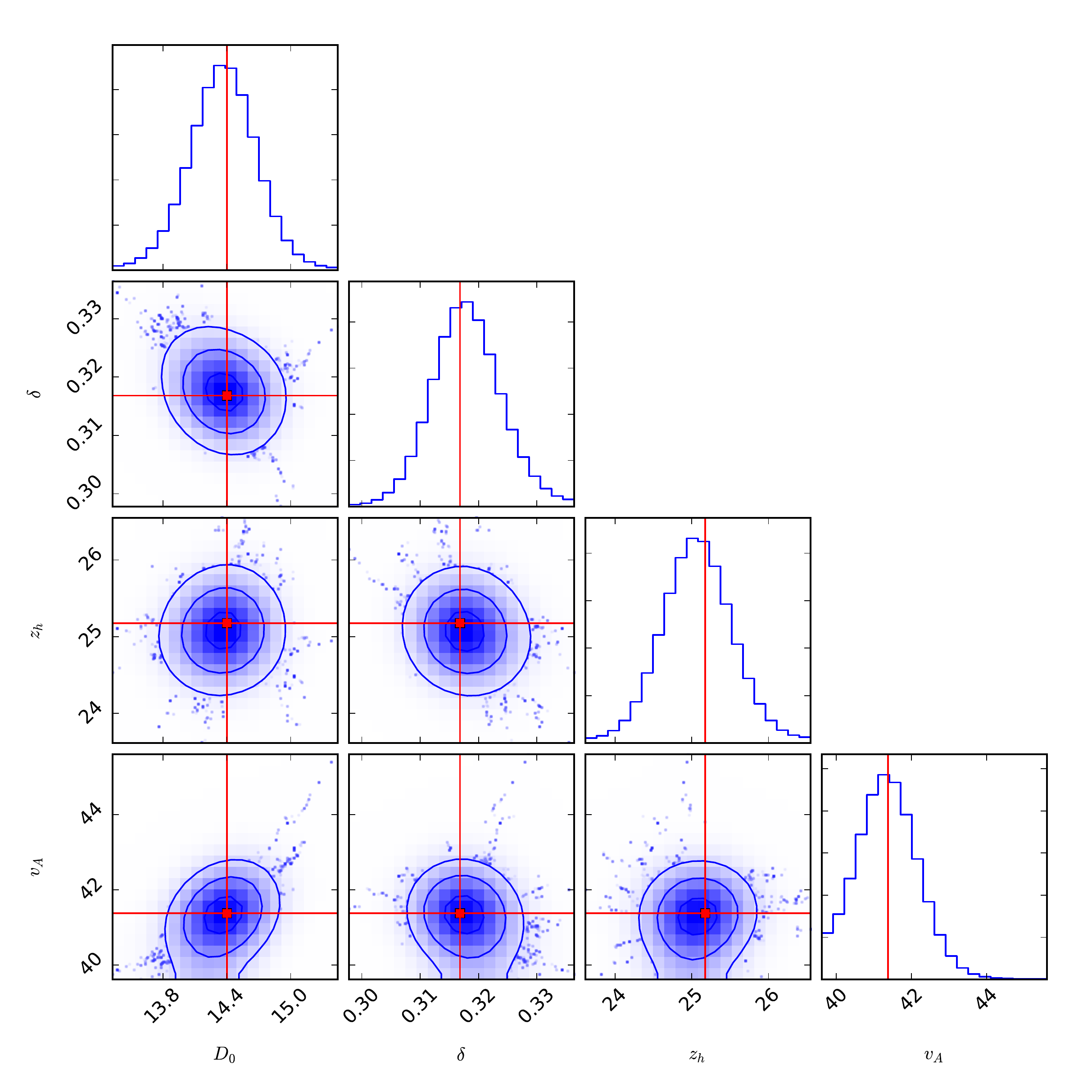}
\caption{Fitting 1D probability and 2D credible regions of posterior PDFs for the combinations of all propagation parameters for pulsar scenario. The regions enclosing $\sigma$, $2\sigma$ and $3\sigma$ CL are shown in step by step lighter blue. The red cross lines and marks in each plot indicates the best-fit value (largest likelihood).}
\label{fig:para_prop_psr}
\end{figure}

\begin{figure}
\centering
\includegraphics[width=0.5\textwidth]{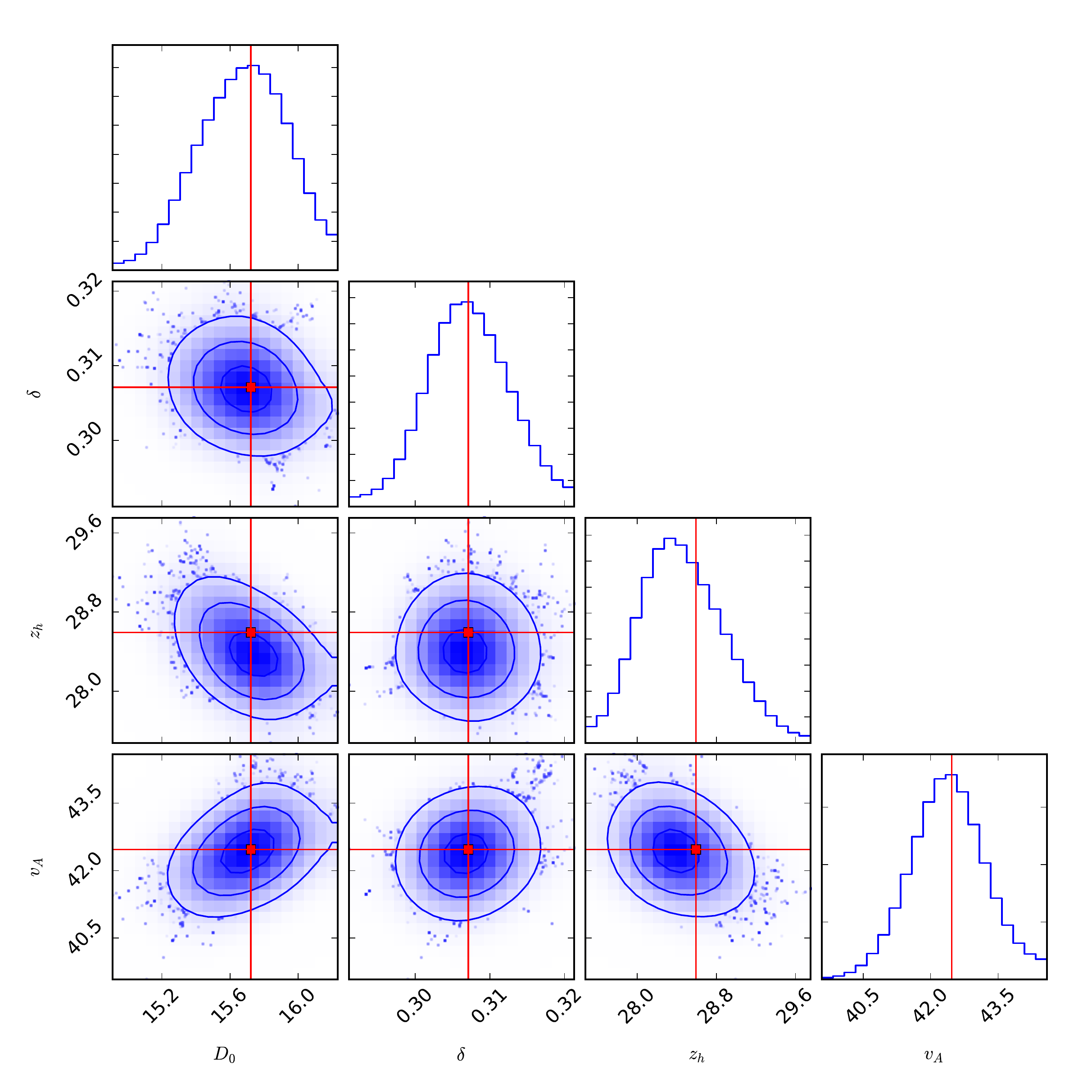}
\caption{Same as Fig. \ref{fig:para_prop_psr} but for DM scenario.}
\label{fig:para_prop_dm}
\end{figure}

The results of posterior probability distributions of the propagation parameters are shown in  Fig. \ref{fig:para_prop_psr} (for pulsar scenario) and Fig. \ref{fig:para_prop_dm} (for DM scenario).

In this work, we adapt the widely used diffusion-reacceleration model to describe the propagation process, and the relevant propagation parameter are $D_{0}$, $\delta$, $z_{h}$, and $v_{A}$. The obtained posterior PDFs are different from previous works to some extent. The classical degeneracy between $D_{0}$ and $z_{h}$ is not obvious due to the data set in this work, but both of them get larger best fit values than previous works. This is because (i) the $D_{0}$ defined in the {\sc dragon} (which represents the perpendicular diffusion coefficient $D_{\perp}$) is not the same as that in {\sc galprop} (which represents the isotropic diffusion coefficient); (ii) the sensitivity region which could breaks the degeneracy between $D_{0}$ and $z_{h}$ is different between $\pbarp$ (10 - 100 GeV) and B/C ($\lesssim 10 \GeV$). The observed AMS-02 $\pbarp$ ratio favors larger $D_{0}$ and $z_{h}$ values.

The $\delta$ value obtained in this work is smaller than some of the previous works because we use one more break in the primary source injection of proton ($\sim 240 \GV$) and helium ($\sim 420 - 500 \GV$) to account for the observed hardening in their observed spectra, other than use only one break and let $\delta$ compromise the different slopes in high energy regions ($\gtrsim 240 - 500  \GV$) (see, e.g., \citet{Niu2017}). In such configuration, we also got smaller fitting uncertainties on $\delta$ ($\sim 0.03$).

Moreover, the fitting results favor relative large values of $v_{A}$, which may not only comes from the constraints of nuclei data in low energy regions, but also the positron data as well.

\subsection{Primary source injection parameters}

\begin{figure*}
\centering
\includegraphics[width=0.9\textwidth]{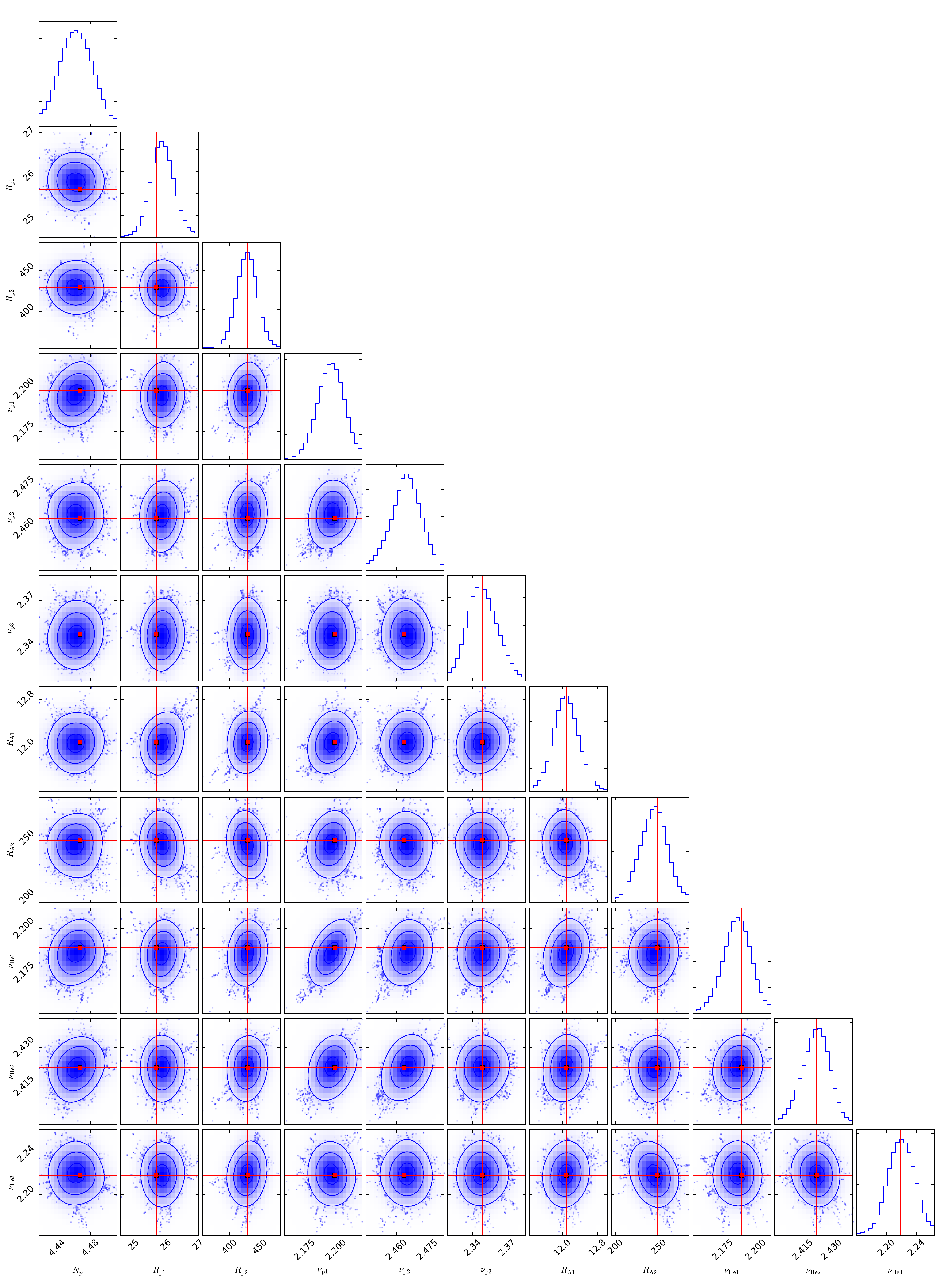}
\caption{Fitting 1D probability and 2D credible regions of posterior PDFs for the combinations of nuclei primary source injection parameters for pulsar scenario. The regions enclosing $\sigma$, $2\sigma$ and $3\sigma$ CL are shown in step by step lighter blue. The red cross lines and marks in each plot indicates the best-fit value (largest likelihood).}
\label{fig:para_pri_nuc_psr}
\end{figure*}

\begin{figure*}
\centering
\includegraphics[width=0.9\textwidth]{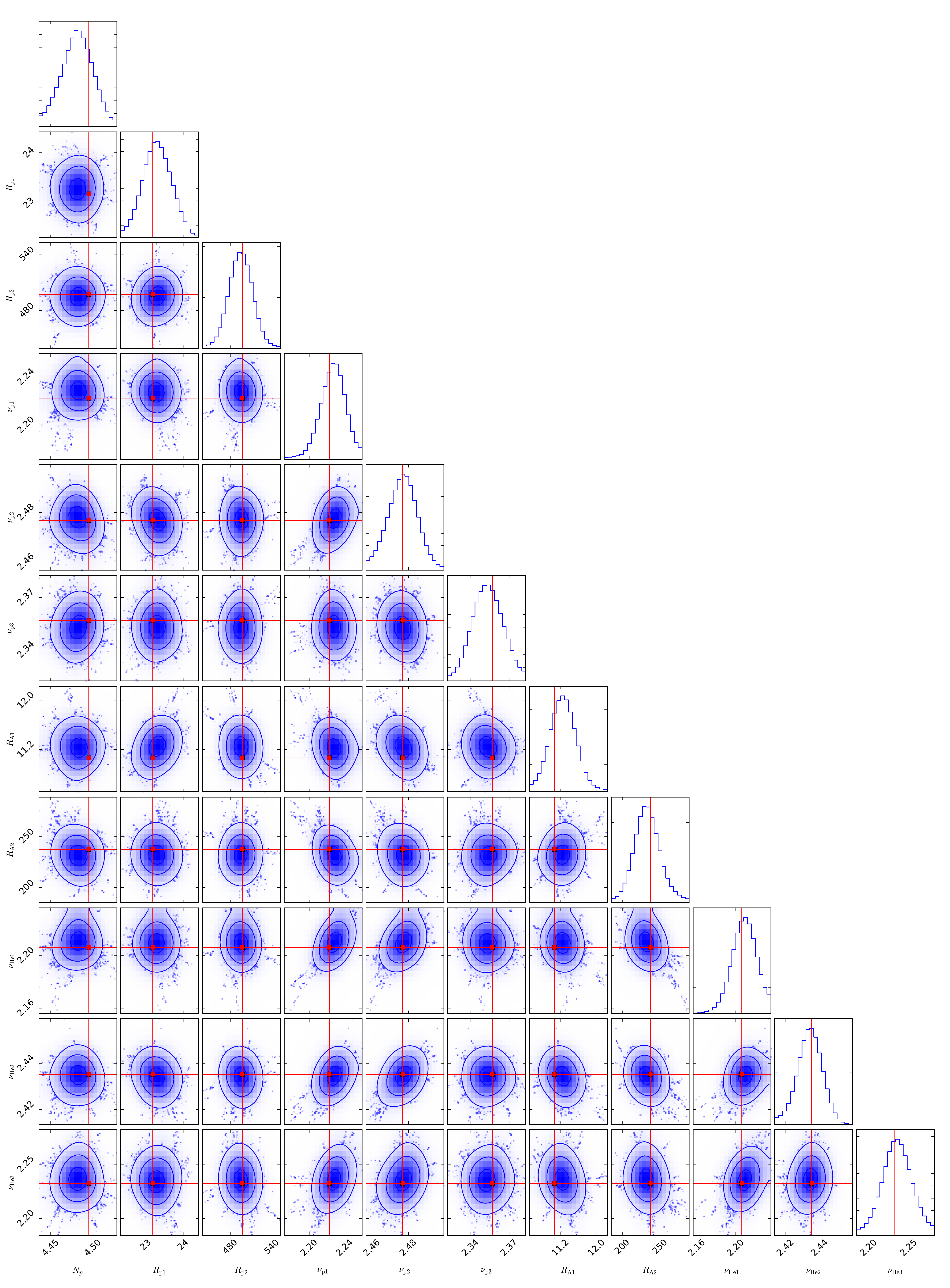}
\caption{Same as Fig. \ref{fig:para_pri_nuc_psr} but for DM scenario.}
\label{fig:para_pri_nuc_dm}
\end{figure*}

\begin{figure}
\centering
\includegraphics[width=0.5\textwidth]{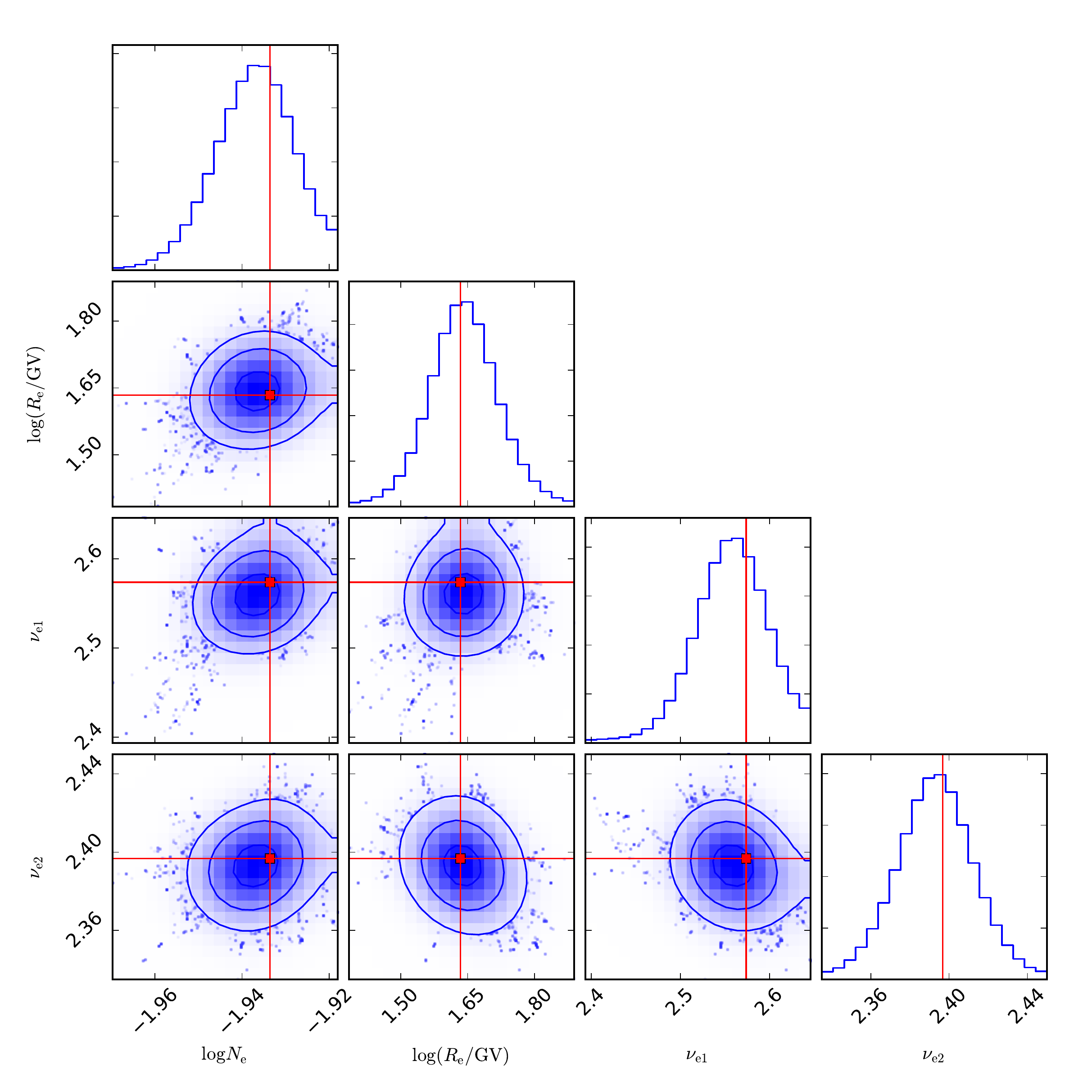}
\caption{Fitting 1D probability and 2D credible regions of posterior PDFs for the combinations of electron primary source injection parameters for pulsar scenario. The regions enclosing $\sigma$, $2\sigma$ and $3\sigma$ CL are shown in step by step lighter blue. The red cross lines and marks in each plot indicates the best-fit value (largest likelihood).}
\label{fig:para_pri_lep_psr}
\end{figure}

\begin{figure}
\centering
\includegraphics[width=0.5\textwidth]{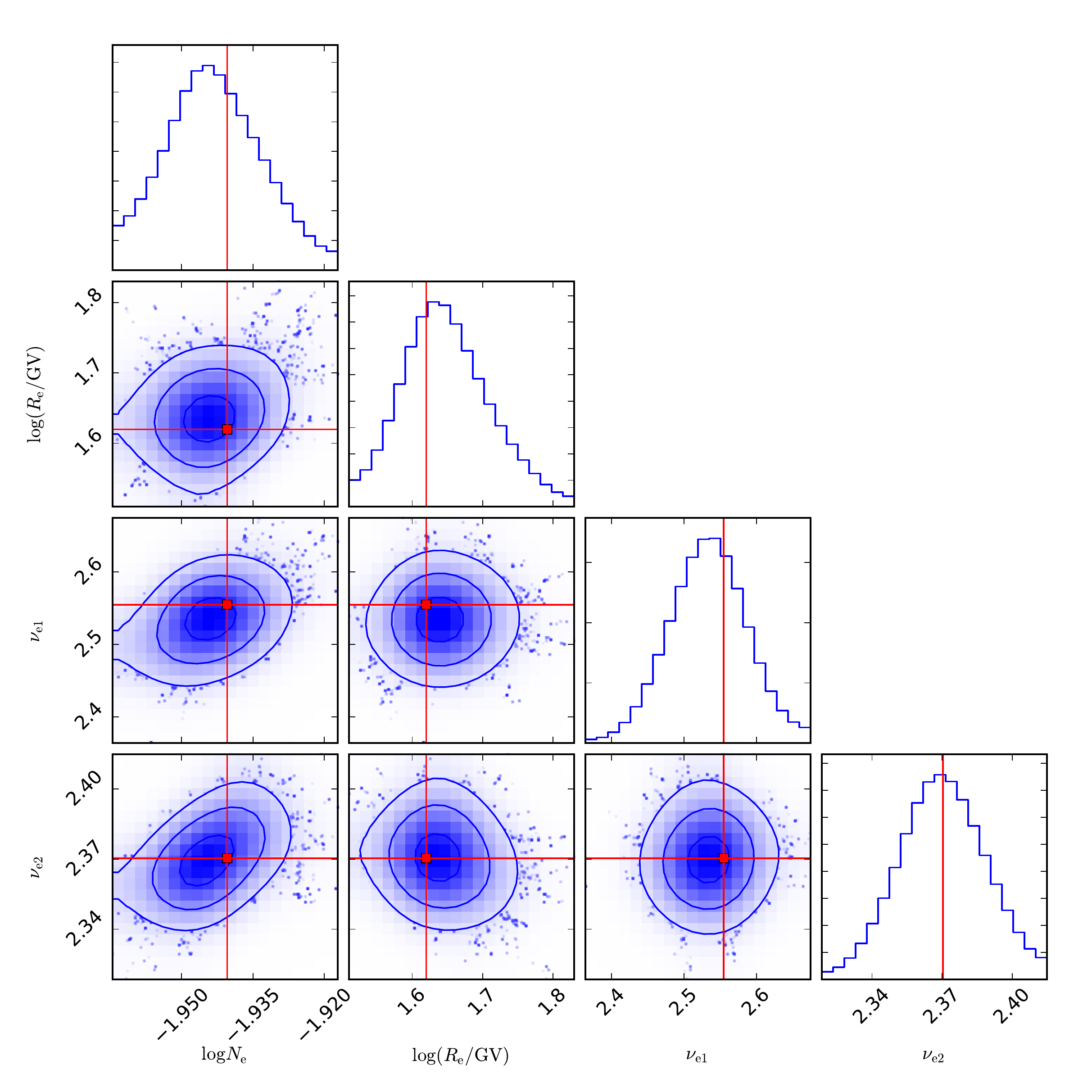}
\caption{Same as Fig. \ref{fig:para_pri_lep_psr} but for DM scenario.}
\label{fig:para_pri_lep_dm}
\end{figure}

The results of posterior probability distributions of the primary source parameters are shown in Figs. \ref{fig:para_pri_nuc_psr} (proton and helium, for pulsar scenario),  \ref{fig:para_pri_nuc_dm} (proton and helium, for DM scenario), and Figs. \ref{fig:para_pri_lep_psr} (electron, for pulsar scenario),  \ref{fig:para_pri_lep_dm} (electron, for DM scenario).

Benefited from the 2 independent  breaks injection spectra for proton and helium, the observed data has been reproduced perfectly. The fitting result shows that the rigidity breaks and the slopes are obviously different between proton and helium spectra. This indicates that the cosmic ray physics has entered a precision-driven era and all these differences should be treated carefully in future studies.  On the other hand, we want to point out that the hardening of the nuclei spectra $\sim 300 \GeV$ could also be reproduced by other proposals, which focus on the propagation and diffusion effects rather than ascribing it to the acceleration near the source. These solutions include proposing a spatial dependent diffusion coefficient \citep{Tomassetti2012,Tomassetti2015prd,Feng2016}, or adding a high-rigidity break in the diffusion coefficient \citep{Genolini2017,Blasi2017,Reinert2018}. With the precise data obtained in future extending to higher energy regions, we would expect more details can be revealed on this theme.

Additionally, the electron primary source injection spectra can be described by a break power-law from 20 GeV to $10^{4}$ GeV (DAMPE data), with $\nu_{\e1} \in [2.54,2.57]$, $\nu_{\e2} \in [2.37,2.39]$, and $R_{\e} \in [38, 47] \GV$.

\subsection{Extra source parameters}

\begin{figure}
\centering
\includegraphics[width=0.5\textwidth]{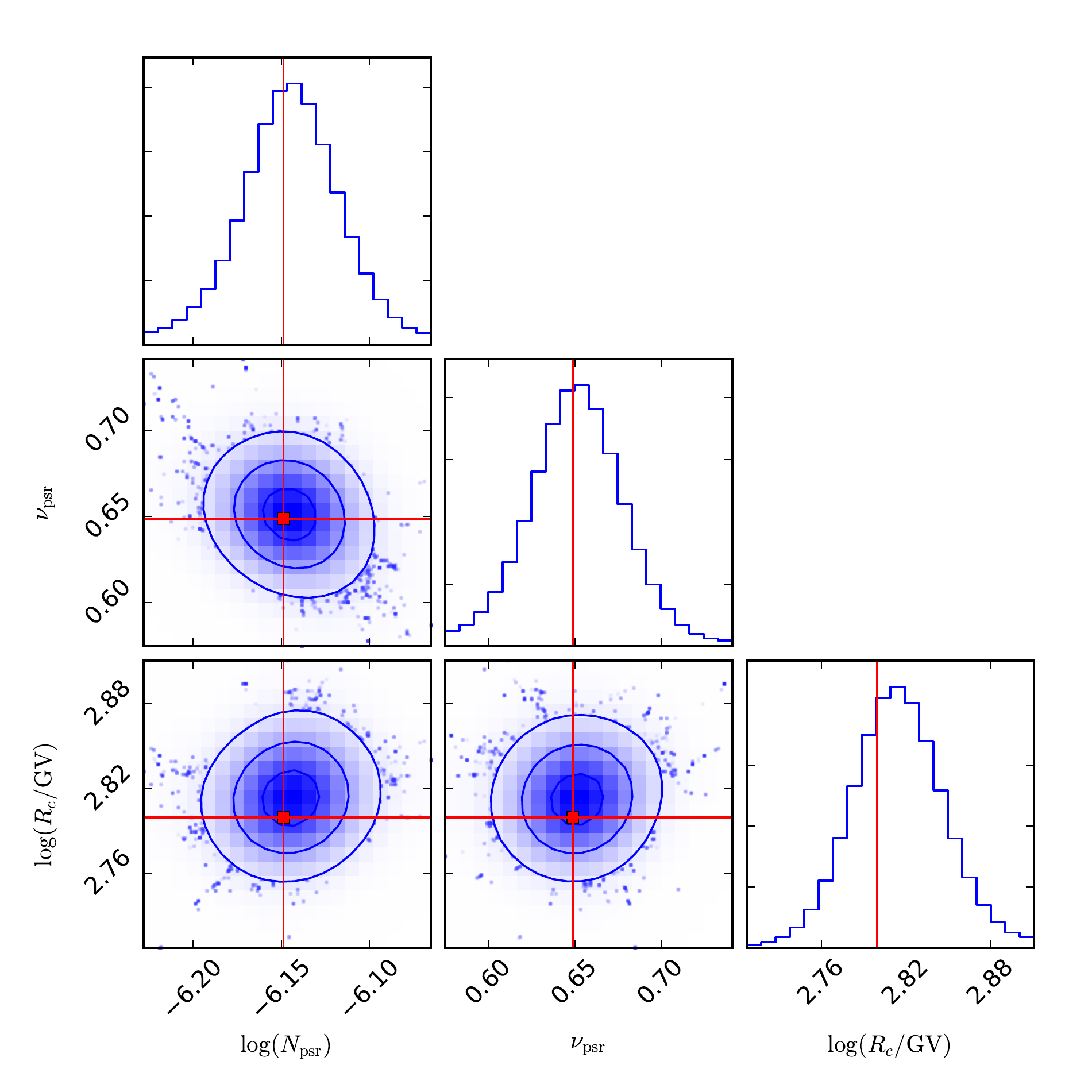}
\caption{Fitting 1D probability and 2D credible regions of posterior PDFs for the combinations of extra lepton soruce parameters from for pulsar scenario. The regions enclosing $\sigma$, $2\sigma$ and $3\sigma$ CL are shown in step by step lighter blue. The red cross lines and marks in each plot indicates the best-fit value (largest likelihood). }
\label{fig:para_extra_psr}
\end{figure}

\begin{figure}
\centering
\includegraphics[width=0.5\textwidth]{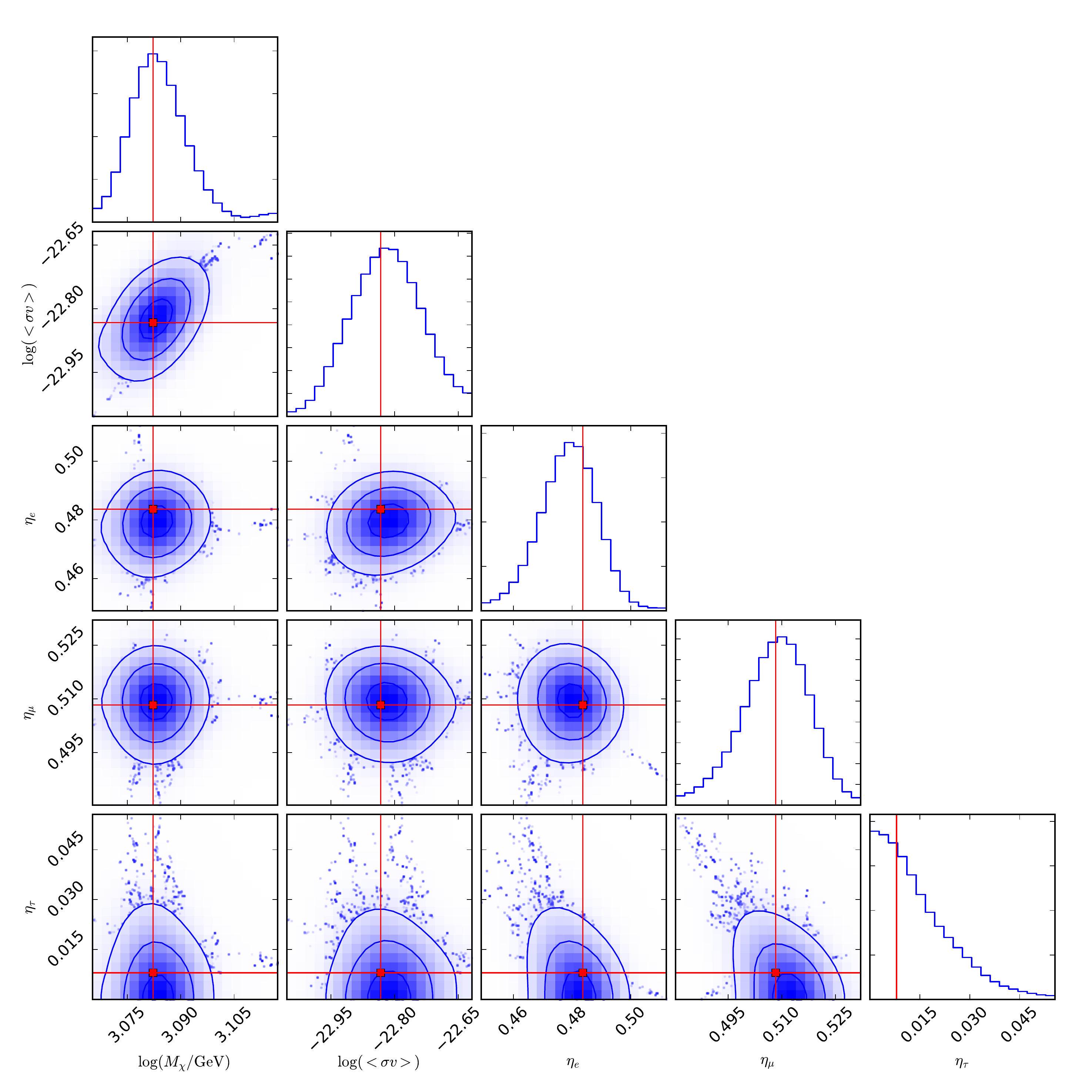}
\caption{Same as Fig. \ref{fig:para_extra_psr} but for DM scenario.}
\label{fig:para_extra_dm}
\end{figure}

The results for posterior probability distributions of the extra source parameters are shown in Figs. \ref{fig:para_extra_psr} (for pulsar scenario),  \ref{fig:para_extra_dm} (for DM scenario).

For the pulsar scenario, the fitting results give $\nu_{\psr} \simeq 0.65$, which is obviously different from the fitting results in previous works (see for e.g., \citep{Profumo2012}). In standard pulsar models, the injection spectrum indices of CREs from pulsars are always in the range $\nu_{\psr} \in [1.0,2.4]$ \citep{Reynolds1988,Thompson1994,Fierro1995}. As a result, more attention should be paid in future researches. This may indicate: (i) there is something wrong or inaccuracy with the classical pulsar CRE injection model; (ii) the CRE excess is not contributed dominatly by pulsars. Moreover, the rigidity cut-off is $R_c \simeq 646$~GV. 

For the DM scenario, we obtain $\sigv \simeq 1.48 \times 10^{-23} \cm^{2} \s^{-1}$ and $\Mdm \simeq 1208 \GeV$. The value of $\sigv$ is about 3 orders larger than that of thermal DM \citep{Jungman1996}. Moreover, we have $\etae \simeq 0.484$, $\etamu \simeq 0.508 $, and $\etatau \simeq 0.008 $, which is obviously different from the fitting results obtained from AMS-02 lepton data alone (see for e.g., \citet{Lin2015}). Consequently, the DM annihilation into $\tautaubar$ is highly suppressed, which provides some hints to construct an appropriate DM model (see for e.g., \citep{Chao2017}).

Because we have $\etae \simeq 0.484$, $\etamu \simeq 0.508 $, and $\etatau \simeq 0.008 $, the constraints from the Fermi-LAT observations on dwarf spheroidal galaxies~\citep{Ackermann:2011wa, GeringerSameth:2011iw, Tsai:2012cs, Ackermann:2015zua, Li:2015kag, Profumo2017} can be avoided~\citep{Yuan2017_dampe}. In order to escape the constraints from the Planck observations of CMB anisotropies~\citep{Ade:2015xua}, the Breit-Wigner mechanism~\citep{Feldman:2008xs, Ibe:2008ye, Guo:2009aj, Bi:2009uj, Bi:2011qm, Hisano:2011dc, Bai:2017fav, Xiang:2017jou} could be employed and the dark $U(1)_D$ model (where the SM fermions and Higgs fields are neutral under it) is considered. We introduce one SM singlet field $S$, one chiral fermionic dark matter particle $\chi$, and three pairs of the vector-like particles (${\widehat{XE}}_i, {\widehat{XE}}_i^c)$, whose quantum numbers under the $SU(3)_C\times SU(2)_L \times U(1)_Y \times U(1)_D$ are
\begin{eqnarray}
& S: (\mathbf{1}, \mathbf{1}, \mathbf{0}, \mathbf{2})~,~~~
\chi: (\mathbf{1}, \mathbf{1}, \mathbf{0}, \mathbf{-1}) \nonumber \\
& {\widehat{XE}}_i: (\mathbf{1}, \mathbf{1}, \mathbf{-1}, \mathbf{-2})~,~~~
{\widehat{XE}}^c_i: (\mathbf{1}, \mathbf{1}, \mathbf{1}, \mathbf{2}) ~.~\,
\end{eqnarray}
The relevant Lagrangian is 
\begin{eqnarray}
-{\cal L} &=& -m_S^2 |S|^2 + \frac{\lambda}{2} |S|^4 + \left(M^V_{ij} {\widehat{XE}}^c_i {\widehat{XE}}_j
\right. \nonumber \\  && \left. + y_{ij} S {\widehat E}_i^c {\widehat{XE}}_j +y S  \chi \chi + {\rm H.C.}\right) ~,~\,
\end{eqnarray}
where ${\widehat E}_i^c$ are the right-handed charged leptons.

For simplicity, we choose $M^V_{ij} = M^V_{i} \delta_{ij}$ and $y_{ij} = y_i \delta_{ij}$. After $S$ acquires a Vacuum Expectation Value (VEV), the $U(1)_D$ gauge symmetry is broken down to a $Z_2$ symmetry under which $\chi$ is odd. Thus, $\chi$ is a DM matter candidate. For simplicity, we assume that the mass of $U(1)_D$ gauge boson is about twice of $\chi$ mass, {\it i.e.}, $M_{Z'} \simeq 2 m_{\chi}$, while the Higgs field $S$ and vector-like particles are heavier than $M_{Z'}$. Moreover,  ${\widehat E}_i^c$ and ${\widehat{XE}}_i^c$ will be mixed due to the $M^V_{i} {\widehat{XE}}^c_i {\widehat{XE}}_i$ and $y_{i} S {\widehat E}_i^c {\widehat{XE}}_i$ terms, and we obtain the mass eigenstates $E_i^c$ and $XE_i^c$ by neglecting the tiny charged lepton masses
\begin{eqnarray}
\left(
\begin{array}{c}
E_i^c \\
XE_i^c
\end{array} \right)=
\left(
\begin{array}{cc}
\cos\theta_i & \sin\theta_i \\
-\sin\theta_i & \cos\theta_i
\end{array}
\right)
\left(
\begin{array}{c}
{\widehat E}_i^{c} \\
{\widehat{XE}}_i^{c \prime}
\end{array} \right)
~,~\,
\end{eqnarray}
where $\tan\theta_i = -y\langle S \rangle/M^V_i$.

Neglecting the charged lepton masses again, we obtain
\begin{eqnarray}
\sigma v = \sum_{i=1}^3\frac{g'^4\sin^2\theta_i}{6\pi}\frac{s-m_\chi^2}{(s-m_{Z'}^2)^2+(m_{Z'}\Gamma_{Z'})^2}~,~\,
\end{eqnarray}
where $m_\chi = y \langle S \rangle$, and $g'$ and $M_{Z'}$ are the gauge coupling and gauge boson mass
for $U(1)_D$ gauge symmetry.

For $m_{Z'}\simeq 2m_\chi$, $Z'$ decays dominantly into leptons, and the decay width is 
\begin{eqnarray}
\Gamma_{Z'}=\sum_{i=1}^3 \frac{g'^2\sin^2\theta_i}{6\pi} m_{Z'}~.~\,
\end{eqnarray}

To explain the DM best fit results, we can choose proper values of $g'$, $\frac{m_{Z'}-2m_\chi}{m_{Z'}}$, $\sin\theta_e$, $\sin\theta_\mu$, and $\sin\theta_\tau$ to reproduce the values of $\Mdm$, $\sigv$ and $\eta_e :\eta_\mu : \eta_\tau$ like that in \citet{Niu:2017lts}.

\subsection{Nuisance parameters}

\begin{figure}
\centering
\includegraphics[width=0.5\textwidth]{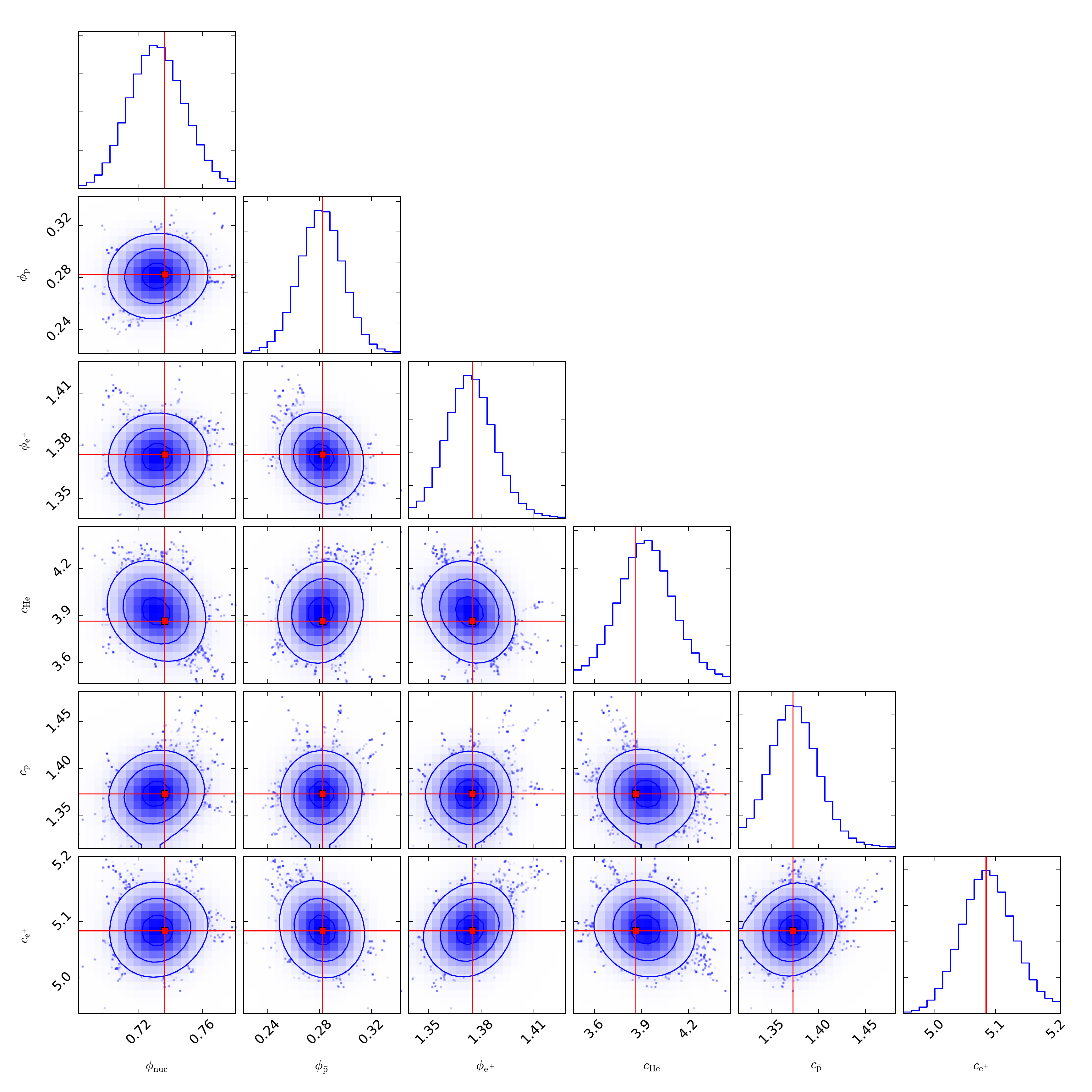}
\caption{Fitting 1D probability and 2D credible regions of posterior PDFs for the combinations of nuisance parameters for pulsar scenario. The regions enclosing $\sigma$, $2\sigma$ and $3\sigma$ CL are shown in step by step lighter blue. The red cross lines and marks in each plot indicates the best-fit value (largest likelihood). }
\label{fig:para_nui_psr}
\end{figure}

\begin{figure}
\centering
\includegraphics[width=0.5\textwidth]{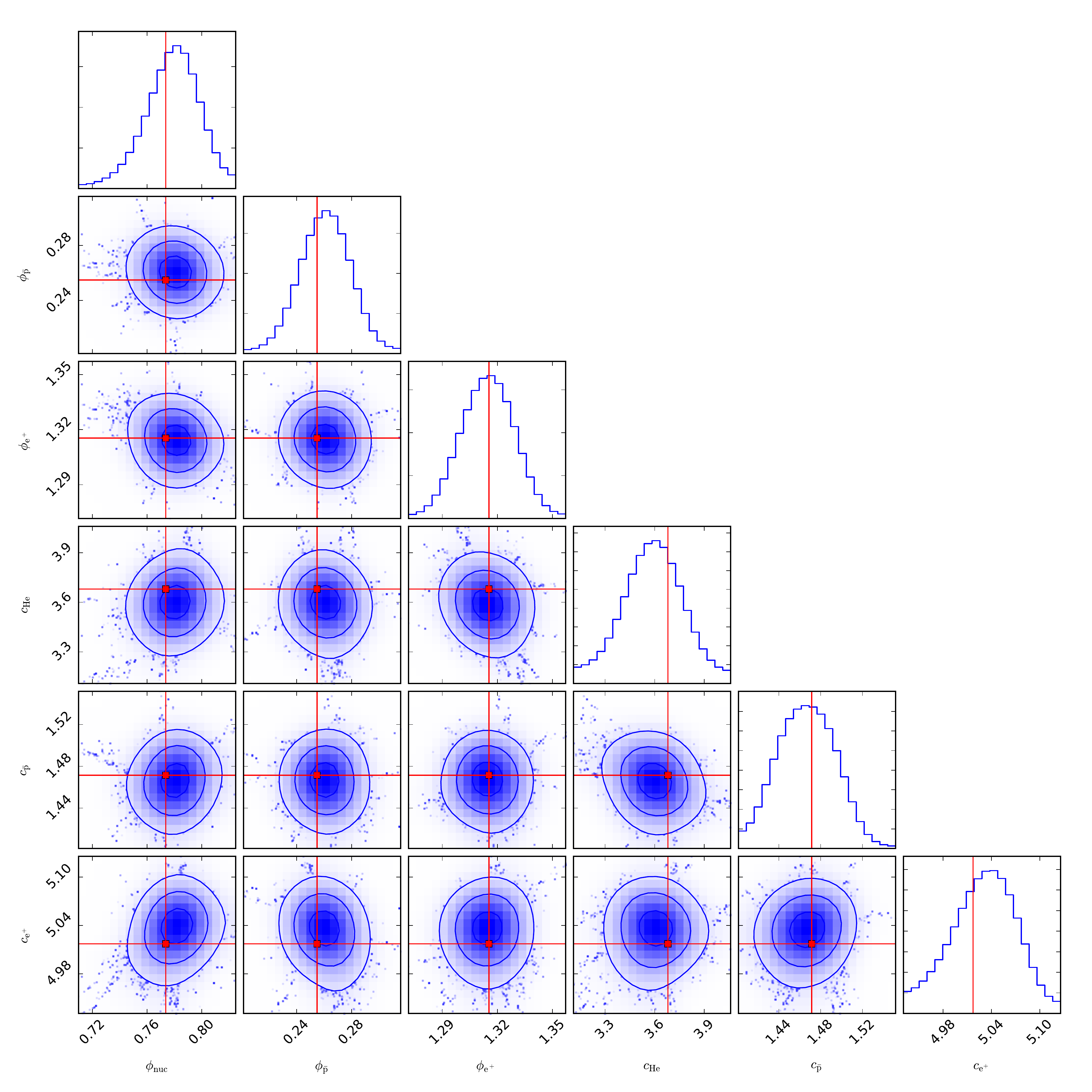}
\caption{Same as Fig. \ref{fig:para_nui_psr} but for DM scenario.}
\label{fig:para_nui_dm}
\end{figure}

In Figs \ref{fig:para_nui_psr} and \ref{fig:para_nui_dm}, the results of posterior probability distributions  represent the necessity to introduce them in the global fitting.

The different values of $\phinuc$, $\phipbar$, and $\phipos$ from the best-fit results represent not only the charge-sign dependent solar modulation (which has also been claimed by some previous works, see, e.g., \citet{Clem1996,Boella2001,Niu2017}), but also a species dependent solar modulation to some extent. As claimed in our previous works \citep{Niu2017}, the force field approximation could not describe the effects of solar modulation to all the species by a single $\phi$, but as an effective model, we can use an independent $\phi$ for each of the species. \footnote{In this work, we use a single $\phinuc$ to modulate the spectra of proton and helium simultaneously. Because a single $\phinuc$ could reproduce the low energy proton and helium spectra precisely under the precision of current data.} The different values of the $\phi$s for different species could reveal the hints to improve the propagation mechanisms of them in the heliosphere. Additionally, the proton, helium, and positron data have been collected from AMS-02 in the same period with a suggested $\phi$ from 0.50 - 0.62 GV \citep{AMS02_proton,AMS02_helium,AMS02_lepton}, which is based on data from the world network of sea level neutron monitors \citep{Usoskin2011}. More details in this field can be gotten in \citet{Corti2016}.

The value of $c_{\pbar} \sim 1.4 - 1.5$ could be explained by the uncertainties on the antiproton production cross section \citep{Tan1983,Duperray2003,Kappl2014,diMauro2014,Lin2016}.

The {\sc dragon} primary source isotopic abundances are inherited from {\sc galprop}, which are taken as the solar system abundances and iterated to achieve and agreement with the propagated abundances as provided by ACE at $\sim 200 \MeV$ nucleon. It is naturally that the normalized factor is different in different energy regions. On the other hand, we always focus on the shape of the spectrum, and $c_{\He}$ could be considered as an independent normalized factor as $N_{p}$, which is just identified as an nuisance parameter to get a better fitting result and not that important in this work.

For $\cpos$, there are several reasons which could ascribe its relative large values: (i) the cross section comes from \citet{Kamae:2004xx,Kamae:2006bf}, which needed a scale factor to correlate its values \citep{Evoli:2017vim}; (ii) the systematics between DAMPE CREs spectrum and AMS-02 positron spectrum is also partially accounted for in the parameter $\cpos$, which lead $\cpos$ not just a indicator of rescale factor on cross section. Moreover, we would like to point our that in this work, we focus on the extra sources which would reproduce the break at $\sim 1 \TeV$ in DAMPE CREs data. Some nuisance parameters ($c_{\pbar}$, $c_{\He}$, and $\cpos$) are employed to fit all the data consistently and precisely (especially the primary source and background, see for e.g., \citet{Lin2015}), which may not have clear physical meanings, but could also give us some hints to improve the details in CR physics in future research.

\section{Conclusion}

In this work, we did Bayesian analysis on the newly released CREs flux (exclude the peak signal at $\sim 1.4 \TeV$) from DAMPE to study the extra source properties in it. In order to deduct the primary electrons, secondary leptons in CREs flux consistently and precisely, we did a global fitting to reproduce the proton flux (from AMS-02 and CREAM), helium flux (from AMS-02 and CREAM), $\pbarp$ ratio (from AMS-02), positron flux (from AMS-02) and CREs flux (from DAMPE) simultaneously. Two independent extra source scenarios are considered, which account the excess of leptons to continuously distributed pulsars in the galaxy and dark matter annihilation (via leptonic channels) in the galactic halo. Both of these scenarios can fit the DAMPE CREs flux within the fitting uncertainties, while DM scenario gave a smaller $\chi^{2}$ and a obvious break at $\sim 1 \TeV$.

Additionally, in the DM scenario, the fitting result gives a dark matter particle's mass $\Mdm \sim 1208 \GeV$ and a cross section $\sigv \sim 1.48 \times 10^{-23} \cm^{3} \s^{-1}$. This is benefited from the break at $\sim 1 \TeV$. In such situations, the cross section in this work still should have a suppress factor to meet the value $\sigv \sim 3 \times 10^{-26} \cm^{3} \s^{-1}$. This discrepancy can be resolved by some proposed mechanisms like the non-thermal production of the DM \citep{Jeannerot1999,Lin2001,Yuan2012}, the Sommerfeld enhancement mechanism \citep{Sommerfeld1931,Hisano2005,Arkani-Hamed2009}, and Breit-Wigner type resonance of the annihilation interaction \citep{Griest1990,Gondolo1990}. 
What's more interesting, the constraints on the annihilation branching fraction shows the $\tautaubar$ annihilation channel is strongly suppressed, while the $\eebar$ and $\mumubar$ channels are almost equally weighted ($\etae = 0.484$, $\etamu = 0.508$, and $\etatau = 0.008$).  This would give some hints for constructing DM models, and we tried to build one in this work to meet the fitting results. 

{\bf Note:} In this work, we can see that the CREs spectrum from DAMPE without the peak can be reproduced by DM scenarios precisely. On the other hand, the spectrum with the peak also can be reproduced by DM annihilation from a local DM sub-structure~\citep{Yuan2017_dampe,Athron2017,Fan2017,Duan2017,Gu2017,Liu2017,Cao2017,Jin2017_dampe,Huang2017,Yang2017,Ge2017}. Both of these situations call for DM particles with $\Mdm \sim 1 - 2 \TeV$. Other independent detection strategy is needed to distinguish the excess in the CREs spectrum which can also be produced from some astrophysical sources~\citep{Fang2017,Yuan2017_dampe,Cholis2017}. Our recent works~\citep{Niu2017_DAV} proposed a novel scenario to probe the interaction between DM particles and electrons with $5 \GeV \lesssim \Mdm \lesssim 10 \TeV$.

\section*{ACKNOWLEDGMENTS}
We would like to thank \citet{Maurin2014} to collect database and associated online tools  for charged cosmic-ray measurements, and  \citet{corner} to provide us the tool to visualize multidimensional samples using a scatterplot matrix. Many thanks for the referees’ valuable and detailed suggestions, which led to a great progress in this work. This research was supported in part by the Projects 11475238 and 11647601 supported by National  Science Foundation of China,
and by Key Research Program of Frontier Sciences, CAS. 
The calculation in this paper are supported by HPC Cluster of SKLTP/ITP-CAS.


\begin{thebibliography}{130}%
\makeatletter
\providecommand \@ifxundefined [1]{%
 \@ifx{#1\undefined}
}%
\providecommand \@ifnum [1]{%
 \ifnum #1\expandafter \@firstoftwo
 \else \expandafter \@secondoftwo
 \fi
}%
\providecommand \@ifx [1]{%
 \ifx #1\expandafter \@firstoftwo
 \else \expandafter \@secondoftwo
 \fi
}%
\providecommand \natexlab [1]{#1}%
\providecommand \enquote  [1]{``#1''}%
\providecommand \bibnamefont  [1]{#1}%
\providecommand \bibfnamefont [1]{#1}%
\providecommand \citenamefont [1]{#1}%
\providecommand \href@noop [0]{\@secondoftwo}%
\providecommand \href [0]{\begingroup \@sanitize@url \@href}%
\providecommand \@href[1]{\@@startlink{#1}\@@href}%
\providecommand \@@href[1]{\endgroup#1\@@endlink}%
\providecommand \@sanitize@url [0]{\catcode `\\12\catcode `\$12\catcode
  `\&12\catcode `\#12\catcode `\^12\catcode `\_12\catcode `\%12\relax}%
\providecommand \@@startlink[1]{}%
\providecommand \@@endlink[0]{}%
\providecommand \url  [0]{\begingroup\@sanitize@url \@url }%
\providecommand \@url [1]{\endgroup\@href {#1}{\urlprefix }}%
\providecommand \urlprefix  [0]{URL }%
\providecommand \Eprint [0]{\href }%
\providecommand \doibase [0]{http://dx.doi.org/}%
\providecommand \selectlanguage [0]{\@gobble}%
\providecommand \bibinfo  [0]{\@secondoftwo}%
\providecommand \bibfield  [0]{\@secondoftwo}%
\providecommand \translation [1]{[#1]}%
\providecommand \BibitemOpen [0]{}%
\providecommand \bibitemStop [0]{}%
\providecommand \bibitemNoStop [0]{.\EOS\space}%
\providecommand \EOS [0]{\spacefactor3000\relax}%
\providecommand \BibitemShut  [1]{\csname bibitem#1\endcsname}%
\let\auto@bib@innerbib\@empty
\bibitem [{\citenamefont {Chang}(2014)}]{Chang2014}%
  \BibitemOpen
  \bibfield  {author} {\bibinfo {author} {\bibfnamefont {J.}\ \bibnamefont
  {Chang}},\ }\bibfield  {title} {\enquote {\bibinfo {title} {Dark matter
  particle explorer: The first chinese cosmic ray and hard gamma-ray detector
  in space},}\ }\href {\doibase 10.11728/cjss2014.05.550} {\bibfield  {journal}
  {\bibinfo  {journal} {Chinese Journal of Space Science}\ }\textbf {\bibinfo
  {volume} {34}},\ \bibinfo {eid} {550} (\bibinfo {year} {2014})}\BibitemShut
  {NoStop}%
\bibitem [{\citenamefont {Chang}\ \emph {et~al.}(2017)\citenamefont {Chang}
  \emph {et~al.}}]{Chang2017}%
  \BibitemOpen
  \bibfield  {author} {\bibinfo {author} {\bibfnamefont {J.}~\bibnamefont
  {Chang}} \emph {et~al.} (\bibinfo {collaboration} {DAMPE}),\ }\bibfield
  {title} {\enquote {\bibinfo {title} {{The DArk Matter Particle Explorer
  mission}},}\ }\href {\doibase 10.1016/j.astropartphys.2017.08.005} {\bibfield
   {journal} {\bibinfo  {journal} {Astropart. Phys.}\ }\textbf {\bibinfo
  {volume} {95}},\ \bibinfo {pages} {6--24} (\bibinfo {year} {2017})},\ \Eprint
  {http://arxiv.org/abs/1706.08453} {arXiv:1706.08453 [astro-ph.IM]}
  \BibitemShut {NoStop}%
\bibitem [{\citenamefont {Ambrosi}\ \emph {et~al.}(2017)\citenamefont {Ambrosi}
  \emph {et~al.}}]{DAMPE2017}%
  \BibitemOpen
  \bibfield  {author} {\bibinfo {author} {\bibfnamefont {G.}~\bibnamefont
  {Ambrosi}} \emph {et~al.} (\bibinfo {collaboration} {DAMPE}),\ }\bibfield
  {title} {\enquote {\bibinfo {title} {{Direct detection of a break in the
  teraelectronvolt cosmic-ray spectrum of electrons and positrons}},}\ }\href
  {\doibase 10.1038/nature24475} {\bibfield  {journal} {\bibinfo  {journal}
  {Nature}\ } (\bibinfo {year} {2017}),\ 10.1038/nature24475},\ \Eprint
  {http://arxiv.org/abs/1711.10981} {arXiv:1711.10981 [astro-ph.HE]}
  \BibitemShut {NoStop}%
\bibitem [{\citenamefont {{Adriani}}\ \emph {et~al.}(2009)\citenamefont
  {{Adriani}}, \citenamefont {{Barbarino}}, \citenamefont {{Bazilevskaya}},
  \citenamefont {{Bellotti}}, \citenamefont {{Boezio}}, \citenamefont
  {{Bogomolov}}, \citenamefont {{Bonechi}}, \citenamefont {{Bongi}},
  \citenamefont {{Bonvicini}}, \citenamefont {{Bottai}},\ and\ \citenamefont
  {et~al.}}]{Adriani2009}%
  \BibitemOpen
  \bibfield  {author} {\bibinfo {author} {\bibfnamefont {O.}~\bibnamefont
  {{Adriani}}}, \bibinfo {author} {\bibfnamefont {G.~C.}\ \bibnamefont
  {{Barbarino}}}, \bibinfo {author} {\bibfnamefont {G.~A.}\ \bibnamefont
  {{Bazilevskaya}}}, \bibinfo {author} {\bibfnamefont {R.}~\bibnamefont
  {{Bellotti}}}, \bibinfo {author} {\bibfnamefont {M.}~\bibnamefont
  {{Boezio}}}, \bibinfo {author} {\bibfnamefont {E.~A.}\ \bibnamefont
  {{Bogomolov}}}, \bibinfo {author} {\bibfnamefont {L.}~\bibnamefont
  {{Bonechi}}}, \bibinfo {author} {\bibfnamefont {M.}~\bibnamefont {{Bongi}}},
  \bibinfo {author} {\bibfnamefont {V.}~\bibnamefont {{Bonvicini}}}, \bibinfo
  {author} {\bibfnamefont {S.}~\bibnamefont {{Bottai}}}, \ and\ \bibinfo
  {author} {\bibnamefont {et~al.}},\ }\bibfield  {title} {\enquote {\bibinfo
  {title} {{An anomalous positron abundance in cosmic rays with energies
  1.5-100GeV}},}\ }\href {\doibase 10.1038/nature07942} {\bibfield  {journal}
  {\bibinfo  {journal} {\nat}\ }\textbf {\bibinfo {volume} {458}},\ \bibinfo
  {pages} {607--609} (\bibinfo {year} {2009})},\ \Eprint
  {http://arxiv.org/abs/0810.4995} {arXiv:0810.4995} \BibitemShut {NoStop}%
\bibitem [{\citenamefont {{PAMELA collaboration}}\ \emph
  {et~al.}(2010)\citenamefont {{PAMELA collaboration}}, \citenamefont
  {{Adriani}}, \citenamefont {{Barbarino}}, \citenamefont {{Bazilevskaya}},
  \citenamefont {{Bellotti}}, \citenamefont {{Boezio}}, \citenamefont
  {{Bogomolov}}, \citenamefont {{Bonechi}}, \citenamefont {{Bongi}},
  \citenamefont {{Bonvicini}}, \citenamefont {{Borisov}},\ and\ \citenamefont
  {el~al.}}]{Adriani2010lepton}%
  \BibitemOpen
  \bibfield  {author} {\bibinfo {author} {\bibnamefont {{PAMELA
  collaboration}}}, \bibinfo {author} {\bibfnamefont {O.}~\bibnamefont
  {{Adriani}}}, \bibinfo {author} {\bibfnamefont {G.~C.}\ \bibnamefont
  {{Barbarino}}}, \bibinfo {author} {\bibfnamefont {G.~A.}\ \bibnamefont
  {{Bazilevskaya}}}, \bibinfo {author} {\bibfnamefont {R.}~\bibnamefont
  {{Bellotti}}}, \bibinfo {author} {\bibfnamefont {M.}~\bibnamefont
  {{Boezio}}}, \bibinfo {author} {\bibfnamefont {E.~A.}\ \bibnamefont
  {{Bogomolov}}}, \bibinfo {author} {\bibfnamefont {L.}~\bibnamefont
  {{Bonechi}}}, \bibinfo {author} {\bibfnamefont {M.}~\bibnamefont {{Bongi}}},
  \bibinfo {author} {\bibfnamefont {V.}~\bibnamefont {{Bonvicini}}}, \bibinfo
  {author} {\bibfnamefont {S.}~\bibnamefont {{Borisov}}}, \ and\ \bibinfo
  {author} {\bibnamefont {el~al.}},\ }\bibfield  {title} {\enquote {\bibinfo
  {title} {{A statistical procedure for the identification of positrons in the
  PAMELA experiment}},}\ }\href {\doibase 10.1016/j.astropartphys.2010.04.007}
  {\bibfield  {journal} {\bibinfo  {journal} {Astroparticle Physics}\ }\textbf
  {\bibinfo {volume} {34}},\ \bibinfo {pages} {1--11} (\bibinfo {year}
  {2010})},\ \Eprint {http://arxiv.org/abs/1001.3522} {arXiv:1001.3522
  [astro-ph.HE]} \BibitemShut {NoStop}%
\bibitem [{\citenamefont {{AMS collaboration}}\ \emph
  {et~al.}(2014{\natexlab{a}})\citenamefont {{AMS collaboration}},
  \citenamefont {{Aguilar}}, \citenamefont {{Aisa}}, \citenamefont {{Alpat}},
  \citenamefont {{Alvino}}, \citenamefont {{Ambrosi}}, \citenamefont
  {{Andeen}}, \citenamefont {{Arruda}}, \citenamefont {{Attig}}, \citenamefont
  {{Azzarello}}, \citenamefont {{Bachlechner}},\ and\ \citenamefont
  {et~al.}}]{AMS02_lepton_sum}%
  \BibitemOpen
  \bibfield  {author} {\bibinfo {author} {\bibnamefont {{AMS collaboration}}},
  \bibinfo {author} {\bibfnamefont {M.}~\bibnamefont {{Aguilar}}}, \bibinfo
  {author} {\bibfnamefont {D.}~\bibnamefont {{Aisa}}}, \bibinfo {author}
  {\bibfnamefont {B.}~\bibnamefont {{Alpat}}}, \bibinfo {author} {\bibfnamefont
  {A.}~\bibnamefont {{Alvino}}}, \bibinfo {author} {\bibfnamefont
  {G.}~\bibnamefont {{Ambrosi}}}, \bibinfo {author} {\bibfnamefont
  {K.}~\bibnamefont {{Andeen}}}, \bibinfo {author} {\bibfnamefont
  {L.}~\bibnamefont {{Arruda}}}, \bibinfo {author} {\bibfnamefont
  {N.}~\bibnamefont {{Attig}}}, \bibinfo {author} {\bibfnamefont
  {P.}~\bibnamefont {{Azzarello}}}, \bibinfo {author} {\bibfnamefont
  {A.}~\bibnamefont {{Bachlechner}}}, \ and\ \bibinfo {author} {\bibnamefont
  {et~al.}},\ }\bibfield  {title} {\enquote {\bibinfo {title} {{Precision
  Measurement of the (e$^{+}$+e$^{-}$) Flux in Primary Cosmic Rays from 0.5 GeV
  to 1 TeV with the Alpha Magnetic Spectrometer on the International Space
  Station}},}\ }\href {\doibase 10.1103/PhysRevLett.113.221102} {\bibfield
  {journal} {\bibinfo  {journal} {Physical Review Letters}\ }\textbf {\bibinfo
  {volume} {113}},\ \bibinfo {eid} {221102} (\bibinfo {year}
  {2014}{\natexlab{a}})}\BibitemShut {NoStop}%
\bibitem [{\citenamefont {{Fermi-LAT collaboration}}\ \emph
  {et~al.}(2012)\citenamefont {{Fermi-LAT collaboration}}, \citenamefont
  {{Ackermann}}, \citenamefont {{Ajello}}, \citenamefont {{Allafort}},
  \citenamefont {{Atwood}}, \citenamefont {{Baldini}}, \citenamefont
  {{Barbiellini}}, \citenamefont {{Bastieri}}, \citenamefont {{Bechtol}},
  \citenamefont {{Bellazzini}}, \citenamefont {{Berenji}},\ and\ \citenamefont
  {el~al.}}]{Ackermann2012}%
  \BibitemOpen
  \bibfield  {author} {\bibinfo {author} {\bibnamefont {{Fermi-LAT
  collaboration}}}, \bibinfo {author} {\bibfnamefont {M.}~\bibnamefont
  {{Ackermann}}}, \bibinfo {author} {\bibfnamefont {M.}~\bibnamefont
  {{Ajello}}}, \bibinfo {author} {\bibfnamefont {A.}~\bibnamefont
  {{Allafort}}}, \bibinfo {author} {\bibfnamefont {W.~B.}\ \bibnamefont
  {{Atwood}}}, \bibinfo {author} {\bibfnamefont {L.}~\bibnamefont {{Baldini}}},
  \bibinfo {author} {\bibfnamefont {G.}~\bibnamefont {{Barbiellini}}}, \bibinfo
  {author} {\bibfnamefont {D.}~\bibnamefont {{Bastieri}}}, \bibinfo {author}
  {\bibfnamefont {K.}~\bibnamefont {{Bechtol}}}, \bibinfo {author}
  {\bibfnamefont {R.}~\bibnamefont {{Bellazzini}}}, \bibinfo {author}
  {\bibfnamefont {B.}~\bibnamefont {{Berenji}}}, \ and\ \bibinfo {author}
  {\bibnamefont {el~al.}},\ }\bibfield  {title} {\enquote {\bibinfo {title}
  {{Measurement of Separate Cosmic-Ray Electron and Positron Spectra with the
  Fermi Large Area Telescope}},}\ }\href {\doibase
  10.1103/PhysRevLett.108.011103} {\bibfield  {journal} {\bibinfo  {journal}
  {Physical Review Letters}\ }\textbf {\bibinfo {volume} {108}},\ \bibinfo
  {eid} {011103} (\bibinfo {year} {2012})},\ \Eprint
  {http://arxiv.org/abs/1109.0521} {arXiv:1109.0521 [astro-ph.HE]} \BibitemShut
  {NoStop}%
\bibitem [{\citenamefont {collaboration}\ \emph
  {et~al.}(2017{\natexlab{a}})\citenamefont {collaboration}, \citenamefont
  {Adriani}, \citenamefont {Akaike}, \citenamefont {Asano}, \citenamefont
  {Asaoka}, \citenamefont {Bagliesi}, \citenamefont {Bigongiari}, \citenamefont
  {Binns}, \citenamefont {Bonechi}, \citenamefont {Bongi}, \citenamefont
  {Brogi},\ and\ \citenamefont {et~al.}}]{CALET2017}%
  \BibitemOpen
  \bibfield  {author} {\bibinfo {author} {\bibfnamefont {CALET}\ \bibnamefont
  {collaboration}}, \bibinfo {author} {\bibfnamefont {O.}~\bibnamefont
  {Adriani}}, \bibinfo {author} {\bibfnamefont {Y.}~\bibnamefont {Akaike}},
  \bibinfo {author} {\bibfnamefont {K.}~\bibnamefont {Asano}}, \bibinfo
  {author} {\bibfnamefont {Y.}~\bibnamefont {Asaoka}}, \bibinfo {author}
  {\bibfnamefont {M.~G.}\ \bibnamefont {Bagliesi}}, \bibinfo {author}
  {\bibfnamefont {G.}~\bibnamefont {Bigongiari}}, \bibinfo {author}
  {\bibfnamefont {W.~R.}\ \bibnamefont {Binns}}, \bibinfo {author}
  {\bibfnamefont {S.}~\bibnamefont {Bonechi}}, \bibinfo {author} {\bibfnamefont
  {M.}~\bibnamefont {Bongi}}, \bibinfo {author} {\bibfnamefont
  {P.}~\bibnamefont {Brogi}}, \ and\ \bibinfo {author} {\bibnamefont
  {et~al.}},\ }\bibfield  {title} {\enquote {\bibinfo {title} {Energy spectrum
  of cosmic-ray electron and positron from 10 gev to 3 tev observed with the
  calorimetric electron telescope on the international space station},}\ }\href
  {\doibase 10.1103/PhysRevLett.119.181101} {\bibfield  {journal} {\bibinfo
  {journal} {Phys. Rev. Lett.}\ }\textbf {\bibinfo {volume} {119}},\ \bibinfo
  {pages} {181101} (\bibinfo {year} {2017}{\natexlab{a}})}\BibitemShut
  {NoStop}%
\bibitem [{\citenamefont {collaboration}\ \emph
  {et~al.}(2017{\natexlab{b}})\citenamefont {collaboration}, \citenamefont
  {Abdollahi}, \citenamefont {Ackermann}, \citenamefont {Ajello}, \citenamefont
  {Atwood}, \citenamefont {Baldini}, \citenamefont {Barbiellini}, \citenamefont
  {Bastieri}, \citenamefont {Bellazzini}, \citenamefont {Bloom},\ and\
  \citenamefont {et~al.}}]{Fermi2017}%
  \BibitemOpen
  \bibfield  {author} {\bibinfo {author} {\bibfnamefont {Fermi-LAT}\
  \bibnamefont {collaboration}}, \bibinfo {author} {\bibfnamefont
  {S.}~\bibnamefont {Abdollahi}}, \bibinfo {author} {\bibfnamefont
  {M.}~\bibnamefont {Ackermann}}, \bibinfo {author} {\bibfnamefont
  {M.}~\bibnamefont {Ajello}}, \bibinfo {author} {\bibfnamefont {W.~B.}\
  \bibnamefont {Atwood}}, \bibinfo {author} {\bibfnamefont {L.}~\bibnamefont
  {Baldini}}, \bibinfo {author} {\bibfnamefont {G.}~\bibnamefont
  {Barbiellini}}, \bibinfo {author} {\bibfnamefont {D.}~\bibnamefont
  {Bastieri}}, \bibinfo {author} {\bibfnamefont {R.}~\bibnamefont
  {Bellazzini}}, \bibinfo {author} {\bibfnamefont {E.~D.}\ \bibnamefont
  {Bloom}}, \ and\ \bibinfo {author} {\bibnamefont {et~al.}},\ }\bibfield
  {title} {\enquote {\bibinfo {title} {Cosmic-ray electron-positron spectrum
  from 7 gev to 2 tev with the fermi large area telescope},}\ }\href {\doibase
  10.1103/PhysRevD.95.082007} {\bibfield  {journal} {\bibinfo  {journal} {Phys.
  Rev. D}\ }\textbf {\bibinfo {volume} {95}},\ \bibinfo {pages} {082007}
  (\bibinfo {year} {2017}{\natexlab{b}})}\BibitemShut {NoStop}%
\bibitem [{\citenamefont {{Malyshev}}\ \emph {et~al.}(2009)\citenamefont
  {{Malyshev}}, \citenamefont {{Cholis}},\ and\ \citenamefont
  {{Gelfand}}}]{Malyshev2009}%
  \BibitemOpen
  \bibfield  {author} {\bibinfo {author} {\bibfnamefont {D.}~\bibnamefont
  {{Malyshev}}}, \bibinfo {author} {\bibfnamefont {I.}~\bibnamefont
  {{Cholis}}}, \ and\ \bibinfo {author} {\bibfnamefont {J.}~\bibnamefont
  {{Gelfand}}},\ }\bibfield  {title} {\enquote {\bibinfo {title} {{Pulsars
  versus dark matter interpretation of ATIC/PAMELA}},}\ }\href {\doibase
  10.1103/PhysRevD.80.063005} {\bibfield  {journal} {\bibinfo  {journal}
  {\prd}\ }\textbf {\bibinfo {volume} {80}},\ \bibinfo {eid} {063005} (\bibinfo
  {year} {2009})},\ \Eprint {http://arxiv.org/abs/0903.1310} {arXiv:0903.1310
  [astro-ph.HE]} \BibitemShut {NoStop}%
\bibitem [{\citenamefont {{Kuhlen}}\ and\ \citenamefont
  {{Malyshev}}(2009)}]{Kuhlen2009}%
  \BibitemOpen
  \bibfield  {author} {\bibinfo {author} {\bibfnamefont {M.}~\bibnamefont
  {{Kuhlen}}}\ and\ \bibinfo {author} {\bibfnamefont {D.}~\bibnamefont
  {{Malyshev}}},\ }\bibfield  {title} {\enquote {\bibinfo {title} {{ATIC,
  PAMELA, HESS, and Fermi data and nearby dark matter subhalos}},}\ }\href
  {\doibase 10.1103/PhysRevD.79.123517} {\bibfield  {journal} {\bibinfo
  {journal} {\prd}\ }\textbf {\bibinfo {volume} {79}},\ \bibinfo {eid} {123517}
  (\bibinfo {year} {2009})},\ \Eprint {http://arxiv.org/abs/0904.3378}
  {arXiv:0904.3378 [hep-ph]} \BibitemShut {NoStop}%
\bibitem [{\citenamefont {{Brun}}\ \emph {et~al.}(2009)\citenamefont {{Brun}},
  \citenamefont {{Delahaye}}, \citenamefont {{Diemand}}, \citenamefont
  {{Profumo}},\ and\ \citenamefont {{Salati}}}]{Brun2009}%
  \BibitemOpen
  \bibfield  {author} {\bibinfo {author} {\bibfnamefont {P.}~\bibnamefont
  {{Brun}}}, \bibinfo {author} {\bibfnamefont {T.}~\bibnamefont {{Delahaye}}},
  \bibinfo {author} {\bibfnamefont {J.}~\bibnamefont {{Diemand}}}, \bibinfo
  {author} {\bibfnamefont {S.}~\bibnamefont {{Profumo}}}, \ and\ \bibinfo
  {author} {\bibfnamefont {P.}~\bibnamefont {{Salati}}},\ }\bibfield  {title}
  {\enquote {\bibinfo {title} {{Cosmic ray lepton puzzle in the light of
  cosmological N-body simulations}},}\ }\href {\doibase
  10.1103/PhysRevD.80.035023} {\bibfield  {journal} {\bibinfo  {journal}
  {\prd}\ }\textbf {\bibinfo {volume} {80}},\ \bibinfo {eid} {035023} (\bibinfo
  {year} {2009})},\ \Eprint {http://arxiv.org/abs/0904.0812} {arXiv:0904.0812
  [astro-ph.HE]} \BibitemShut {NoStop}%
\bibitem [{\citenamefont {{Gendelev}}\ \emph {et~al.}(2010)\citenamefont
  {{Gendelev}}, \citenamefont {{Profumo}},\ and\ \citenamefont
  {{Dormody}}}]{Gendelev2010}%
  \BibitemOpen
  \bibfield  {author} {\bibinfo {author} {\bibfnamefont {L.}~\bibnamefont
  {{Gendelev}}}, \bibinfo {author} {\bibfnamefont {S.}~\bibnamefont
  {{Profumo}}}, \ and\ \bibinfo {author} {\bibfnamefont {M.}~\bibnamefont
  {{Dormody}}},\ }\bibfield  {title} {\enquote {\bibinfo {title} {{The
  contribution of Fermi gamma-ray pulsars to the local flux of cosmic-ray
  electrons and positrons}},}\ }\href {\doibase 10.1088/1475-7516/2010/02/016}
  {\bibfield  {journal} {\bibinfo  {journal} {\jcap}\ }\textbf {\bibinfo
  {volume} {2}},\ \bibinfo {eid} {016} (\bibinfo {year} {2010})},\ \Eprint
  {http://arxiv.org/abs/1001.4540} {arXiv:1001.4540 [astro-ph.HE]} \BibitemShut
  {NoStop}%
\bibitem [{\citenamefont {{Profumo}}(2012)}]{Profumo2012}%
  \BibitemOpen
  \bibfield  {author} {\bibinfo {author} {\bibfnamefont {S.}~\bibnamefont
  {{Profumo}}},\ }\bibfield  {title} {\enquote {\bibinfo {title} {{Dissecting
  cosmic-ray electron-positron data with Occam's razor: the role of known
  pulsars}},}\ }\href {\doibase 10.2478/s11534-011-0099-z} {\bibfield
  {journal} {\bibinfo  {journal} {Central European Journal of Physics}\
  }\textbf {\bibinfo {volume} {10}},\ \bibinfo {pages} {1--31} (\bibinfo {year}
  {2012})},\ \Eprint {http://arxiv.org/abs/0812.4457} {arXiv:0812.4457}
  \BibitemShut {NoStop}%
\bibitem [{\citenamefont {{Panov}}(2013)}]{Panov2013}%
  \BibitemOpen
  \bibfield  {author} {\bibinfo {author} {\bibfnamefont {A.~D.}\ \bibnamefont
  {{Panov}}},\ }\bibfield  {title} {\enquote {\bibinfo {title} {{Electrons and
  Positrons in Cosmic Rays}},}\ }in\ \href {\doibase
  10.1088/1742-6596/409/1/012004} {\emph {\bibinfo {booktitle} {Journal of
  Physics Conference Series}}},\ \bibinfo {series} {Journal of Physics
  Conference Series}, Vol.\ \bibinfo {volume} {409}\ (\bibinfo {year} {2013})\
  p.\ \bibinfo {pages} {012004},\ \Eprint {http://arxiv.org/abs/1303.6118}
  {arXiv:1303.6118 [astro-ph.HE]} \BibitemShut {NoStop}%
\bibitem [{\citenamefont {{Fang}}\ \emph {et~al.}(2017)\citenamefont {{Fang}},
  \citenamefont {{Bi}},\ and\ \citenamefont {{Yin}}}]{Fang2017}%
  \BibitemOpen
  \bibfield  {author} {\bibinfo {author} {\bibfnamefont {K.}~\bibnamefont
  {{Fang}}}, \bibinfo {author} {\bibfnamefont {X.-J.}\ \bibnamefont {{Bi}}}, \
  and\ \bibinfo {author} {\bibfnamefont {P.-F.}\ \bibnamefont {{Yin}}},\
  }\bibfield  {title} {\enquote {\bibinfo {title} {{Explanation of the
  knee-like feature in the DAMPE cosmic $e^-+e^+$ energy spectrum}},}\
  }\href@noop {} {\bibfield  {journal} {\bibinfo  {journal} {ArXiv e-prints}\ }
  (\bibinfo {year} {2017})},\ \Eprint {http://arxiv.org/abs/1711.10996}
  {arXiv:1711.10996 [astro-ph.HE]} \BibitemShut {NoStop}%
\bibitem [{\citenamefont {{Yuan}}\ \emph
  {et~al.}(2017{\natexlab{a}})\citenamefont {{Yuan}}, \citenamefont {{Feng}},
  \citenamefont {{Yin}}, \citenamefont {{Fan}}, \citenamefont {{Bi}},
  \citenamefont {{Cui}}, \citenamefont {{Dong}}, \citenamefont {{Guo}},
  \citenamefont {{Fang}}, \citenamefont {{Hu}}, \citenamefont {{Huang}},
  \citenamefont {{Lei}}, \citenamefont {{Li}}, \citenamefont {{Lin}},
  \citenamefont {{Liu}}, \citenamefont {{Ma}}, \citenamefont {{Peng}},
  \citenamefont {{Qiao}}, \citenamefont {{Shen}}, \citenamefont {{Su}},
  \citenamefont {{Wei}}, \citenamefont {{Xu}}, \citenamefont {{Yue}},
  \citenamefont {{Zang}}, \citenamefont {{Zhang}}, \citenamefont {{Zhang}},
  \citenamefont {{Zhang}}, \citenamefont {{Zhang}},\ and\ \citenamefont
  {{Zhang}}}]{Yuan2017_dampe}%
  \BibitemOpen
  \bibfield  {author} {\bibinfo {author} {\bibfnamefont {Q.}~\bibnamefont
  {{Yuan}}}, \bibinfo {author} {\bibfnamefont {L.}~\bibnamefont {{Feng}}},
  \bibinfo {author} {\bibfnamefont {P.-F.}\ \bibnamefont {{Yin}}}, \bibinfo
  {author} {\bibfnamefont {Y.-Z.}\ \bibnamefont {{Fan}}}, \bibinfo {author}
  {\bibfnamefont {X.-J.}\ \bibnamefont {{Bi}}}, \bibinfo {author}
  {\bibfnamefont {M.-Y.}\ \bibnamefont {{Cui}}}, \bibinfo {author}
  {\bibfnamefont {T.-K.}\ \bibnamefont {{Dong}}}, \bibinfo {author}
  {\bibfnamefont {Y.-Q.}\ \bibnamefont {{Guo}}}, \bibinfo {author}
  {\bibfnamefont {K.}~\bibnamefont {{Fang}}}, \bibinfo {author} {\bibfnamefont
  {H.-B.}\ \bibnamefont {{Hu}}}, \bibinfo {author} {\bibfnamefont
  {X.}~\bibnamefont {{Huang}}}, \bibinfo {author} {\bibfnamefont {S.-J.}\
  \bibnamefont {{Lei}}}, \bibinfo {author} {\bibfnamefont {X.}~\bibnamefont
  {{Li}}}, \bibinfo {author} {\bibfnamefont {S.-J.}\ \bibnamefont {{Lin}}},
  \bibinfo {author} {\bibfnamefont {H.}~\bibnamefont {{Liu}}}, \bibinfo
  {author} {\bibfnamefont {P.-X.}\ \bibnamefont {{Ma}}}, \bibinfo {author}
  {\bibfnamefont {W.-X.}\ \bibnamefont {{Peng}}}, \bibinfo {author}
  {\bibfnamefont {R.}~\bibnamefont {{Qiao}}}, \bibinfo {author} {\bibfnamefont
  {Z.-Q.}\ \bibnamefont {{Shen}}}, \bibinfo {author} {\bibfnamefont
  {M.}~\bibnamefont {{Su}}}, \bibinfo {author} {\bibfnamefont {Y.-F.}\
  \bibnamefont {{Wei}}}, \bibinfo {author} {\bibfnamefont {Z.-L.}\ \bibnamefont
  {{Xu}}}, \bibinfo {author} {\bibfnamefont {C.}~\bibnamefont {{Yue}}},
  \bibinfo {author} {\bibfnamefont {J.-J.}\ \bibnamefont {{Zang}}}, \bibinfo
  {author} {\bibfnamefont {C.}~\bibnamefont {{Zhang}}}, \bibinfo {author}
  {\bibfnamefont {X.}~\bibnamefont {{Zhang}}}, \bibinfo {author} {\bibfnamefont
  {Y.-P.}\ \bibnamefont {{Zhang}}}, \bibinfo {author} {\bibfnamefont {Y.-J.}\
  \bibnamefont {{Zhang}}}, \ and\ \bibinfo {author} {\bibfnamefont {Y.-L.}\
  \bibnamefont {{Zhang}}},\ }\bibfield  {title} {\enquote {\bibinfo {title}
  {{Interpretations of the DAMPE electron data}},}\ }\href@noop {} {\bibfield
  {journal} {\bibinfo  {journal} {ArXiv e-prints}\ } (\bibinfo {year}
  {2017}{\natexlab{a}})},\ \Eprint {http://arxiv.org/abs/1711.10989}
  {arXiv:1711.10989 [astro-ph.HE]} \BibitemShut {NoStop}%
\bibitem [{\citenamefont {{Athron}}\ \emph {et~al.}(2017)\citenamefont
  {{Athron}}, \citenamefont {{Balazs}}, \citenamefont {{Fowlie}},\ and\
  \citenamefont {{Zhang}}}]{Athron2017}%
  \BibitemOpen
  \bibfield  {author} {\bibinfo {author} {\bibfnamefont {P.}~\bibnamefont
  {{Athron}}}, \bibinfo {author} {\bibfnamefont {C.}~\bibnamefont {{Balazs}}},
  \bibinfo {author} {\bibfnamefont {A.}~\bibnamefont {{Fowlie}}}, \ and\
  \bibinfo {author} {\bibfnamefont {Y.}~\bibnamefont {{Zhang}}},\ }\bibfield
  {title} {\enquote {\bibinfo {title} {{Model-independent analysis of the DAMPE
  excess}},}\ }\href@noop {} {\bibfield  {journal} {\bibinfo  {journal} {ArXiv
  e-prints}\ } (\bibinfo {year} {2017})},\ \Eprint
  {http://arxiv.org/abs/1711.11376} {arXiv:1711.11376 [hep-ph]} \BibitemShut
  {NoStop}%
\bibitem [{\citenamefont {{Fan}}\ \emph {et~al.}(2017)\citenamefont {{Fan}},
  \citenamefont {{Huang}}, \citenamefont {{Spinrath}}, \citenamefont {{Sming
  Tsai}},\ and\ \citenamefont {{Yuan}}}]{Fan2017}%
  \BibitemOpen
  \bibfield  {author} {\bibinfo {author} {\bibfnamefont {Y.-Z.}\ \bibnamefont
  {{Fan}}}, \bibinfo {author} {\bibfnamefont {W.-C.}\ \bibnamefont {{Huang}}},
  \bibinfo {author} {\bibfnamefont {M.}~\bibnamefont {{Spinrath}}}, \bibinfo
  {author} {\bibfnamefont {Y.-L.}\ \bibnamefont {{Sming Tsai}}}, \ and\
  \bibinfo {author} {\bibfnamefont {Q.}~\bibnamefont {{Yuan}}},\ }\bibfield
  {title} {\enquote {\bibinfo {title} {{A model explaining neutrino masses and
  the DAMPE cosmic ray electron excess}},}\ }\href@noop {} {\bibfield
  {journal} {\bibinfo  {journal} {ArXiv e-prints}\ } (\bibinfo {year}
  {2017})},\ \Eprint {http://arxiv.org/abs/1711.10995} {arXiv:1711.10995
  [hep-ph]} \BibitemShut {NoStop}%
\bibitem [{\citenamefont {{Duan}}\ \emph {et~al.}(2017)\citenamefont {{Duan}},
  \citenamefont {{Feng}}, \citenamefont {{Wang}}, \citenamefont {{Wu}},
  \citenamefont {{Yang}},\ and\ \citenamefont {{Zheng}}}]{Duan2017}%
  \BibitemOpen
  \bibfield  {author} {\bibinfo {author} {\bibfnamefont {G.~H.}\ \bibnamefont
  {{Duan}}}, \bibinfo {author} {\bibfnamefont {L.}~\bibnamefont {{Feng}}},
  \bibinfo {author} {\bibfnamefont {F.}~\bibnamefont {{Wang}}}, \bibinfo
  {author} {\bibfnamefont {L.}~\bibnamefont {{Wu}}}, \bibinfo {author}
  {\bibfnamefont {J.~M.}\ \bibnamefont {{Yang}}}, \ and\ \bibinfo {author}
  {\bibfnamefont {R.}~\bibnamefont {{Zheng}}},\ }\bibfield  {title} {\enquote
  {\bibinfo {title} {{Simplified TeV leptophilic dark matter in light of DAMPE
  data}},}\ }\href@noop {} {\bibfield  {journal} {\bibinfo  {journal} {ArXiv
  e-prints}\ } (\bibinfo {year} {2017})},\ \Eprint
  {http://arxiv.org/abs/1711.11012} {arXiv:1711.11012 [hep-ph]} \BibitemShut
  {NoStop}%
\bibitem [{\citenamefont {{Gu}}\ and\ \citenamefont {{He}}(2017)}]{Gu2017}%
  \BibitemOpen
  \bibfield  {author} {\bibinfo {author} {\bibfnamefont {P.-H.}\ \bibnamefont
  {{Gu}}}\ and\ \bibinfo {author} {\bibfnamefont {X.-G.}\ \bibnamefont
  {{He}}},\ }\bibfield  {title} {\enquote {\bibinfo {title} {{Electrophilic
  dark matter with dark photon: from DAMPE to direct detection}},}\ }\href@noop
  {} {\bibfield  {journal} {\bibinfo  {journal} {ArXiv e-prints}\ } (\bibinfo
  {year} {2017})},\ \Eprint {http://arxiv.org/abs/1711.11000} {arXiv:1711.11000
  [hep-ph]} \BibitemShut {NoStop}%
\bibitem [{\citenamefont {{Liu}}\ and\ \citenamefont {{Liu}}(2017)}]{Liu2017}%
  \BibitemOpen
  \bibfield  {author} {\bibinfo {author} {\bibfnamefont {X.}~\bibnamefont
  {{Liu}}}\ and\ \bibinfo {author} {\bibfnamefont {Z.}~\bibnamefont {{Liu}}},\
  }\bibfield  {title} {\enquote {\bibinfo {title} {{TeV dark matter and the
  DAMPE electron excess}},}\ }\href@noop {} {\bibfield  {journal} {\bibinfo
  {journal} {ArXiv e-prints}\ } (\bibinfo {year} {2017})},\ \Eprint
  {http://arxiv.org/abs/1711.11579} {arXiv:1711.11579 [hep-ph]} \BibitemShut
  {NoStop}%
\bibitem [{\citenamefont {{Cao}}\ \emph {et~al.}(2017)\citenamefont {{Cao}},
  \citenamefont {{Feng}}, \citenamefont {{Guo}}, \citenamefont {{Shang}},
  \citenamefont {{Wang}},\ and\ \citenamefont {{Wu}}}]{Cao2017}%
  \BibitemOpen
  \bibfield  {author} {\bibinfo {author} {\bibfnamefont {J.}~\bibnamefont
  {{Cao}}}, \bibinfo {author} {\bibfnamefont {L.}~\bibnamefont {{Feng}}},
  \bibinfo {author} {\bibfnamefont {X.}~\bibnamefont {{Guo}}}, \bibinfo
  {author} {\bibfnamefont {L.}~\bibnamefont {{Shang}}}, \bibinfo {author}
  {\bibfnamefont {F.}~\bibnamefont {{Wang}}}, \ and\ \bibinfo {author}
  {\bibfnamefont {P.}~\bibnamefont {{Wu}}},\ }\bibfield  {title} {\enquote
  {\bibinfo {title} {{Scalar dark matter interpretation of the DAMPE data with
  U(1) gauge interactions}},}\ }\href@noop {} {\bibfield  {journal} {\bibinfo
  {journal} {ArXiv e-prints}\ } (\bibinfo {year} {2017})},\ \Eprint
  {http://arxiv.org/abs/1711.11452} {arXiv:1711.11452 [hep-ph]} \BibitemShut
  {NoStop}%
\bibitem [{\citenamefont {{Profumo}}\ \emph {et~al.}(2017)\citenamefont
  {{Profumo}}, \citenamefont {{Queiroz}}, \citenamefont {{Silk}},\ and\
  \citenamefont {{Siqueira}}}]{Profumo2017}%
  \BibitemOpen
  \bibfield  {author} {\bibinfo {author} {\bibfnamefont {S.}~\bibnamefont
  {{Profumo}}}, \bibinfo {author} {\bibfnamefont {F.~S.}\ \bibnamefont
  {{Queiroz}}}, \bibinfo {author} {\bibfnamefont {J.}~\bibnamefont {{Silk}}}, \
  and\ \bibinfo {author} {\bibfnamefont {C.}~\bibnamefont {{Siqueira}}},\
  }\bibfield  {title} {\enquote {\bibinfo {title} {{Searching for Secluded Dark
  Matter with H.E.S.S., Fermi-LAT, and Planck}},}\ }\href@noop {} {\bibfield
  {journal} {\bibinfo  {journal} {ArXiv e-prints}\ } (\bibinfo {year}
  {2017})},\ \Eprint {http://arxiv.org/abs/1711.03133} {arXiv:1711.03133
  [hep-ph]} \BibitemShut {NoStop}%
\bibitem [{\citenamefont {{AMS collaboration}}\ \emph
  {et~al.}(2013{\natexlab{a}})\citenamefont {{AMS collaboration}},
  \citenamefont {{Aguilar}}, \citenamefont {{Alberti}}, \citenamefont
  {{Alpat}}, \citenamefont {{Alvino}}, \citenamefont {{Ambrosi}}, \citenamefont
  {{Andeen}}, \citenamefont {{Anderhub}}, \citenamefont {{Arruda}},
  \citenamefont {{Azzarello}}, \citenamefont {{Bachlechner}},\ and\
  \citenamefont {et~al.}}]{AMS2013}%
  \BibitemOpen
  \bibfield  {author} {\bibinfo {author} {\bibnamefont {{AMS collaboration}}},
  \bibinfo {author} {\bibfnamefont {M.}~\bibnamefont {{Aguilar}}}, \bibinfo
  {author} {\bibfnamefont {G.}~\bibnamefont {{Alberti}}}, \bibinfo {author}
  {\bibfnamefont {B.}~\bibnamefont {{Alpat}}}, \bibinfo {author} {\bibfnamefont
  {A.}~\bibnamefont {{Alvino}}}, \bibinfo {author} {\bibfnamefont
  {G.}~\bibnamefont {{Ambrosi}}}, \bibinfo {author} {\bibfnamefont
  {K.}~\bibnamefont {{Andeen}}}, \bibinfo {author} {\bibfnamefont
  {H.}~\bibnamefont {{Anderhub}}}, \bibinfo {author} {\bibfnamefont
  {L.}~\bibnamefont {{Arruda}}}, \bibinfo {author} {\bibfnamefont
  {P.}~\bibnamefont {{Azzarello}}}, \bibinfo {author} {\bibfnamefont
  {A.}~\bibnamefont {{Bachlechner}}}, \ and\ \bibinfo {author} {\bibnamefont
  {et~al.}},\ }\bibfield  {title} {\enquote {\bibinfo {title} {{First Result
  from the Alpha Magnetic Spectrometer on the International Space Station:
  Precision Measurement of the Positron Fraction in Primary Cosmic Rays of
  0.5-350 GeV}},}\ }\href {\doibase 10.1103/PhysRevLett.110.141102} {\bibfield
  {journal} {\bibinfo  {journal} {Physical Review Letters}\ }\textbf {\bibinfo
  {volume} {110}},\ \bibinfo {eid} {141102} (\bibinfo {year}
  {2013}{\natexlab{a}})}\BibitemShut {NoStop}%
\bibitem [{\citenamefont {{Barwick}}\ \emph {et~al.}(1997)\citenamefont
  {{Barwick}}, \citenamefont {{Beatty}}, \citenamefont {{Bhattacharyya}},
  \citenamefont {{Bower}}, \citenamefont {{Chaput}}, \citenamefont {{Coutu}},
  \citenamefont {{de Nolfo}}, \citenamefont {{Knapp}}, \citenamefont
  {{Lowder}}, \citenamefont {{McKee}}, \citenamefont {{Mueller}}, \citenamefont
  {{Musser}}, \citenamefont {{Nutter}}, \citenamefont {{Schneider}},
  \citenamefont {{Swordy}}, \citenamefont {{Tarle}}, \citenamefont {{Tomasch}},
  \citenamefont {{Torbet}},\ and\ \citenamefont {{HEAT
  collaboration}}}]{Barwick1997}%
  \BibitemOpen
  \bibfield  {author} {\bibinfo {author} {\bibfnamefont {S.~W.}\ \bibnamefont
  {{Barwick}}}, \bibinfo {author} {\bibfnamefont {J.~J.}\ \bibnamefont
  {{Beatty}}}, \bibinfo {author} {\bibfnamefont {A.}~\bibnamefont
  {{Bhattacharyya}}}, \bibinfo {author} {\bibfnamefont {C.~R.}\ \bibnamefont
  {{Bower}}}, \bibinfo {author} {\bibfnamefont {C.~J.}\ \bibnamefont
  {{Chaput}}}, \bibinfo {author} {\bibfnamefont {S.}~\bibnamefont {{Coutu}}},
  \bibinfo {author} {\bibfnamefont {G.~A.}\ \bibnamefont {{de Nolfo}}},
  \bibinfo {author} {\bibfnamefont {J.}~\bibnamefont {{Knapp}}}, \bibinfo
  {author} {\bibfnamefont {D.~M.}\ \bibnamefont {{Lowder}}}, \bibinfo {author}
  {\bibfnamefont {S.}~\bibnamefont {{McKee}}}, \bibinfo {author} {\bibfnamefont
  {D.}~\bibnamefont {{Mueller}}}, \bibinfo {author} {\bibfnamefont {J.~A.}\
  \bibnamefont {{Musser}}}, \bibinfo {author} {\bibfnamefont {S.~L.}\
  \bibnamefont {{Nutter}}}, \bibinfo {author} {\bibfnamefont {E.}~\bibnamefont
  {{Schneider}}}, \bibinfo {author} {\bibfnamefont {S.~P.}\ \bibnamefont
  {{Swordy}}}, \bibinfo {author} {\bibfnamefont {G.}~\bibnamefont {{Tarle}}},
  \bibinfo {author} {\bibfnamefont {A.~D.}\ \bibnamefont {{Tomasch}}}, \bibinfo
  {author} {\bibfnamefont {E.}~\bibnamefont {{Torbet}}}, \ and\ \bibinfo
  {author} {\bibnamefont {{HEAT collaboration}}},\ }\bibfield  {title}
  {\enquote {\bibinfo {title} {{Measurements of the Cosmic-Ray Positron
  Fraction from 1 to 50 GeV}},}\ }\href {\doibase 10.1086/310706} {\bibfield
  {journal} {\bibinfo  {journal} {\apjl}\ }\textbf {\bibinfo {volume} {482}},\
  \bibinfo {pages} {L191--L194} (\bibinfo {year} {1997})},\ \Eprint
  {http://arxiv.org/abs/astro-ph/9703192} {astro-ph/9703192} \BibitemShut
  {NoStop}%
\bibitem [{\citenamefont {{AMS-01 collaboration}}(2007)}]{AMS01}%
  \BibitemOpen
  \bibfield  {author} {\bibinfo {author} {\bibnamefont {{AMS-01
  collaboration}}},\ }\bibfield  {title} {\enquote {\bibinfo {title}
  {{Cosmic-ray positron fraction measurement from 1 to 30 GeV with AMS-01}},}\
  }\href {\doibase 10.1016/j.physletb.2007.01.024} {\bibfield  {journal}
  {\bibinfo  {journal} {Physics Letters B}\ }\textbf {\bibinfo {volume}
  {646}},\ \bibinfo {pages} {145--154} (\bibinfo {year} {2007})},\ \Eprint
  {http://arxiv.org/abs/astro-ph/0703154} {astro-ph/0703154} \BibitemShut
  {NoStop}%
\bibitem [{\citenamefont {{AMS collaboration}}\ \emph
  {et~al.}(2013{\natexlab{b}})\citenamefont {{AMS collaboration}},
  \citenamefont {{Aguilar}}, \citenamefont {{Alberti}}, \citenamefont
  {{Alpat}}, \citenamefont {{Alvino}}, \citenamefont {{Ambrosi}}, \citenamefont
  {{Andeen}}, \citenamefont {{Anderhub}}, \citenamefont {{Arruda}},
  \citenamefont {{Azzarello}}, \citenamefont {{Bachlechner}},\ and\
  \citenamefont {et~al.}}]{AMS02_fraction01}%
  \BibitemOpen
  \bibfield  {author} {\bibinfo {author} {\bibnamefont {{AMS collaboration}}},
  \bibinfo {author} {\bibfnamefont {M.}~\bibnamefont {{Aguilar}}}, \bibinfo
  {author} {\bibfnamefont {G.}~\bibnamefont {{Alberti}}}, \bibinfo {author}
  {\bibfnamefont {B.}~\bibnamefont {{Alpat}}}, \bibinfo {author} {\bibfnamefont
  {A.}~\bibnamefont {{Alvino}}}, \bibinfo {author} {\bibfnamefont
  {G.}~\bibnamefont {{Ambrosi}}}, \bibinfo {author} {\bibfnamefont
  {K.}~\bibnamefont {{Andeen}}}, \bibinfo {author} {\bibfnamefont
  {H.}~\bibnamefont {{Anderhub}}}, \bibinfo {author} {\bibfnamefont
  {L.}~\bibnamefont {{Arruda}}}, \bibinfo {author} {\bibfnamefont
  {P.}~\bibnamefont {{Azzarello}}}, \bibinfo {author} {\bibfnamefont
  {A.}~\bibnamefont {{Bachlechner}}}, \ and\ \bibinfo {author} {\bibnamefont
  {et~al.}},\ }\bibfield  {title} {\enquote {\bibinfo {title} {{First Result
  from the Alpha Magnetic Spectrometer on the International Space Station:
  Precision Measurement of the Positron Fraction in Primary Cosmic Rays of
  0.5-350 GeV}},}\ }\href {\doibase 10.1103/PhysRevLett.110.141102} {\bibfield
  {journal} {\bibinfo  {journal} {Physical Review Letters}\ }\textbf {\bibinfo
  {volume} {110}},\ \bibinfo {eid} {141102} (\bibinfo {year}
  {2013}{\natexlab{b}})}\BibitemShut {NoStop}%
\bibitem [{\citenamefont {{AMS collaboration}}\ \emph
  {et~al.}(2014{\natexlab{b}})\citenamefont {{AMS collaboration}},
  \citenamefont {{Accardo}}, \citenamefont {{Aguilar}}, \citenamefont {{Aisa}},
  \citenamefont {{Alvino}}, \citenamefont {{Ambrosi}}, \citenamefont
  {{Andeen}}, \citenamefont {{Arruda}}, \citenamefont {{Attig}}, \citenamefont
  {{Azzarello}}, \citenamefont {{Bachlechner}},\ and\ \citenamefont
  {et~al.}}]{AMS02_fraction02}%
  \BibitemOpen
  \bibfield  {author} {\bibinfo {author} {\bibnamefont {{AMS collaboration}}},
  \bibinfo {author} {\bibfnamefont {L.}~\bibnamefont {{Accardo}}}, \bibinfo
  {author} {\bibfnamefont {M.}~\bibnamefont {{Aguilar}}}, \bibinfo {author}
  {\bibfnamefont {D.}~\bibnamefont {{Aisa}}}, \bibinfo {author} {\bibfnamefont
  {A.}~\bibnamefont {{Alvino}}}, \bibinfo {author} {\bibfnamefont
  {G.}~\bibnamefont {{Ambrosi}}}, \bibinfo {author} {\bibfnamefont
  {K.}~\bibnamefont {{Andeen}}}, \bibinfo {author} {\bibfnamefont
  {L.}~\bibnamefont {{Arruda}}}, \bibinfo {author} {\bibfnamefont
  {N.}~\bibnamefont {{Attig}}}, \bibinfo {author} {\bibfnamefont
  {P.}~\bibnamefont {{Azzarello}}}, \bibinfo {author} {\bibfnamefont
  {A.}~\bibnamefont {{Bachlechner}}}, \ and\ \bibinfo {author} {\bibnamefont
  {et~al.}},\ }\bibfield  {title} {\enquote {\bibinfo {title} {{High Statistics
  Measurement of the Positron Fraction in Primary Cosmic Rays of 0.5-500 GeV
  with the Alpha Magnetic Spectrometer on the International Space Station}},}\
  }\href {\doibase 10.1103/PhysRevLett.113.121101} {\bibfield  {journal}
  {\bibinfo  {journal} {Physical Review Letters}\ }\textbf {\bibinfo {volume}
  {113}},\ \bibinfo {eid} {121101} (\bibinfo {year}
  {2014}{\natexlab{b}})}\BibitemShut {NoStop}%
\bibitem [{\citenamefont {{AMS collaboration}}\ \emph
  {et~al.}(2014{\natexlab{c}})\citenamefont {{AMS collaboration}},
  \citenamefont {{Aguilar}}, \citenamefont {{Aisa}}, \citenamefont {{Alvino}},
  \citenamefont {{Ambrosi}}, \citenamefont {{Andeen}}, \citenamefont
  {{Arruda}}, \citenamefont {{Attig}}, \citenamefont {{Azzarello}},
  \citenamefont {{Bachlechner}}, \citenamefont {{Barao}},\ and\ \citenamefont
  {et~al.}}]{AMS02_lepton}%
  \BibitemOpen
  \bibfield  {author} {\bibinfo {author} {\bibnamefont {{AMS collaboration}}},
  \bibinfo {author} {\bibfnamefont {M.}~\bibnamefont {{Aguilar}}}, \bibinfo
  {author} {\bibfnamefont {D.}~\bibnamefont {{Aisa}}}, \bibinfo {author}
  {\bibfnamefont {A.}~\bibnamefont {{Alvino}}}, \bibinfo {author}
  {\bibfnamefont {G.}~\bibnamefont {{Ambrosi}}}, \bibinfo {author}
  {\bibfnamefont {K.}~\bibnamefont {{Andeen}}}, \bibinfo {author}
  {\bibfnamefont {L.}~\bibnamefont {{Arruda}}}, \bibinfo {author}
  {\bibfnamefont {N.}~\bibnamefont {{Attig}}}, \bibinfo {author} {\bibfnamefont
  {P.}~\bibnamefont {{Azzarello}}}, \bibinfo {author} {\bibfnamefont
  {A.}~\bibnamefont {{Bachlechner}}}, \bibinfo {author} {\bibfnamefont
  {F.}~\bibnamefont {{Barao}}}, \ and\ \bibinfo {author} {\bibnamefont
  {et~al.}},\ }\bibfield  {title} {\enquote {\bibinfo {title} {{Electron and
  Positron Fluxes in Primary Cosmic Rays Measured with the Alpha Magnetic
  Spectrometer on the International Space Station}},}\ }\href {\doibase
  10.1103/PhysRevLett.113.121102} {\bibfield  {journal} {\bibinfo  {journal}
  {Physical Review Letters}\ }\textbf {\bibinfo {volume} {113}},\ \bibinfo
  {eid} {121102} (\bibinfo {year} {2014}{\natexlab{c}})}\BibitemShut {NoStop}%
\bibitem [{\citenamefont {{Shen}}(1970)}]{Shen1970}%
  \BibitemOpen
  \bibfield  {author} {\bibinfo {author} {\bibfnamefont {C.~S.}\ \bibnamefont
  {{Shen}}},\ }\bibfield  {title} {\enquote {\bibinfo {title} {{Pulsars and
  Very High-Energy Cosmic-Ray Electrons}},}\ }\href {\doibase 10.1086/180650}
  {\bibfield  {journal} {\bibinfo  {journal} {\apjl}\ }\textbf {\bibinfo
  {volume} {162}},\ \bibinfo {pages} {L181} (\bibinfo {year}
  {1970})}\BibitemShut {NoStop}%
\bibitem [{\citenamefont {{Zhang}}\ and\ \citenamefont
  {{Cheng}}(2001)}]{Zhang2001}%
  \BibitemOpen
  \bibfield  {author} {\bibinfo {author} {\bibfnamefont {L.}~\bibnamefont
  {{Zhang}}}\ and\ \bibinfo {author} {\bibfnamefont {K.~S.}\ \bibnamefont
  {{Cheng}}},\ }\bibfield  {title} {\enquote {\bibinfo {title} {{Cosmic-ray
  positrons from mature gamma-ray pulsars}},}\ }\href {\doibase
  10.1051/0004-6361:20010021} {\bibfield  {journal} {\bibinfo  {journal}
  {\aap}\ }\textbf {\bibinfo {volume} {368}},\ \bibinfo {pages} {1063--1070}
  (\bibinfo {year} {2001})}\BibitemShut {NoStop}%
\bibitem [{\citenamefont {{Y{\"u}ksel}}\ \emph {et~al.}(2009)\citenamefont
  {{Y{\"u}ksel}}, \citenamefont {{Kistler}},\ and\ \citenamefont
  {{Stanev}}}]{Yuksel2009}%
  \BibitemOpen
  \bibfield  {author} {\bibinfo {author} {\bibfnamefont {H.}~\bibnamefont
  {{Y{\"u}ksel}}}, \bibinfo {author} {\bibfnamefont {M.~D.}\ \bibnamefont
  {{Kistler}}}, \ and\ \bibinfo {author} {\bibfnamefont {T.}~\bibnamefont
  {{Stanev}}},\ }\bibfield  {title} {\enquote {\bibinfo {title} {{TeV Gamma
  Rays from Geminga and the Origin of the GeV Positron Excess}},}\ }\href
  {\doibase 10.1103/PhysRevLett.103.051101} {\bibfield  {journal} {\bibinfo
  {journal} {Physical Review Letters}\ }\textbf {\bibinfo {volume} {103}},\
  \bibinfo {eid} {051101} (\bibinfo {year} {2009})},\ \Eprint
  {http://arxiv.org/abs/0810.2784} {arXiv:0810.2784} \BibitemShut {NoStop}%
\bibitem [{\citenamefont {{Hooper}}\ \emph {et~al.}(2009)\citenamefont
  {{Hooper}}, \citenamefont {{Blasi}},\ and\ \citenamefont {{Dario
  Serpico}}}]{Hooper2009}%
  \BibitemOpen
  \bibfield  {author} {\bibinfo {author} {\bibfnamefont {D.}~\bibnamefont
  {{Hooper}}}, \bibinfo {author} {\bibfnamefont {P.}~\bibnamefont {{Blasi}}}, \
  and\ \bibinfo {author} {\bibfnamefont {P.}~\bibnamefont {{Dario Serpico}}},\
  }\bibfield  {title} {\enquote {\bibinfo {title} {{Pulsars as the sources of
  high energy cosmic ray positrons}},}\ }\href {\doibase
  10.1088/1475-7516/2009/01/025} {\bibfield  {journal} {\bibinfo  {journal}
  {\jcap}\ }\textbf {\bibinfo {volume} {1}},\ \bibinfo {eid} {025} (\bibinfo
  {year} {2009})},\ \Eprint {http://arxiv.org/abs/0810.1527} {arXiv:0810.1527}
  \BibitemShut {NoStop}%
\bibitem [{\citenamefont {{Blasi}}(2009)}]{Blasi2009}%
  \BibitemOpen
  \bibfield  {author} {\bibinfo {author} {\bibfnamefont {P.}~\bibnamefont
  {{Blasi}}},\ }\bibfield  {title} {\enquote {\bibinfo {title} {{Origin of the
  Positron Excess in Cosmic Rays}},}\ }\href {\doibase
  10.1103/PhysRevLett.103.051104} {\bibfield  {journal} {\bibinfo  {journal}
  {Physical Review Letters}\ }\textbf {\bibinfo {volume} {103}},\ \bibinfo
  {eid} {051104} (\bibinfo {year} {2009})},\ \Eprint
  {http://arxiv.org/abs/0903.2794} {arXiv:0903.2794 [astro-ph.HE]} \BibitemShut
  {NoStop}%
\bibitem [{\citenamefont {{Hu}}\ \emph {et~al.}(2009)\citenamefont {{Hu}},
  \citenamefont {{Yuan}}, \citenamefont {{Wang}}, \citenamefont {{Fan}},
  \citenamefont {{Zhang}},\ and\ \citenamefont {{Bi}}}]{Hu2009}%
  \BibitemOpen
  \bibfield  {author} {\bibinfo {author} {\bibfnamefont {H.-B.}\ \bibnamefont
  {{Hu}}}, \bibinfo {author} {\bibfnamefont {Q.}~\bibnamefont {{Yuan}}},
  \bibinfo {author} {\bibfnamefont {B.}~\bibnamefont {{Wang}}}, \bibinfo
  {author} {\bibfnamefont {C.}~\bibnamefont {{Fan}}}, \bibinfo {author}
  {\bibfnamefont {J.-L.}\ \bibnamefont {{Zhang}}}, \ and\ \bibinfo {author}
  {\bibfnamefont {X.-J.}\ \bibnamefont {{Bi}}},\ }\bibfield  {title} {\enquote
  {\bibinfo {title} {{On the e$^{+}$e$^{-}$ Excesses and the Knee of the Cosmic
  Ray Spectra Hints of Cosmic Ray Acceleration in Young Supernova Remnants}},}\
  }\href {\doibase 10.1088/0004-637X/700/2/L170} {\bibfield  {journal}
  {\bibinfo  {journal} {\apjl}\ }\textbf {\bibinfo {volume} {700}},\ \bibinfo
  {pages} {L170--L173} (\bibinfo {year} {2009})},\ \Eprint
  {http://arxiv.org/abs/0901.1520} {arXiv:0901.1520 [astro-ph.HE]} \BibitemShut
  {NoStop}%
\bibitem [{\citenamefont {{Fujita}}\ \emph {et~al.}(2009)\citenamefont
  {{Fujita}}, \citenamefont {{Kohri}}, \citenamefont {{Yamazaki}},\ and\
  \citenamefont {{Ioka}}}]{Fujita2009}%
  \BibitemOpen
  \bibfield  {author} {\bibinfo {author} {\bibfnamefont {Y.}~\bibnamefont
  {{Fujita}}}, \bibinfo {author} {\bibfnamefont {K.}~\bibnamefont {{Kohri}}},
  \bibinfo {author} {\bibfnamefont {R.}~\bibnamefont {{Yamazaki}}}, \ and\
  \bibinfo {author} {\bibfnamefont {K.}~\bibnamefont {{Ioka}}},\ }\bibfield
  {title} {\enquote {\bibinfo {title} {{Is the PAMELA anomaly caused by
  supernova explosions near the Earth?}}}\ }\href {\doibase
  10.1103/PhysRevD.80.063003} {\bibfield  {journal} {\bibinfo  {journal}
  {\prd}\ }\textbf {\bibinfo {volume} {80}},\ \bibinfo {eid} {063003} (\bibinfo
  {year} {2009})},\ \Eprint {http://arxiv.org/abs/0903.5298} {arXiv:0903.5298
  [astro-ph.HE]} \BibitemShut {NoStop}%
\bibitem [{\citenamefont {{Bergstr{\"o}m}}\ \emph {et~al.}(2008)\citenamefont
  {{Bergstr{\"o}m}}, \citenamefont {{Bringmann}},\ and\ \citenamefont
  {{Edsj{\"o}}}}]{Bergstrom2008}%
  \BibitemOpen
  \bibfield  {author} {\bibinfo {author} {\bibfnamefont {L.}~\bibnamefont
  {{Bergstr{\"o}m}}}, \bibinfo {author} {\bibfnamefont {T.}~\bibnamefont
  {{Bringmann}}}, \ and\ \bibinfo {author} {\bibfnamefont {J.}~\bibnamefont
  {{Edsj{\"o}}}},\ }\bibfield  {title} {\enquote {\bibinfo {title} {{New
  positron spectral features from supersymmetric dark matter: A way to explain
  the PAMELA data?}}}\ }\href {\doibase 10.1103/PhysRevD.78.103520} {\bibfield
  {journal} {\bibinfo  {journal} {\prd}\ }\textbf {\bibinfo {volume} {78}},\
  \bibinfo {eid} {103520} (\bibinfo {year} {2008})},\ \Eprint
  {http://arxiv.org/abs/0808.3725} {arXiv:0808.3725} \BibitemShut {NoStop}%
\bibitem [{\citenamefont {{Barger}}\ \emph {et~al.}(2009)\citenamefont
  {{Barger}}, \citenamefont {{Keung}}, \citenamefont {{Marfatia}},\ and\
  \citenamefont {{Shaughnessy}}}]{Barger2009}%
  \BibitemOpen
  \bibfield  {author} {\bibinfo {author} {\bibfnamefont {V.}~\bibnamefont
  {{Barger}}}, \bibinfo {author} {\bibfnamefont {W.-Y.}\ \bibnamefont
  {{Keung}}}, \bibinfo {author} {\bibfnamefont {D.}~\bibnamefont {{Marfatia}}},
  \ and\ \bibinfo {author} {\bibfnamefont {G.}~\bibnamefont {{Shaughnessy}}},\
  }\bibfield  {title} {\enquote {\bibinfo {title} {{PAMELA and dark matter}},}\
  }\href {\doibase 10.1016/j.physletb.2009.01.016} {\bibfield  {journal}
  {\bibinfo  {journal} {Physics Letters B}\ }\textbf {\bibinfo {volume}
  {672}},\ \bibinfo {pages} {141--146} (\bibinfo {year} {2009})},\ \Eprint
  {http://arxiv.org/abs/0809.0162} {arXiv:0809.0162 [hep-ph]} \BibitemShut
  {NoStop}%
\bibitem [{\citenamefont {{Cirelli}}\ \emph {et~al.}(2009)\citenamefont
  {{Cirelli}}, \citenamefont {{Kadastik}}, \citenamefont {{Raidal}},\ and\
  \citenamefont {{Strumia}}}]{Cirelli2009}%
  \BibitemOpen
  \bibfield  {author} {\bibinfo {author} {\bibfnamefont {M.}~\bibnamefont
  {{Cirelli}}}, \bibinfo {author} {\bibfnamefont {M.}~\bibnamefont
  {{Kadastik}}}, \bibinfo {author} {\bibfnamefont {M.}~\bibnamefont
  {{Raidal}}}, \ and\ \bibinfo {author} {\bibfnamefont {A.}~\bibnamefont
  {{Strumia}}},\ }\bibfield  {title} {\enquote {\bibinfo {title}
  {{Model-independent implications of the e$^{+}$, e$^{-}$, anti-proton cosmic
  ray spectra on properties of Dark Matter}},}\ }\href {\doibase
  10.1016/j.nuclphysb.2008.11.031} {\bibfield  {journal} {\bibinfo  {journal}
  {Nuclear Physics B}\ }\textbf {\bibinfo {volume} {813}},\ \bibinfo {pages}
  {1--21} (\bibinfo {year} {2009})},\ \Eprint {http://arxiv.org/abs/0809.2409}
  {arXiv:0809.2409 [hep-ph]} \BibitemShut {NoStop}%
\bibitem [{\citenamefont {{Zhang}}\ \emph {et~al.}(2009)\citenamefont
  {{Zhang}}, \citenamefont {{Bi}}, \citenamefont {{Liu}}, \citenamefont
  {{Liu}}, \citenamefont {{Yin}}, \citenamefont {{Yuan}},\ and\ \citenamefont
  {{Zhu}}}]{Zhang2009}%
  \BibitemOpen
  \bibfield  {author} {\bibinfo {author} {\bibfnamefont {J.}~\bibnamefont
  {{Zhang}}}, \bibinfo {author} {\bibfnamefont {X.-J.}\ \bibnamefont {{Bi}}},
  \bibinfo {author} {\bibfnamefont {J.}~\bibnamefont {{Liu}}}, \bibinfo
  {author} {\bibfnamefont {S.-M.}\ \bibnamefont {{Liu}}}, \bibinfo {author}
  {\bibfnamefont {P.-F.}\ \bibnamefont {{Yin}}}, \bibinfo {author}
  {\bibfnamefont {Q.}~\bibnamefont {{Yuan}}}, \ and\ \bibinfo {author}
  {\bibfnamefont {S.-H.}\ \bibnamefont {{Zhu}}},\ }\bibfield  {title} {\enquote
  {\bibinfo {title} {{Discriminating different scenarios to account for the
  cosmic e$^{±}$ excess by synchrotron and inverse Compton radiation}},}\
  }\href {\doibase 10.1103/PhysRevD.80.023007} {\bibfield  {journal} {\bibinfo
  {journal} {\prd}\ }\textbf {\bibinfo {volume} {80}},\ \bibinfo {eid} {023007}
  (\bibinfo {year} {2009})},\ \Eprint {http://arxiv.org/abs/0812.0522}
  {arXiv:0812.0522} \BibitemShut {NoStop}%
\bibitem [{\citenamefont {{Bergstr{\"o}m}}\ \emph {et~al.}(2009)\citenamefont
  {{Bergstr{\"o}m}}, \citenamefont {{Edsj{\"o}}},\ and\ \citenamefont
  {{Zaharijas}}}]{Bergstrom2009}%
  \BibitemOpen
  \bibfield  {author} {\bibinfo {author} {\bibfnamefont {L.}~\bibnamefont
  {{Bergstr{\"o}m}}}, \bibinfo {author} {\bibfnamefont {J.}~\bibnamefont
  {{Edsj{\"o}}}}, \ and\ \bibinfo {author} {\bibfnamefont {G.}~\bibnamefont
  {{Zaharijas}}},\ }\bibfield  {title} {\enquote {\bibinfo {title} {{Dark
  Matter Interpretation of Recent Electron and Positron Data}},}\ }\href
  {\doibase 10.1103/PhysRevLett.103.031103} {\bibfield  {journal} {\bibinfo
  {journal} {Physical Review Letters}\ }\textbf {\bibinfo {volume} {103}},\
  \bibinfo {eid} {031103} (\bibinfo {year} {2009})},\ \Eprint
  {http://arxiv.org/abs/0905.0333} {arXiv:0905.0333 [astro-ph.HE]} \BibitemShut
  {NoStop}%
\bibitem [{\citenamefont {{Yin}}\ \emph {et~al.}(2013)\citenamefont {{Yin}},
  \citenamefont {{Yu}}, \citenamefont {{Yuan}},\ and\ \citenamefont
  {{Bi}}}]{Yin2013}%
  \BibitemOpen
  \bibfield  {author} {\bibinfo {author} {\bibfnamefont {P.-F.}\ \bibnamefont
  {{Yin}}}, \bibinfo {author} {\bibfnamefont {Z.-H.}\ \bibnamefont {{Yu}}},
  \bibinfo {author} {\bibfnamefont {Q.}~\bibnamefont {{Yuan}}}, \ and\ \bibinfo
  {author} {\bibfnamefont {X.-J.}\ \bibnamefont {{Bi}}},\ }\bibfield  {title}
  {\enquote {\bibinfo {title} {{Pulsar interpretation for the AMS-02
  result}},}\ }\href {\doibase 10.1103/PhysRevD.88.023001} {\bibfield
  {journal} {\bibinfo  {journal} {\prd}\ }\textbf {\bibinfo {volume} {88}},\
  \bibinfo {eid} {023001} (\bibinfo {year} {2013})},\ \Eprint
  {http://arxiv.org/abs/1304.4128} {arXiv:1304.4128 [astro-ph.HE]} \BibitemShut
  {NoStop}%
\bibitem [{\citenamefont {{Dev}}\ \emph {et~al.}(2014)\citenamefont {{Dev}},
  \citenamefont {{Ghosh}}, \citenamefont {{Okada}},\ and\ \citenamefont
  {{Saha}}}]{Dev2014}%
  \BibitemOpen
  \bibfield  {author} {\bibinfo {author} {\bibfnamefont {P.~S.~B.}\
  \bibnamefont {{Dev}}}, \bibinfo {author} {\bibfnamefont {D.~K.}\ \bibnamefont
  {{Ghosh}}}, \bibinfo {author} {\bibfnamefont {N.}~\bibnamefont {{Okada}}}, \
  and\ \bibinfo {author} {\bibfnamefont {I.}~\bibnamefont {{Saha}}},\
  }\bibfield  {title} {\enquote {\bibinfo {title} {{Neutrino mass and dark
  matter in light of recent AMS-02 results}},}\ }\href {\doibase
  10.1103/PhysRevD.89.095001} {\bibfield  {journal} {\bibinfo  {journal}
  {\prd}\ }\textbf {\bibinfo {volume} {89}},\ \bibinfo {eid} {095001} (\bibinfo
  {year} {2014})},\ \Eprint {http://arxiv.org/abs/1307.6204} {arXiv:1307.6204
  [hep-ph]} \BibitemShut {NoStop}%
\bibitem [{\citenamefont {{Lewis}}\ and\ \citenamefont
  {{Bridle}}(2002)}]{Lewis2002}%
  \BibitemOpen
  \bibfield  {author} {\bibinfo {author} {\bibfnamefont {A.}~\bibnamefont
  {{Lewis}}}\ and\ \bibinfo {author} {\bibfnamefont {S.}~\bibnamefont
  {{Bridle}}},\ }\bibfield  {title} {\enquote {\bibinfo {title} {{Cosmological
  parameters from CMB and other data: A Monte Carlo approach}},}\ }\href
  {\doibase 10.1103/PhysRevD.66.103511} {\bibfield  {journal} {\bibinfo
  {journal} {\prd}\ }\textbf {\bibinfo {volume} {66}},\ \bibinfo {eid} {103511}
  (\bibinfo {year} {2002})},\ \Eprint {http://arxiv.org/abs/astro-ph/0205436}
  {astro-ph/0205436} \BibitemShut {NoStop}%
\bibitem [{\citenamefont {{Liu}}\ \emph {et~al.}(2010)\citenamefont {{Liu}},
  \citenamefont {{Yuan}}, \citenamefont {{Bi}}, \citenamefont {{Li}},\ and\
  \citenamefont {{Zhang}}}]{Liu2010}%
  \BibitemOpen
  \bibfield  {author} {\bibinfo {author} {\bibfnamefont {J.}~\bibnamefont
  {{Liu}}}, \bibinfo {author} {\bibfnamefont {Q.}~\bibnamefont {{Yuan}}},
  \bibinfo {author} {\bibfnamefont {X.}~\bibnamefont {{Bi}}}, \bibinfo {author}
  {\bibfnamefont {H.}~\bibnamefont {{Li}}}, \ and\ \bibinfo {author}
  {\bibfnamefont {X.}~\bibnamefont {{Zhang}}},\ }\bibfield  {title} {\enquote
  {\bibinfo {title} {{Markov chain Monte Carlo study on dark matter property
  related to the cosmic e$^{±}$ excesses}},}\ }\href {\doibase
  10.1103/PhysRevD.81.023516} {\bibfield  {journal} {\bibinfo  {journal}
  {\prd}\ }\textbf {\bibinfo {volume} {81}},\ \bibinfo {eid} {023516} (\bibinfo
  {year} {2010})},\ \Eprint {http://arxiv.org/abs/0906.3858} {arXiv:0906.3858
  [astro-ph.CO]} \BibitemShut {NoStop}%
\bibitem [{\citenamefont {Lin}\ \emph {et~al.}(2015)\citenamefont {Lin},
  \citenamefont {Yuan},\ and\ \citenamefont {Bi}}]{Lin2015}%
  \BibitemOpen
  \bibfield  {author} {\bibinfo {author} {\bibfnamefont {Su-Jie}\ \bibnamefont
  {Lin}}, \bibinfo {author} {\bibfnamefont {Qiang}\ \bibnamefont {Yuan}}, \
  and\ \bibinfo {author} {\bibfnamefont {Xiao-Jun}\ \bibnamefont {Bi}},\
  }\bibfield  {title} {\enquote {\bibinfo {title} {Quantitative study of the
  ams-02 electron/positron spectra: implications for the pulsar and dark matter
  properties},}\ }\href {\doibase 10.1103/PhysRevD.91.063508} {\bibfield
  {journal} {\bibinfo  {journal} {Physical Review D}\ }\textbf {\bibinfo
  {volume} {91}},\ \bibinfo {pages} {063508} (\bibinfo {year} {2015})},\
  \Eprint {http://arxiv.org/abs/1409.6248} {arXiv:1409.6248 [astro-ph.HE]}
  \BibitemShut {NoStop}%
\bibitem [{\citenamefont {{Yuan}}\ \emph
  {et~al.}(2017{\natexlab{b}})\citenamefont {{Yuan}}, \citenamefont {{Lin}},
  \citenamefont {{Fang}},\ and\ \citenamefont {{Bi}}}]{Yuan2017}%
  \BibitemOpen
  \bibfield  {author} {\bibinfo {author} {\bibfnamefont {Q.}~\bibnamefont
  {{Yuan}}}, \bibinfo {author} {\bibfnamefont {S.-J.}\ \bibnamefont {{Lin}}},
  \bibinfo {author} {\bibfnamefont {K.}~\bibnamefont {{Fang}}}, \ and\ \bibinfo
  {author} {\bibfnamefont {X.-J.}\ \bibnamefont {{Bi}}},\ }\bibfield  {title}
  {\enquote {\bibinfo {title} {{Propagation of cosmic rays in the AMS-02
  era}},}\ }\href@noop {} {\bibfield  {journal} {\bibinfo  {journal} {ArXiv
  e-prints}\ } (\bibinfo {year} {2017}{\natexlab{b}})},\ \Eprint
  {http://arxiv.org/abs/1701.06149} {arXiv:1701.06149 [astro-ph.HE]}
  \BibitemShut {NoStop}%
\bibitem [{\citenamefont {{Niu}}\ and\ \citenamefont {{Li}}(2018)}]{Niu2017}%
  \BibitemOpen
  \bibfield  {author} {\bibinfo {author} {\bibfnamefont {J.-S.}\ \bibnamefont
  {{Niu}}}\ and\ \bibinfo {author} {\bibfnamefont {T.}~\bibnamefont {{Li}}},\
  }\bibfield  {title} {\enquote {\bibinfo {title} {{Galactic cosmic-ray model
  in the light of AMS-02 nuclei data}},}\ }\href {\doibase
  10.1103/PhysRevD.97.023015} {\bibfield  {journal} {\bibinfo  {journal}
  {\prd}\ }\textbf {\bibinfo {volume} {97}},\ \bibinfo {eid} {023015} (\bibinfo
  {year} {2018})},\ \Eprint {http://arxiv.org/abs/1705.11089} {arXiv:1705.11089
  [astro-ph.HE]} \BibitemShut {NoStop}%
\bibitem [{\citenamefont {{Panov}}\ \emph {et~al.}(2006)\citenamefont
  {{Panov}}, \citenamefont {{Adams}}, \citenamefont {{Ahn}}, \citenamefont
  {{Bashindzhagyan}}, \citenamefont {{Batkov}}, \citenamefont {{Chang}},
  \citenamefont {{Christl}}, \citenamefont {{Fazely}}, \citenamefont {{Ganel}},
  \citenamefont {{Gunashingha}}, \citenamefont {{Guzik}}, \citenamefont
  {{Isbert}}, \citenamefont {{Kim}}, \citenamefont {{Kouznetsov}},
  \citenamefont {{Panasyuk}}, \citenamefont {{Schmidt}}, \citenamefont {{Seo}},
  \citenamefont {{Sokolskaya}}, \citenamefont {{Watts}}, \citenamefont
  {{Wefel}}, \citenamefont {{Wu}},\ and\ \citenamefont
  {{Zatsepin}}}]{ATIC2006}%
  \BibitemOpen
  \bibfield  {author} {\bibinfo {author} {\bibfnamefont {A.~D.}\ \bibnamefont
  {{Panov}}}, \bibinfo {author} {\bibfnamefont {J.~H.}\ \bibnamefont
  {{Adams}}}, \bibinfo {author} {\bibfnamefont {H.~S.}\ \bibnamefont {{Ahn}}},
  \bibinfo {author} {\bibfnamefont {G.~L.}\ \bibnamefont {{Bashindzhagyan}}},
  \bibinfo {author} {\bibfnamefont {K.~E.}\ \bibnamefont {{Batkov}}}, \bibinfo
  {author} {\bibfnamefont {J.}~\bibnamefont {{Chang}}}, \bibinfo {author}
  {\bibfnamefont {M.}~\bibnamefont {{Christl}}}, \bibinfo {author}
  {\bibfnamefont {A.~R.}\ \bibnamefont {{Fazely}}}, \bibinfo {author}
  {\bibfnamefont {O.}~\bibnamefont {{Ganel}}}, \bibinfo {author} {\bibfnamefont
  {R.~M.}\ \bibnamefont {{Gunashingha}}}, \bibinfo {author} {\bibfnamefont
  {T.~G.}\ \bibnamefont {{Guzik}}}, \bibinfo {author} {\bibfnamefont
  {J.}~\bibnamefont {{Isbert}}}, \bibinfo {author} {\bibfnamefont {K.~C.}\
  \bibnamefont {{Kim}}}, \bibinfo {author} {\bibfnamefont {E.~N.}\ \bibnamefont
  {{Kouznetsov}}}, \bibinfo {author} {\bibfnamefont {M.~I.}\ \bibnamefont
  {{Panasyuk}}}, \bibinfo {author} {\bibfnamefont {W.~K.~H.}\ \bibnamefont
  {{Schmidt}}}, \bibinfo {author} {\bibfnamefont {E.~S.}\ \bibnamefont
  {{Seo}}}, \bibinfo {author} {\bibfnamefont {N.~V.}\ \bibnamefont
  {{Sokolskaya}}}, \bibinfo {author} {\bibfnamefont {J.~W.}\ \bibnamefont
  {{Watts}}}, \bibinfo {author} {\bibfnamefont {J.~P.}\ \bibnamefont
  {{Wefel}}}, \bibinfo {author} {\bibfnamefont {J.}~\bibnamefont {{Wu}}}, \
  and\ \bibinfo {author} {\bibfnamefont {V.~I.}\ \bibnamefont {{Zatsepin}}},\
  }\bibfield  {title} {\enquote {\bibinfo {title} {{The results of ATIC-2
  experiment for elemental spectra of cosmic rays}},}\ }\href@noop {}
  {\bibfield  {journal} {\bibinfo  {journal} {ArXiv Astrophysics e-prints}\ }
  (\bibinfo {year} {2006})},\ \Eprint {http://arxiv.org/abs/astro-ph/0612377}
  {astro-ph/0612377} \BibitemShut {NoStop}%
\bibitem [{\citenamefont {{Ahn}}\ \emph {et~al.}(2010)\citenamefont {{Ahn}},
  \citenamefont {{Allison}}, \citenamefont {{Bagliesi}}, \citenamefont
  {{Beatty}}, \citenamefont {{Bigongiari}}, \citenamefont {{Childers}},
  \citenamefont {{Conklin}}, \citenamefont {{Coutu}}, \citenamefont
  {{DuVernois}}, \citenamefont {{Ganel}}, \citenamefont {{Han}}, \citenamefont
  {{Jeon}}, \citenamefont {{Kim}}, \citenamefont {{Lee}}, \citenamefont
  {{Lutz}}, \citenamefont {{Maestro}}, \citenamefont {{Malinin}}, \citenamefont
  {{Marrocchesi}}, \citenamefont {{Minnick}}, \citenamefont {{Mognet}},
  \citenamefont {{Nam}}, \citenamefont {{Nam}}, \citenamefont {{Nutter}},
  \citenamefont {{Park}}, \citenamefont {{Park}}, \citenamefont {{Seo}},
  \citenamefont {{Sina}}, \citenamefont {{Wu}}, \citenamefont {{Yang}},
  \citenamefont {{Yoon}}, \citenamefont {{Zei}},\ and\ \citenamefont
  {{Zinn}}}]{CREAM2010}%
  \BibitemOpen
  \bibfield  {author} {\bibinfo {author} {\bibfnamefont {H.~S.}\ \bibnamefont
  {{Ahn}}}, \bibinfo {author} {\bibfnamefont {P.}~\bibnamefont {{Allison}}},
  \bibinfo {author} {\bibfnamefont {M.~G.}\ \bibnamefont {{Bagliesi}}},
  \bibinfo {author} {\bibfnamefont {J.~J.}\ \bibnamefont {{Beatty}}}, \bibinfo
  {author} {\bibfnamefont {G.}~\bibnamefont {{Bigongiari}}}, \bibinfo {author}
  {\bibfnamefont {J.~T.}\ \bibnamefont {{Childers}}}, \bibinfo {author}
  {\bibfnamefont {N.~B.}\ \bibnamefont {{Conklin}}}, \bibinfo {author}
  {\bibfnamefont {S.}~\bibnamefont {{Coutu}}}, \bibinfo {author} {\bibfnamefont
  {M.~A.}\ \bibnamefont {{DuVernois}}}, \bibinfo {author} {\bibfnamefont
  {O.}~\bibnamefont {{Ganel}}}, \bibinfo {author} {\bibfnamefont {J.~H.}\
  \bibnamefont {{Han}}}, \bibinfo {author} {\bibfnamefont {J.~A.}\ \bibnamefont
  {{Jeon}}}, \bibinfo {author} {\bibfnamefont {K.~C.}\ \bibnamefont {{Kim}}},
  \bibinfo {author} {\bibfnamefont {M.~H.}\ \bibnamefont {{Lee}}}, \bibinfo
  {author} {\bibfnamefont {L.}~\bibnamefont {{Lutz}}}, \bibinfo {author}
  {\bibfnamefont {P.}~\bibnamefont {{Maestro}}}, \bibinfo {author}
  {\bibfnamefont {A.}~\bibnamefont {{Malinin}}}, \bibinfo {author}
  {\bibfnamefont {P.~S.}\ \bibnamefont {{Marrocchesi}}}, \bibinfo {author}
  {\bibfnamefont {S.}~\bibnamefont {{Minnick}}}, \bibinfo {author}
  {\bibfnamefont {S.~I.}\ \bibnamefont {{Mognet}}}, \bibinfo {author}
  {\bibfnamefont {J.}~\bibnamefont {{Nam}}}, \bibinfo {author} {\bibfnamefont
  {S.}~\bibnamefont {{Nam}}}, \bibinfo {author} {\bibfnamefont {S.~L.}\
  \bibnamefont {{Nutter}}}, \bibinfo {author} {\bibfnamefont {I.~H.}\
  \bibnamefont {{Park}}}, \bibinfo {author} {\bibfnamefont {N.~H.}\
  \bibnamefont {{Park}}}, \bibinfo {author} {\bibfnamefont {E.~S.}\
  \bibnamefont {{Seo}}}, \bibinfo {author} {\bibfnamefont {R.}~\bibnamefont
  {{Sina}}}, \bibinfo {author} {\bibfnamefont {J.}~\bibnamefont {{Wu}}},
  \bibinfo {author} {\bibfnamefont {J.}~\bibnamefont {{Yang}}}, \bibinfo
  {author} {\bibfnamefont {Y.~S.}\ \bibnamefont {{Yoon}}}, \bibinfo {author}
  {\bibfnamefont {R.}~\bibnamefont {{Zei}}}, \ and\ \bibinfo {author}
  {\bibfnamefont {S.~Y.}\ \bibnamefont {{Zinn}}},\ }\bibfield  {title}
  {\enquote {\bibinfo {title} {{Discrepant Hardening Observed in Cosmic-ray
  Elemental Spectra}},}\ }\href {\doibase 10.1088/2041-8205/714/1/L89}
  {\bibfield  {journal} {\bibinfo  {journal} {\apjl}\ }\textbf {\bibinfo
  {volume} {714}},\ \bibinfo {pages} {L89--L93} (\bibinfo {year} {2010})},\
  \Eprint {http://arxiv.org/abs/1004.1123} {arXiv:1004.1123 [astro-ph.HE]}
  \BibitemShut {NoStop}%
\bibitem [{\citenamefont {{PAMELA collaboration}}\ \emph
  {et~al.}(2011{\natexlab{a}})\citenamefont {{PAMELA collaboration}},
  \citenamefont {{Adriani}}, \citenamefont {{Barbarino}}, \citenamefont
  {{Bazilevskaya}}, \citenamefont {{Bellotti}}, \citenamefont {{Boezio}},
  \citenamefont {{Bogomolov}}, \citenamefont {{Bonechi}}, \citenamefont
  {{Bongi}}, \citenamefont {{Bonvicini}}, \citenamefont {{Borisov}},\ and\
  \citenamefont {el~al.}}]{PAMELA2011}%
  \BibitemOpen
  \bibfield  {author} {\bibinfo {author} {\bibnamefont {{PAMELA
  collaboration}}}, \bibinfo {author} {\bibfnamefont {O.}~\bibnamefont
  {{Adriani}}}, \bibinfo {author} {\bibfnamefont {G.~C.}\ \bibnamefont
  {{Barbarino}}}, \bibinfo {author} {\bibfnamefont {G.~A.}\ \bibnamefont
  {{Bazilevskaya}}}, \bibinfo {author} {\bibfnamefont {R.}~\bibnamefont
  {{Bellotti}}}, \bibinfo {author} {\bibfnamefont {M.}~\bibnamefont
  {{Boezio}}}, \bibinfo {author} {\bibfnamefont {E.~A.}\ \bibnamefont
  {{Bogomolov}}}, \bibinfo {author} {\bibfnamefont {L.}~\bibnamefont
  {{Bonechi}}}, \bibinfo {author} {\bibfnamefont {M.}~\bibnamefont {{Bongi}}},
  \bibinfo {author} {\bibfnamefont {V.}~\bibnamefont {{Bonvicini}}}, \bibinfo
  {author} {\bibfnamefont {S.}~\bibnamefont {{Borisov}}}, \ and\ \bibinfo
  {author} {\bibnamefont {el~al.}},\ }\bibfield  {title} {\enquote {\bibinfo
  {title} {{PAMELA Measurements of Cosmic-Ray Proton and Helium Spectra}},}\
  }\href {\doibase 10.1126/science.1199172} {\bibfield  {journal} {\bibinfo
  {journal} {Science}\ }\textbf {\bibinfo {volume} {332}},\ \bibinfo {pages}
  {69} (\bibinfo {year} {2011}{\natexlab{a}})},\ \Eprint
  {http://arxiv.org/abs/1103.4055} {arXiv:1103.4055 [astro-ph.HE]} \BibitemShut
  {NoStop}%
\bibitem [{\citenamefont {{AMS collaboration}}\ \emph
  {et~al.}(2015{\natexlab{a}})\citenamefont {{AMS collaboration}},
  \citenamefont {{Aguilar}}, \citenamefont {{Aisa}}, \citenamefont {{Alpat}},
  \citenamefont {{Alvino}}, \citenamefont {{Ambrosi}}, \citenamefont
  {{Andeen}}, \citenamefont {{Arruda}}, \citenamefont {{Attig}}, \citenamefont
  {{Azzarello}}, \citenamefont {{Bachlechner}},\ and\ \citenamefont
  {et~al.}}]{AMS02_proton}%
  \BibitemOpen
  \bibfield  {author} {\bibinfo {author} {\bibnamefont {{AMS collaboration}}},
  \bibinfo {author} {\bibfnamefont {M.}~\bibnamefont {{Aguilar}}}, \bibinfo
  {author} {\bibfnamefont {D.}~\bibnamefont {{Aisa}}}, \bibinfo {author}
  {\bibfnamefont {B.}~\bibnamefont {{Alpat}}}, \bibinfo {author} {\bibfnamefont
  {A.}~\bibnamefont {{Alvino}}}, \bibinfo {author} {\bibfnamefont
  {G.}~\bibnamefont {{Ambrosi}}}, \bibinfo {author} {\bibfnamefont
  {K.}~\bibnamefont {{Andeen}}}, \bibinfo {author} {\bibfnamefont
  {L.}~\bibnamefont {{Arruda}}}, \bibinfo {author} {\bibfnamefont
  {N.}~\bibnamefont {{Attig}}}, \bibinfo {author} {\bibfnamefont
  {P.}~\bibnamefont {{Azzarello}}}, \bibinfo {author} {\bibfnamefont
  {A.}~\bibnamefont {{Bachlechner}}}, \ and\ \bibinfo {author} {\bibnamefont
  {et~al.}},\ }\bibfield  {title} {\enquote {\bibinfo {title} {{Precision
  Measurement of the Proton Flux in Primary Cosmic Rays from Rigidity 1 GV to
  1.8 TV with the Alpha Magnetic Spectrometer on the International Space
  Station}},}\ }\href {\doibase 10.1103/PhysRevLett.114.171103} {\bibfield
  {journal} {\bibinfo  {journal} {Physical Review Letters}\ }\textbf {\bibinfo
  {volume} {114}},\ \bibinfo {eid} {171103} (\bibinfo {year}
  {2015}{\natexlab{a}})}\BibitemShut {NoStop}%
\bibitem [{\citenamefont {{AMS collaboration}}\ \emph
  {et~al.}(2015{\natexlab{b}})\citenamefont {{AMS collaboration}},
  \citenamefont {{Aguilar}}, \citenamefont {{Aisa}}, \citenamefont {{Alpat}},
  \citenamefont {{Alvino}}, \citenamefont {{Ambrosi}}, \citenamefont
  {{Andeen}}, \citenamefont {{Arruda}}, \citenamefont {{Attig}}, \citenamefont
  {{Azzarello}}, \citenamefont {{Bachlechner}},\ and\ \citenamefont
  {et~al.}}]{AMS02_helium}%
  \BibitemOpen
  \bibfield  {author} {\bibinfo {author} {\bibnamefont {{AMS collaboration}}},
  \bibinfo {author} {\bibfnamefont {M.}~\bibnamefont {{Aguilar}}}, \bibinfo
  {author} {\bibfnamefont {D.}~\bibnamefont {{Aisa}}}, \bibinfo {author}
  {\bibfnamefont {B.}~\bibnamefont {{Alpat}}}, \bibinfo {author} {\bibfnamefont
  {A.}~\bibnamefont {{Alvino}}}, \bibinfo {author} {\bibfnamefont
  {G.}~\bibnamefont {{Ambrosi}}}, \bibinfo {author} {\bibfnamefont
  {K.}~\bibnamefont {{Andeen}}}, \bibinfo {author} {\bibfnamefont
  {L.}~\bibnamefont {{Arruda}}}, \bibinfo {author} {\bibfnamefont
  {N.}~\bibnamefont {{Attig}}}, \bibinfo {author} {\bibfnamefont
  {P.}~\bibnamefont {{Azzarello}}}, \bibinfo {author} {\bibfnamefont
  {A.}~\bibnamefont {{Bachlechner}}}, \ and\ \bibinfo {author} {\bibnamefont
  {et~al.}},\ }\bibfield  {title} {\enquote {\bibinfo {title} {{Precision
  Measurement of the Helium Flux in Primary Cosmic Rays of Rigidities 1.9 GV to
  3 TV with the Alpha Magnetic Spectrometer on the International Space
  Station}},}\ }\href {\doibase 10.1103/PhysRevLett.115.211101} {\bibfield
  {journal} {\bibinfo  {journal} {Physical Review Letters}\ }\textbf {\bibinfo
  {volume} {115}},\ \bibinfo {eid} {211101} (\bibinfo {year}
  {2015}{\natexlab{b}})}\BibitemShut {NoStop}%
\bibitem [{\citenamefont {{Korsmeier}}\ and\ \citenamefont
  {{Cuoco}}(2016)}]{Korsmeier2016}%
  \BibitemOpen
  \bibfield  {author} {\bibinfo {author} {\bibfnamefont {M.}~\bibnamefont
  {{Korsmeier}}}\ and\ \bibinfo {author} {\bibfnamefont {A.}~\bibnamefont
  {{Cuoco}}},\ }\bibfield  {title} {\enquote {\bibinfo {title} {{Galactic
  cosmic-ray propagation in the light of AMS-02: I. Analysis of protons,
  helium, and antiprotons}},}\ }\href@noop {} {\bibfield  {journal} {\bibinfo
  {journal} {ArXiv e-prints}\ } (\bibinfo {year} {2016})},\ \Eprint
  {http://arxiv.org/abs/1607.06093} {arXiv:1607.06093 [astro-ph.HE]}
  \BibitemShut {NoStop}%
\bibitem [{\citenamefont {{Strong}}\ \emph {et~al.}(2000)\citenamefont
  {{Strong}}, \citenamefont {{Moskalenko}},\ and\ \citenamefont
  {{Reimer}}}]{Strong2000}%
  \BibitemOpen
  \bibfield  {author} {\bibinfo {author} {\bibfnamefont {A.~W.}\ \bibnamefont
  {{Strong}}}, \bibinfo {author} {\bibfnamefont {I.~V.}\ \bibnamefont
  {{Moskalenko}}}, \ and\ \bibinfo {author} {\bibfnamefont {O.}~\bibnamefont
  {{Reimer}}},\ }\bibfield  {title} {\enquote {\bibinfo {title} {{Diffuse
  Continuum Gamma Rays from the Galaxy}},}\ }\href {\doibase 10.1086/309038}
  {\bibfield  {journal} {\bibinfo  {journal} {\apj}\ }\textbf {\bibinfo
  {volume} {537}},\ \bibinfo {pages} {763--784} (\bibinfo {year} {2000})},\
  \Eprint {http://arxiv.org/abs/astro-ph/9811296} {astro-ph/9811296}
  \BibitemShut {NoStop}%
\bibitem [{\citenamefont {{PAMELA collaboration}}\ \emph
  {et~al.}(2011{\natexlab{b}})\citenamefont {{PAMELA collaboration}},
  \citenamefont {{Adriani}}, \citenamefont {{Barbarino}}, \citenamefont
  {{Bazilevskaya}}, \citenamefont {{Bellotti}}, \citenamefont {{Boezio}},
  \citenamefont {{Bogomolov}}, \citenamefont {{Bonechi}}, \citenamefont
  {{Bongi}}, \citenamefont {{Bonvicini}}, \citenamefont {{Borisov}},\ and\
  \citenamefont {el~al.}}]{Adriani2011}%
  \BibitemOpen
  \bibfield  {author} {\bibinfo {author} {\bibnamefont {{PAMELA
  collaboration}}}, \bibinfo {author} {\bibfnamefont {O.}~\bibnamefont
  {{Adriani}}}, \bibinfo {author} {\bibfnamefont {G.~C.}\ \bibnamefont
  {{Barbarino}}}, \bibinfo {author} {\bibfnamefont {G.~A.}\ \bibnamefont
  {{Bazilevskaya}}}, \bibinfo {author} {\bibfnamefont {R.}~\bibnamefont
  {{Bellotti}}}, \bibinfo {author} {\bibfnamefont {M.}~\bibnamefont
  {{Boezio}}}, \bibinfo {author} {\bibfnamefont {E.~A.}\ \bibnamefont
  {{Bogomolov}}}, \bibinfo {author} {\bibfnamefont {L.}~\bibnamefont
  {{Bonechi}}}, \bibinfo {author} {\bibfnamefont {M.}~\bibnamefont {{Bongi}}},
  \bibinfo {author} {\bibfnamefont {V.}~\bibnamefont {{Bonvicini}}}, \bibinfo
  {author} {\bibfnamefont {S.}~\bibnamefont {{Borisov}}}, \ and\ \bibinfo
  {author} {\bibnamefont {el~al.}},\ }\bibfield  {title} {\enquote {\bibinfo
  {title} {{PAMELA Measurements of Cosmic-Ray Proton and Helium Spectra}},}\
  }\href {\doibase 10.1126/science.1199172} {\bibfield  {journal} {\bibinfo
  {journal} {Science}\ }\textbf {\bibinfo {volume} {332}},\ \bibinfo {pages}
  {69} (\bibinfo {year} {2011}{\natexlab{b}})},\ \Eprint
  {http://arxiv.org/abs/1103.4055} {arXiv:1103.4055 [astro-ph.HE]} \BibitemShut
  {NoStop}%
\bibitem [{\citenamefont {{Tan}}\ and\ \citenamefont {{Ng}}(1983)}]{Tan1983}%
  \BibitemOpen
  \bibfield  {author} {\bibinfo {author} {\bibfnamefont {L.~C.}\ \bibnamefont
  {{Tan}}}\ and\ \bibinfo {author} {\bibfnamefont {L.~K.}\ \bibnamefont
  {{Ng}}},\ }\bibfield  {title} {\enquote {\bibinfo {title} {{Parametrisation
  of hadron inclusive cross sections in p-p collisions extended to very low
  energies}},}\ }\href {\doibase 10.1088/0305-4616/9/10/015} {\bibfield
  {journal} {\bibinfo  {journal} {Journal of Physics G Nuclear Physics}\
  }\textbf {\bibinfo {volume} {9}},\ \bibinfo {pages} {1289--1308} (\bibinfo
  {year} {1983})}\BibitemShut {NoStop}%
\bibitem [{\citenamefont {{Duperray}}\ \emph {et~al.}(2003)\citenamefont
  {{Duperray}}, \citenamefont {{Huang}}, \citenamefont {{Protasov}},\ and\
  \citenamefont {{Bu{\'e}nerd}}}]{Duperray2003}%
  \BibitemOpen
  \bibfield  {author} {\bibinfo {author} {\bibfnamefont {R.~P.}\ \bibnamefont
  {{Duperray}}}, \bibinfo {author} {\bibfnamefont {C.-Y.}\ \bibnamefont
  {{Huang}}}, \bibinfo {author} {\bibfnamefont {K.~V.}\ \bibnamefont
  {{Protasov}}}, \ and\ \bibinfo {author} {\bibfnamefont {M.}~\bibnamefont
  {{Bu{\'e}nerd}}},\ }\bibfield  {title} {\enquote {\bibinfo {title}
  {{Parametrization of the antiproton inclusive production cross section on
  nuclei}},}\ }\href {\doibase 10.1103/PhysRevD.68.094017} {\bibfield
  {journal} {\bibinfo  {journal} {\prd}\ }\textbf {\bibinfo {volume} {68}},\
  \bibinfo {eid} {094017} (\bibinfo {year} {2003})},\ \Eprint
  {http://arxiv.org/abs/astro-ph/0305274} {astro-ph/0305274} \BibitemShut
  {NoStop}%
\bibitem [{\citenamefont {{Kappl}}\ and\ \citenamefont
  {{Winkler}}(2014)}]{Kappl2014}%
  \BibitemOpen
  \bibfield  {author} {\bibinfo {author} {\bibfnamefont {R.}~\bibnamefont
  {{Kappl}}}\ and\ \bibinfo {author} {\bibfnamefont {M.~W.}\ \bibnamefont
  {{Winkler}}},\ }\bibfield  {title} {\enquote {\bibinfo {title} {{The cosmic
  ray antiproton background for AMS-02}},}\ }\href {\doibase
  10.1088/1475-7516/2014/09/051} {\bibfield  {journal} {\bibinfo  {journal}
  {\jcap}\ }\textbf {\bibinfo {volume} {9}},\ \bibinfo {eid} {051} (\bibinfo
  {year} {2014})},\ \Eprint {http://arxiv.org/abs/1408.0299} {arXiv:1408.0299
  [hep-ph]} \BibitemShut {NoStop}%
\bibitem [{\citenamefont {{di Mauro}}\ \emph {et~al.}(2014)\citenamefont {{di
  Mauro}}, \citenamefont {{Donato}}, \citenamefont {{Goudelis}},\ and\
  \citenamefont {{Serpico}}}]{diMauro2014}%
  \BibitemOpen
  \bibfield  {author} {\bibinfo {author} {\bibfnamefont {M.}~\bibnamefont {{di
  Mauro}}}, \bibinfo {author} {\bibfnamefont {F.}~\bibnamefont {{Donato}}},
  \bibinfo {author} {\bibfnamefont {A.}~\bibnamefont {{Goudelis}}}, \ and\
  \bibinfo {author} {\bibfnamefont {P.~D.}\ \bibnamefont {{Serpico}}},\
  }\bibfield  {title} {\enquote {\bibinfo {title} {{New evaluation of the
  antiproton production cross section for cosmic ray studies}},}\ }\href
  {\doibase 10.1103/PhysRevD.90.085017} {\bibfield  {journal} {\bibinfo
  {journal} {\prd}\ }\textbf {\bibinfo {volume} {90}},\ \bibinfo {eid} {085017}
  (\bibinfo {year} {2014})},\ \Eprint {http://arxiv.org/abs/1408.0288}
  {arXiv:1408.0288 [hep-ph]} \BibitemShut {NoStop}%
\bibitem [{\citenamefont {{Delahaye}}\ \emph {et~al.}(2009)\citenamefont
  {{Delahaye}}, \citenamefont {{Lineros}}, \citenamefont {{Donato}},
  \citenamefont {{Fornengo}}, \citenamefont {{Lavalle}}, \citenamefont
  {{Salati}},\ and\ \citenamefont {{Taillet}}}]{Delahaye2009}%
  \BibitemOpen
  \bibfield  {author} {\bibinfo {author} {\bibfnamefont {T.}~\bibnamefont
  {{Delahaye}}}, \bibinfo {author} {\bibfnamefont {R.}~\bibnamefont
  {{Lineros}}}, \bibinfo {author} {\bibfnamefont {F.}~\bibnamefont {{Donato}}},
  \bibinfo {author} {\bibfnamefont {N.}~\bibnamefont {{Fornengo}}}, \bibinfo
  {author} {\bibfnamefont {J.}~\bibnamefont {{Lavalle}}}, \bibinfo {author}
  {\bibfnamefont {P.}~\bibnamefont {{Salati}}}, \ and\ \bibinfo {author}
  {\bibfnamefont {R.}~\bibnamefont {{Taillet}}},\ }\bibfield  {title} {\enquote
  {\bibinfo {title} {{Galactic secondary positron flux at the Earth}},}\ }\href
  {\doibase 10.1051/0004-6361/200811130} {\bibfield  {journal} {\bibinfo
  {journal} {\aap}\ }\textbf {\bibinfo {volume} {501}},\ \bibinfo {pages}
  {821--833} (\bibinfo {year} {2009})},\ \Eprint
  {http://arxiv.org/abs/0809.5268} {arXiv:0809.5268} \BibitemShut {NoStop}%
\bibitem [{\citenamefont {{Mori}}(2009)}]{Mori2009}%
  \BibitemOpen
  \bibfield  {author} {\bibinfo {author} {\bibfnamefont {M.}~\bibnamefont
  {{Mori}}},\ }\bibfield  {title} {\enquote {\bibinfo {title} {{Nuclear
  enhancement factor in calculation of Galactic diffuse gamma-rays: A new
  estimate with DPMJET-3}},}\ }\href {\doibase
  10.1016/j.astropartphys.2009.03.004} {\bibfield  {journal} {\bibinfo
  {journal} {Astroparticle Physics}\ }\textbf {\bibinfo {volume} {31}},\
  \bibinfo {pages} {341--343} (\bibinfo {year} {2009})},\ \Eprint
  {http://arxiv.org/abs/0903.3260} {arXiv:0903.3260 [astro-ph.HE]} \BibitemShut
  {NoStop}%
\bibitem [{\citenamefont {{Cirelli}}\ \emph {et~al.}(2011)\citenamefont
  {{Cirelli}}, \citenamefont {{Corcella}}, \citenamefont {{Hektor}},
  \citenamefont {{H{\"u}tsi}}, \citenamefont {{Kadastik}}, \citenamefont
  {{Panci}}, \citenamefont {{Raidal}}, \citenamefont {{Sala}},\ and\
  \citenamefont {{Strumia}}}]{Cirelli2011}%
  \BibitemOpen
  \bibfield  {author} {\bibinfo {author} {\bibfnamefont {M.}~\bibnamefont
  {{Cirelli}}}, \bibinfo {author} {\bibfnamefont {G.}~\bibnamefont
  {{Corcella}}}, \bibinfo {author} {\bibfnamefont {A.}~\bibnamefont
  {{Hektor}}}, \bibinfo {author} {\bibfnamefont {G.}~\bibnamefont
  {{H{\"u}tsi}}}, \bibinfo {author} {\bibfnamefont {M.}~\bibnamefont
  {{Kadastik}}}, \bibinfo {author} {\bibfnamefont {P.}~\bibnamefont {{Panci}}},
  \bibinfo {author} {\bibfnamefont {M.}~\bibnamefont {{Raidal}}}, \bibinfo
  {author} {\bibfnamefont {F.}~\bibnamefont {{Sala}}}, \ and\ \bibinfo {author}
  {\bibfnamefont {A.}~\bibnamefont {{Strumia}}},\ }\bibfield  {title} {\enquote
  {\bibinfo {title} {{PPPC 4 DM ID: a poor particle physicist cookbook for dark
  matter indirect detection}},}\ }\href {\doibase
  10.1088/1475-7516/2011/03/051} {\bibfield  {journal} {\bibinfo  {journal}
  {\jcap}\ }\textbf {\bibinfo {volume} {3}},\ \bibinfo {eid} {051} (\bibinfo
  {year} {2011})},\ \Eprint {http://arxiv.org/abs/1012.4515} {arXiv:1012.4515
  [hep-ph]} \BibitemShut {NoStop}%
\bibitem [{\citenamefont {{Ciafaloni}}\ \emph {et~al.}(2011)\citenamefont
  {{Ciafaloni}}, \citenamefont {{Comelli}}, \citenamefont {{Riotto}},
  \citenamefont {{Sala}}, \citenamefont {{Strumia}},\ and\ \citenamefont
  {{Urbano}}}]{Ciafaloni2011}%
  \BibitemOpen
  \bibfield  {author} {\bibinfo {author} {\bibfnamefont {P.}~\bibnamefont
  {{Ciafaloni}}}, \bibinfo {author} {\bibfnamefont {D.}~\bibnamefont
  {{Comelli}}}, \bibinfo {author} {\bibfnamefont {A.}~\bibnamefont {{Riotto}}},
  \bibinfo {author} {\bibfnamefont {F.}~\bibnamefont {{Sala}}}, \bibinfo
  {author} {\bibfnamefont {A.}~\bibnamefont {{Strumia}}}, \ and\ \bibinfo
  {author} {\bibfnamefont {A.}~\bibnamefont {{Urbano}}},\ }\bibfield  {title}
  {\enquote {\bibinfo {title} {{Weak corrections are relevant for dark matter
  indirect detection}},}\ }\href {\doibase 10.1088/1475-7516/2011/03/019}
  {\bibfield  {journal} {\bibinfo  {journal} {\jcap}\ }\textbf {\bibinfo
  {volume} {3}},\ \bibinfo {eid} {019} (\bibinfo {year} {2011})},\ \Eprint
  {http://arxiv.org/abs/1009.0224} {arXiv:1009.0224 [hep-ph]} \BibitemShut
  {NoStop}%
\bibitem [{\citenamefont {{Navarro}}\ \emph {et~al.}(2004)\citenamefont
  {{Navarro}}, \citenamefont {{Hayashi}}, \citenamefont {{Power}},
  \citenamefont {{Jenkins}}, \citenamefont {{Frenk}}, \citenamefont {{White}},
  \citenamefont {{Springel}}, \citenamefont {{Stadel}},\ and\ \citenamefont
  {{Quinn}}}]{Navarro2004}%
  \BibitemOpen
  \bibfield  {author} {\bibinfo {author} {\bibfnamefont {J.~F.}\ \bibnamefont
  {{Navarro}}}, \bibinfo {author} {\bibfnamefont {E.}~\bibnamefont
  {{Hayashi}}}, \bibinfo {author} {\bibfnamefont {C.}~\bibnamefont {{Power}}},
  \bibinfo {author} {\bibfnamefont {A.~R.}\ \bibnamefont {{Jenkins}}}, \bibinfo
  {author} {\bibfnamefont {C.~S.}\ \bibnamefont {{Frenk}}}, \bibinfo {author}
  {\bibfnamefont {S.~D.~M.}\ \bibnamefont {{White}}}, \bibinfo {author}
  {\bibfnamefont {V.}~\bibnamefont {{Springel}}}, \bibinfo {author}
  {\bibfnamefont {J.}~\bibnamefont {{Stadel}}}, \ and\ \bibinfo {author}
  {\bibfnamefont {T.~R.}\ \bibnamefont {{Quinn}}},\ }\bibfield  {title}
  {\enquote {\bibinfo {title} {{The inner structure of {$\Lambda$}CDM haloes -
  III. Universality and asymptotic slopes}},}\ }\href {\doibase
  10.1111/j.1365-2966.2004.07586.x} {\bibfield  {journal} {\bibinfo  {journal}
  {\mnras}\ }\textbf {\bibinfo {volume} {349}},\ \bibinfo {pages} {1039--1051}
  (\bibinfo {year} {2004})},\ \Eprint {http://arxiv.org/abs/astro-ph/0311231}
  {astro-ph/0311231} \BibitemShut {NoStop}%
\bibitem [{\citenamefont {{Merritt}}\ \emph {et~al.}(2006)\citenamefont
  {{Merritt}}, \citenamefont {{Graham}}, \citenamefont {{Moore}}, \citenamefont
  {{Diemand}},\ and\ \citenamefont {{Terzi{\'c}}}}]{Merritt2006}%
  \BibitemOpen
  \bibfield  {author} {\bibinfo {author} {\bibfnamefont {D.}~\bibnamefont
  {{Merritt}}}, \bibinfo {author} {\bibfnamefont {A.~W.}\ \bibnamefont
  {{Graham}}}, \bibinfo {author} {\bibfnamefont {B.}~\bibnamefont {{Moore}}},
  \bibinfo {author} {\bibfnamefont {J.}~\bibnamefont {{Diemand}}}, \ and\
  \bibinfo {author} {\bibfnamefont {B.}~\bibnamefont {{Terzi{\'c}}}},\
  }\bibfield  {title} {\enquote {\bibinfo {title} {{Empirical Models for Dark
  Matter Halos. I. Nonparametric Construction of Density Profiles and
  Comparison with Parametric Models}},}\ }\href {\doibase 10.1086/508988}
  {\bibfield  {journal} {\bibinfo  {journal} {\aj}\ }\textbf {\bibinfo {volume}
  {132}},\ \bibinfo {pages} {2685--2700} (\bibinfo {year} {2006})},\ \Eprint
  {http://arxiv.org/abs/astro-ph/0509417} {astro-ph/0509417} \BibitemShut
  {NoStop}%
\bibitem [{\citenamefont {{Einasto}}(2009)}]{Einasto2009}%
  \BibitemOpen
  \bibfield  {author} {\bibinfo {author} {\bibfnamefont {J.}~\bibnamefont
  {{Einasto}}},\ }\bibfield  {title} {\enquote {\bibinfo {title} {{Dark
  Matter}},}\ }\href@noop {} {\bibfield  {journal} {\bibinfo  {journal} {ArXiv
  e-prints}\ } (\bibinfo {year} {2009})},\ \Eprint
  {http://arxiv.org/abs/0901.0632} {arXiv:0901.0632 [astro-ph.CO]} \BibitemShut
  {NoStop}%
\bibitem [{\citenamefont {{Navarro}}\ \emph {et~al.}(2010)\citenamefont
  {{Navarro}}, \citenamefont {{Ludlow}}, \citenamefont {{Springel}},
  \citenamefont {{Wang}}, \citenamefont {{Vogelsberger}}, \citenamefont
  {{White}}, \citenamefont {{Jenkins}}, \citenamefont {{Frenk}},\ and\
  \citenamefont {{Helmi}}}]{Navarro2010}%
  \BibitemOpen
  \bibfield  {author} {\bibinfo {author} {\bibfnamefont {J.~F.}\ \bibnamefont
  {{Navarro}}}, \bibinfo {author} {\bibfnamefont {A.}~\bibnamefont {{Ludlow}}},
  \bibinfo {author} {\bibfnamefont {V.}~\bibnamefont {{Springel}}}, \bibinfo
  {author} {\bibfnamefont {J.}~\bibnamefont {{Wang}}}, \bibinfo {author}
  {\bibfnamefont {M.}~\bibnamefont {{Vogelsberger}}}, \bibinfo {author}
  {\bibfnamefont {S.~D.~M.}\ \bibnamefont {{White}}}, \bibinfo {author}
  {\bibfnamefont {A.}~\bibnamefont {{Jenkins}}}, \bibinfo {author}
  {\bibfnamefont {C.~S.}\ \bibnamefont {{Frenk}}}, \ and\ \bibinfo {author}
  {\bibfnamefont {A.}~\bibnamefont {{Helmi}}},\ }\bibfield  {title} {\enquote
  {\bibinfo {title} {{The diversity and similarity of simulated cold dark
  matter haloes}},}\ }\href {\doibase 10.1111/j.1365-2966.2009.15878.x}
  {\bibfield  {journal} {\bibinfo  {journal} {\mnras}\ }\textbf {\bibinfo
  {volume} {402}},\ \bibinfo {pages} {21--34} (\bibinfo {year} {2010})},\
  \Eprint {http://arxiv.org/abs/0810.1522} {arXiv:0810.1522} \BibitemShut
  {NoStop}%
\bibitem [{\citenamefont {{Catena}}\ and\ \citenamefont
  {{Ullio}}(2010)}]{Catena2010}%
  \BibitemOpen
  \bibfield  {author} {\bibinfo {author} {\bibfnamefont {R.}~\bibnamefont
  {{Catena}}}\ and\ \bibinfo {author} {\bibfnamefont {P.}~\bibnamefont
  {{Ullio}}},\ }\bibfield  {title} {\enquote {\bibinfo {title} {{A novel
  determination of the local dark matter density}},}\ }\href {\doibase
  10.1088/1475-7516/2010/08/004} {\bibfield  {journal} {\bibinfo  {journal}
  {\jcap}\ }\textbf {\bibinfo {volume} {8}},\ \bibinfo {eid} {004} (\bibinfo
  {year} {2010})},\ \Eprint {http://arxiv.org/abs/0907.0018} {arXiv:0907.0018}
  \BibitemShut {NoStop}%
\bibitem [{\citenamefont {{Weber}}\ and\ \citenamefont {{de
  Boer}}(2010)}]{Weber2010}%
  \BibitemOpen
  \bibfield  {author} {\bibinfo {author} {\bibfnamefont {M.}~\bibnamefont
  {{Weber}}}\ and\ \bibinfo {author} {\bibfnamefont {W.}~\bibnamefont {{de
  Boer}}},\ }\bibfield  {title} {\enquote {\bibinfo {title} {{Determination of
  the local dark matter density in our Galaxy}},}\ }\href {\doibase
  10.1051/0004-6361/200913381} {\bibfield  {journal} {\bibinfo  {journal}
  {\aap}\ }\textbf {\bibinfo {volume} {509}},\ \bibinfo {eid} {A25} (\bibinfo
  {year} {2010})},\ \Eprint {http://arxiv.org/abs/0910.4272} {arXiv:0910.4272
  [astro-ph.CO]} \BibitemShut {NoStop}%
\bibitem [{\citenamefont {{Salucci}}\ \emph {et~al.}(2010)\citenamefont
  {{Salucci}}, \citenamefont {{Nesti}}, \citenamefont {{Gentile}},\ and\
  \citenamefont {{Frigerio Martins}}}]{Salucci2010}%
  \BibitemOpen
  \bibfield  {author} {\bibinfo {author} {\bibfnamefont {P.}~\bibnamefont
  {{Salucci}}}, \bibinfo {author} {\bibfnamefont {F.}~\bibnamefont {{Nesti}}},
  \bibinfo {author} {\bibfnamefont {G.}~\bibnamefont {{Gentile}}}, \ and\
  \bibinfo {author} {\bibfnamefont {C.}~\bibnamefont {{Frigerio Martins}}},\
  }\bibfield  {title} {\enquote {\bibinfo {title} {{The dark matter density at
  the Sun's location}},}\ }\href {\doibase 10.1051/0004-6361/201014385}
  {\bibfield  {journal} {\bibinfo  {journal} {\aap}\ }\textbf {\bibinfo
  {volume} {523}},\ \bibinfo {eid} {A83} (\bibinfo {year} {2010})},\ \Eprint
  {http://arxiv.org/abs/1003.3101} {arXiv:1003.3101} \BibitemShut {NoStop}%
\bibitem [{\citenamefont {{Pato}}\ \emph {et~al.}(2010)\citenamefont {{Pato}},
  \citenamefont {{Agertz}}, \citenamefont {{Bertone}}, \citenamefont
  {{Moore}},\ and\ \citenamefont {{Teyssier}}}]{Pato2010}%
  \BibitemOpen
  \bibfield  {author} {\bibinfo {author} {\bibfnamefont {M.}~\bibnamefont
  {{Pato}}}, \bibinfo {author} {\bibfnamefont {O.}~\bibnamefont {{Agertz}}},
  \bibinfo {author} {\bibfnamefont {G.}~\bibnamefont {{Bertone}}}, \bibinfo
  {author} {\bibfnamefont {B.}~\bibnamefont {{Moore}}}, \ and\ \bibinfo
  {author} {\bibfnamefont {R.}~\bibnamefont {{Teyssier}}},\ }\bibfield  {title}
  {\enquote {\bibinfo {title} {{Systematic uncertainties in the determination
  of the local dark matter density}},}\ }\href {\doibase
  10.1103/PhysRevD.82.023531} {\bibfield  {journal} {\bibinfo  {journal}
  {\prd}\ }\textbf {\bibinfo {volume} {82}},\ \bibinfo {eid} {023531} (\bibinfo
  {year} {2010})},\ \Eprint {http://arxiv.org/abs/1006.1322} {arXiv:1006.1322
  [astro-ph.HE]} \BibitemShut {NoStop}%
\bibitem [{\citenamefont {{Iocco}}\ \emph {et~al.}(2011)\citenamefont
  {{Iocco}}, \citenamefont {{Pato}}, \citenamefont {{Bertone}},\ and\
  \citenamefont {{Jetzer}}}]{Iocco2011}%
  \BibitemOpen
  \bibfield  {author} {\bibinfo {author} {\bibfnamefont {F.}~\bibnamefont
  {{Iocco}}}, \bibinfo {author} {\bibfnamefont {M.}~\bibnamefont {{Pato}}},
  \bibinfo {author} {\bibfnamefont {G.}~\bibnamefont {{Bertone}}}, \ and\
  \bibinfo {author} {\bibfnamefont {P.}~\bibnamefont {{Jetzer}}},\ }\bibfield
  {title} {\enquote {\bibinfo {title} {{Dark Matter distribution in the Milky
  Way: microlensing and dynamical constraints}},}\ }\href {\doibase
  10.1088/1475-7516/2011/11/029} {\bibfield  {journal} {\bibinfo  {journal}
  {\jcap}\ }\textbf {\bibinfo {volume} {11}},\ \bibinfo {eid} {029} (\bibinfo
  {year} {2011})},\ \Eprint {http://arxiv.org/abs/1107.5810} {arXiv:1107.5810
  [astro-ph.GA]} \BibitemShut {NoStop}%
\bibitem [{\citenamefont {{Gleeson}}\ and\ \citenamefont
  {{Axford}}(1968)}]{Gleeson1968}%
  \BibitemOpen
  \bibfield  {author} {\bibinfo {author} {\bibfnamefont {L.~J.}\ \bibnamefont
  {{Gleeson}}}\ and\ \bibinfo {author} {\bibfnamefont {W.~I.}\ \bibnamefont
  {{Axford}}},\ }\bibfield  {title} {\enquote {\bibinfo {title} {{Solar
  Modulation of Galactic Cosmic Rays}},}\ }\href {\doibase 10.1086/149822}
  {\bibfield  {journal} {\bibinfo  {journal} {\apj}\ }\textbf {\bibinfo
  {volume} {154}},\ \bibinfo {pages} {1011} (\bibinfo {year}
  {1968})}\BibitemShut {NoStop}%
\bibitem [{\citenamefont {{Evoli}}\ \emph {et~al.}(2008)\citenamefont
  {{Evoli}}, \citenamefont {{Gaggero}}, \citenamefont {{Grasso}},\ and\
  \citenamefont {{Maccione}}}]{Evoli2008}%
  \BibitemOpen
  \bibfield  {author} {\bibinfo {author} {\bibfnamefont {C.}~\bibnamefont
  {{Evoli}}}, \bibinfo {author} {\bibfnamefont {D.}~\bibnamefont {{Gaggero}}},
  \bibinfo {author} {\bibfnamefont {D.}~\bibnamefont {{Grasso}}}, \ and\
  \bibinfo {author} {\bibfnamefont {L.}~\bibnamefont {{Maccione}}},\ }\bibfield
   {title} {\enquote {\bibinfo {title} {{Cosmic ray nuclei, antiprotons and
  gamma rays in the galaxy: a new diffusion model}},}\ }\href {\doibase
  10.1088/1475-7516/2008/10/018} {\bibfield  {journal} {\bibinfo  {journal}
  {\jcap}\ }\textbf {\bibinfo {volume} {10}},\ \bibinfo {eid} {018} (\bibinfo
  {year} {2008})},\ \Eprint {http://arxiv.org/abs/0807.4730} {arXiv:0807.4730}
  \BibitemShut {NoStop}%
\bibitem [{\citenamefont {{J{\'o}hannesson}}\ \emph {et~al.}(2016)\citenamefont
  {{J{\'o}hannesson}}, \citenamefont {{Ruiz de Austri}}, \citenamefont
  {{Vincent}}, \citenamefont {{Moskalenko}}, \citenamefont {{Orlando}},
  \citenamefont {{Porter}}, \citenamefont {{Strong}}, \citenamefont {{Trotta}},
  \citenamefont {{Feroz}}, \citenamefont {{Graff}},\ and\ \citenamefont
  {{Hobson}}}]{Johannesson2016}%
  \BibitemOpen
  \bibfield  {author} {\bibinfo {author} {\bibfnamefont {G.}~\bibnamefont
  {{J{\'o}hannesson}}}, \bibinfo {author} {\bibfnamefont {R.}~\bibnamefont
  {{Ruiz de Austri}}}, \bibinfo {author} {\bibfnamefont {A.~C.}\ \bibnamefont
  {{Vincent}}}, \bibinfo {author} {\bibfnamefont {I.~V.}\ \bibnamefont
  {{Moskalenko}}}, \bibinfo {author} {\bibfnamefont {E.}~\bibnamefont
  {{Orlando}}}, \bibinfo {author} {\bibfnamefont {T.~A.}\ \bibnamefont
  {{Porter}}}, \bibinfo {author} {\bibfnamefont {A.~W.}\ \bibnamefont
  {{Strong}}}, \bibinfo {author} {\bibfnamefont {R.}~\bibnamefont {{Trotta}}},
  \bibinfo {author} {\bibfnamefont {F.}~\bibnamefont {{Feroz}}}, \bibinfo
  {author} {\bibfnamefont {P.}~\bibnamefont {{Graff}}}, \ and\ \bibinfo
  {author} {\bibfnamefont {M.~P.}\ \bibnamefont {{Hobson}}},\ }\bibfield
  {title} {\enquote {\bibinfo {title} {{Bayesian Analysis of Cosmic Ray
  Propagation: Evidence against Homogeneous Diffusion}},}\ }\href {\doibase
  10.3847/0004-637X/824/1/16} {\bibfield  {journal} {\bibinfo  {journal}
  {\apj}\ }\textbf {\bibinfo {volume} {824}},\ \bibinfo {eid} {16} (\bibinfo
  {year} {2016})},\ \Eprint {http://arxiv.org/abs/1602.02243} {arXiv:1602.02243
  [astro-ph.HE]} \BibitemShut {NoStop}%
\bibitem [{\citenamefont {{Goodman}}\ and\ \citenamefont
  {{Weare}}(2010)}]{Goodman2010}%
  \BibitemOpen
  \bibfield  {author} {\bibinfo {author} {\bibfnamefont {J.}~\bibnamefont
  {{Goodman}}}\ and\ \bibinfo {author} {\bibfnamefont {J.}~\bibnamefont
  {{Weare}}},\ }\bibfield  {title} {\enquote {\bibinfo {title} {{Ensemble
  samplers with affine invariance}},}\ }\href {\doibase
  10.2140/camcos.2010.5.65} {\bibfield  {journal} {\bibinfo  {journal}
  {Communications in Applied Mathematics and Computational Science}\ }\textbf
  {\bibinfo {volume} {5}},\ \bibinfo {pages} {65--80} (\bibinfo {year}
  {2010})}\BibitemShut {NoStop}%
\bibitem [{\citenamefont {{Foreman-Mackey}}\ \emph {et~al.}(2013)\citenamefont
  {{Foreman-Mackey}}, \citenamefont {{Hogg}}, \citenamefont {{Lang}},\ and\
  \citenamefont {{Goodman}}}]{Mackey2013}%
  \BibitemOpen
  \bibfield  {author} {\bibinfo {author} {\bibfnamefont {D.}~\bibnamefont
  {{Foreman-Mackey}}}, \bibinfo {author} {\bibfnamefont {D.~W.}\ \bibnamefont
  {{Hogg}}}, \bibinfo {author} {\bibfnamefont {D.}~\bibnamefont {{Lang}}}, \
  and\ \bibinfo {author} {\bibfnamefont {J.}~\bibnamefont {{Goodman}}},\
  }\bibfield  {title} {\enquote {\bibinfo {title} {{emcee: The MCMC Hammer}},}\
  }\href {\doibase 10.1086/670067} {\bibfield  {journal} {\bibinfo  {journal}
  {\pasp}\ }\textbf {\bibinfo {volume} {125}},\ \bibinfo {pages} {306--312}
  (\bibinfo {year} {2013})},\ \Eprint {http://arxiv.org/abs/1202.3665}
  {arXiv:1202.3665 [astro-ph.IM]} \BibitemShut {NoStop}%
\bibitem [{\citenamefont {{AMS collaboration}}\ \emph
  {et~al.}(2016)\citenamefont {{AMS collaboration}}, \citenamefont {{Aguilar}},
  \citenamefont {{Ali Cavasonza}}, \citenamefont {{Alpat}}, \citenamefont
  {{Ambrosi}}, \citenamefont {{Arruda}}, \citenamefont {{Attig}}, \citenamefont
  {{Aupetit}}, \citenamefont {{Azzarello}}, \citenamefont {{Bachlechner}},
  \citenamefont {{Barao}},\ and\ \citenamefont {et~al.}}]{AMS02_pbar_proton}%
  \BibitemOpen
  \bibfield  {author} {\bibinfo {author} {\bibnamefont {{AMS collaboration}}},
  \bibinfo {author} {\bibfnamefont {M.}~\bibnamefont {{Aguilar}}}, \bibinfo
  {author} {\bibfnamefont {L.}~\bibnamefont {{Ali Cavasonza}}}, \bibinfo
  {author} {\bibfnamefont {B.}~\bibnamefont {{Alpat}}}, \bibinfo {author}
  {\bibfnamefont {G.}~\bibnamefont {{Ambrosi}}}, \bibinfo {author}
  {\bibfnamefont {L.}~\bibnamefont {{Arruda}}}, \bibinfo {author}
  {\bibfnamefont {N.}~\bibnamefont {{Attig}}}, \bibinfo {author} {\bibfnamefont
  {S.}~\bibnamefont {{Aupetit}}}, \bibinfo {author} {\bibfnamefont
  {P.}~\bibnamefont {{Azzarello}}}, \bibinfo {author} {\bibfnamefont
  {A.}~\bibnamefont {{Bachlechner}}}, \bibinfo {author} {\bibfnamefont
  {F.}~\bibnamefont {{Barao}}}, \ and\ \bibinfo {author} {\bibnamefont
  {et~al.}},\ }\bibfield  {title} {\enquote {\bibinfo {title} {{Antiproton
  Flux, Antiproton-to-Proton Flux Ratio, and Properties of Elementary Particle
  Fluxes in Primary Cosmic Rays Measured with the Alpha Magnetic Spectrometer
  on the International Space Station}},}\ }\href {\doibase
  10.1103/PhysRevLett.117.091103} {\bibfield  {journal} {\bibinfo  {journal}
  {Physical Review Letters}\ }\textbf {\bibinfo {volume} {117}},\ \bibinfo
  {eid} {091103} (\bibinfo {year} {2016})}\BibitemShut {NoStop}%
\bibitem [{\citenamefont {{Tomassetti}}(2012)}]{Tomassetti2012}%
  \BibitemOpen
  \bibfield  {author} {\bibinfo {author} {\bibfnamefont {N.}~\bibnamefont
  {{Tomassetti}}},\ }\bibfield  {title} {\enquote {\bibinfo {title} {{Origin of
  the Cosmic-Ray Spectral Hardening}},}\ }\href {\doibase
  10.1088/2041-8205/752/1/L13} {\bibfield  {journal} {\bibinfo  {journal}
  {\apjl}\ }\textbf {\bibinfo {volume} {752}},\ \bibinfo {eid} {L13} (\bibinfo
  {year} {2012})},\ \Eprint {http://arxiv.org/abs/1204.4492} {arXiv:1204.4492
  [astro-ph.HE]} \BibitemShut {NoStop}%
\bibitem [{\citenamefont {{Tomassetti}}(2015)}]{Tomassetti2015prd}%
  \BibitemOpen
  \bibfield  {author} {\bibinfo {author} {\bibfnamefont {N.}~\bibnamefont
  {{Tomassetti}}},\ }\bibfield  {title} {\enquote {\bibinfo {title}
  {{Cosmic-ray protons, nuclei, electrons, and antiparticles under a two-halo
  scenario of diffusive propagation}},}\ }\href {\doibase
  10.1103/PhysRevD.92.081301} {\bibfield  {journal} {\bibinfo  {journal}
  {\prd}\ }\textbf {\bibinfo {volume} {92}},\ \bibinfo {eid} {081301} (\bibinfo
  {year} {2015})},\ \Eprint {http://arxiv.org/abs/1509.05775} {arXiv:1509.05775
  [astro-ph.HE]} \BibitemShut {NoStop}%
\bibitem [{\citenamefont {{Feng}}\ \emph {et~al.}(2016)\citenamefont {{Feng}},
  \citenamefont {{Tomassetti}},\ and\ \citenamefont {{Oliva}}}]{Feng2016}%
  \BibitemOpen
  \bibfield  {author} {\bibinfo {author} {\bibfnamefont {J.}~\bibnamefont
  {{Feng}}}, \bibinfo {author} {\bibfnamefont {N.}~\bibnamefont
  {{Tomassetti}}}, \ and\ \bibinfo {author} {\bibfnamefont {A.}~\bibnamefont
  {{Oliva}}},\ }\bibfield  {title} {\enquote {\bibinfo {title} {{Bayesian
  analysis of spatial-dependent cosmic-ray propagation: Astrophysical
  background of antiprotons and positrons}},}\ }\href {\doibase
  10.1103/PhysRevD.94.123007} {\bibfield  {journal} {\bibinfo  {journal}
  {\prd}\ }\textbf {\bibinfo {volume} {94}},\ \bibinfo {eid} {123007} (\bibinfo
  {year} {2016})},\ \Eprint {http://arxiv.org/abs/1610.06182} {arXiv:1610.06182
  [astro-ph.HE]} \BibitemShut {NoStop}%
\bibitem [{\citenamefont {{G{\'e}nolini}}\ \emph {et~al.}(2017)\citenamefont
  {{G{\'e}nolini}}, \citenamefont {{Serpico}}, \citenamefont {{Boudaud}},
  \citenamefont {{Caroff}}, \citenamefont {{Poulin}}, \citenamefont {{Derome}},
  \citenamefont {{Lavalle}}, \citenamefont {{Maurin}}, \citenamefont
  {{Poireau}}, \citenamefont {{Rosier}}, \citenamefont {{Salati}},\ and\
  \citenamefont {{Vecchi}}}]{Genolini2017}%
  \BibitemOpen
  \bibfield  {author} {\bibinfo {author} {\bibfnamefont {Y.}~\bibnamefont
  {{G{\'e}nolini}}}, \bibinfo {author} {\bibfnamefont {P.~D.}\ \bibnamefont
  {{Serpico}}}, \bibinfo {author} {\bibfnamefont {M.}~\bibnamefont
  {{Boudaud}}}, \bibinfo {author} {\bibfnamefont {S.}~\bibnamefont {{Caroff}}},
  \bibinfo {author} {\bibfnamefont {V.}~\bibnamefont {{Poulin}}}, \bibinfo
  {author} {\bibfnamefont {L.}~\bibnamefont {{Derome}}}, \bibinfo {author}
  {\bibfnamefont {J.}~\bibnamefont {{Lavalle}}}, \bibinfo {author}
  {\bibfnamefont {D.}~\bibnamefont {{Maurin}}}, \bibinfo {author}
  {\bibfnamefont {V.}~\bibnamefont {{Poireau}}}, \bibinfo {author}
  {\bibfnamefont {S.}~\bibnamefont {{Rosier}}}, \bibinfo {author}
  {\bibfnamefont {P.}~\bibnamefont {{Salati}}}, \ and\ \bibinfo {author}
  {\bibfnamefont {M.}~\bibnamefont {{Vecchi}}},\ }\bibfield  {title} {\enquote
  {\bibinfo {title} {{Indications for a High-Rigidity Break in the Cosmic-Ray
  Diffusion Coefficient}},}\ }\href {\doibase 10.1103/PhysRevLett.119.241101}
  {\bibfield  {journal} {\bibinfo  {journal} {Physical Review Letters}\
  }\textbf {\bibinfo {volume} {119}},\ \bibinfo {eid} {241101} (\bibinfo {year}
  {2017})},\ \Eprint {http://arxiv.org/abs/1706.09812} {arXiv:1706.09812
  [astro-ph.HE]} \BibitemShut {NoStop}%
\bibitem [{\citenamefont {{Blasi}}(2017)}]{Blasi2017}%
  \BibitemOpen
  \bibfield  {author} {\bibinfo {author} {\bibfnamefont {P.}~\bibnamefont
  {{Blasi}}},\ }\bibfield  {title} {\enquote {\bibinfo {title} {{On the
  spectrum of stable secondary nuclei in cosmic rays}},}\ }\href {\doibase
  10.1093/mnras/stx1696} {\bibfield  {journal} {\bibinfo  {journal} {\mnras}\
  }\textbf {\bibinfo {volume} {471}},\ \bibinfo {pages} {1662--1670} (\bibinfo
  {year} {2017})},\ \Eprint {http://arxiv.org/abs/1707.00525} {arXiv:1707.00525
  [astro-ph.HE]} \BibitemShut {NoStop}%
\bibitem [{\citenamefont {{Reinert}}\ and\ \citenamefont
  {{Winkler}}(2018)}]{Reinert2018}%
  \BibitemOpen
  \bibfield  {author} {\bibinfo {author} {\bibfnamefont {A.}~\bibnamefont
  {{Reinert}}}\ and\ \bibinfo {author} {\bibfnamefont {M.~W.}\ \bibnamefont
  {{Winkler}}},\ }\bibfield  {title} {\enquote {\bibinfo {title} {{A precision
  search for WIMPs with charged cosmic rays}},}\ }\href {\doibase
  10.1088/1475-7516/2018/01/055} {\bibfield  {journal} {\bibinfo  {journal}
  {\jcap}\ }\textbf {\bibinfo {volume} {1}},\ \bibinfo {eid} {055} (\bibinfo
  {year} {2018})},\ \Eprint {http://arxiv.org/abs/1712.00002} {arXiv:1712.00002
  [astro-ph.HE]} \BibitemShut {NoStop}%
\bibitem [{\citenamefont {{Reynolds}}(1988)}]{Reynolds1988}%
  \BibitemOpen
  \bibfield  {author} {\bibinfo {author} {\bibfnamefont {S.~P.}\ \bibnamefont
  {{Reynolds}}},\ }\bibfield  {title} {\enquote {\bibinfo {title} {{Filamentary
  structure in Crab-like supernova remnants}},}\ }\href {\doibase
  10.1086/166243} {\bibfield  {journal} {\bibinfo  {journal} {\apj}\ }\textbf
  {\bibinfo {volume} {327}},\ \bibinfo {pages} {853--858} (\bibinfo {year}
  {1988})}\BibitemShut {NoStop}%
\bibitem [{\citenamefont {{Thompson}}\ \emph {et~al.}(1994)\citenamefont
  {{Thompson}}, \citenamefont {{Arzoumanian}}, \citenamefont {{Bertsch}},
  \citenamefont {{Brazier}}, \citenamefont {{Chiang}}, \citenamefont
  {{D'Amico}}, \citenamefont {{Dingus}}, \citenamefont {{Esposito}},
  \citenamefont {{Fierro}}, \citenamefont {{Fichtel}}, \citenamefont
  {{Hartman}}, \citenamefont {{Hunter}}, \citenamefont {{Johnston}},
  \citenamefont {{Kanbach}}, \citenamefont {{Kaspi}}, \citenamefont
  {{Kniffen}}, \citenamefont {{Lin}}, \citenamefont {{Lyne}}, \citenamefont
  {{Manchester}}, \citenamefont {{Mattox}}, \citenamefont
  {{Mayer-Hasselwander}}, \citenamefont {{Michelson}}, \citenamefont {{von
  Montigny}}, \citenamefont {{Nel}}, \citenamefont {{Nice}}, \citenamefont
  {{Nolan}}, \citenamefont {{Ramanamurthy}}, \citenamefont {{Shemar}},
  \citenamefont {{Schneid}}, \citenamefont {{Sreekumar}},\ and\ \citenamefont
  {{Taylor}}}]{Thompson1994}%
  \BibitemOpen
  \bibfield  {author} {\bibinfo {author} {\bibfnamefont {D.~J.}\ \bibnamefont
  {{Thompson}}}, \bibinfo {author} {\bibfnamefont {Z.}~\bibnamefont
  {{Arzoumanian}}}, \bibinfo {author} {\bibfnamefont {D.~L.}\ \bibnamefont
  {{Bertsch}}}, \bibinfo {author} {\bibfnamefont {K.~T.~S.}\ \bibnamefont
  {{Brazier}}}, \bibinfo {author} {\bibfnamefont {J.}~\bibnamefont {{Chiang}}},
  \bibinfo {author} {\bibfnamefont {N.}~\bibnamefont {{D'Amico}}}, \bibinfo
  {author} {\bibfnamefont {B.~L.}\ \bibnamefont {{Dingus}}}, \bibinfo {author}
  {\bibfnamefont {J.~A.}\ \bibnamefont {{Esposito}}}, \bibinfo {author}
  {\bibfnamefont {J.~M.}\ \bibnamefont {{Fierro}}}, \bibinfo {author}
  {\bibfnamefont {C.~E.}\ \bibnamefont {{Fichtel}}}, \bibinfo {author}
  {\bibfnamefont {R.~C.}\ \bibnamefont {{Hartman}}}, \bibinfo {author}
  {\bibfnamefont {S.~D.}\ \bibnamefont {{Hunter}}}, \bibinfo {author}
  {\bibfnamefont {S.}~\bibnamefont {{Johnston}}}, \bibinfo {author}
  {\bibfnamefont {G.}~\bibnamefont {{Kanbach}}}, \bibinfo {author}
  {\bibfnamefont {V.~M.}\ \bibnamefont {{Kaspi}}}, \bibinfo {author}
  {\bibfnamefont {D.~A.}\ \bibnamefont {{Kniffen}}}, \bibinfo {author}
  {\bibfnamefont {Y.~C.}\ \bibnamefont {{Lin}}}, \bibinfo {author}
  {\bibfnamefont {A.~G.}\ \bibnamefont {{Lyne}}}, \bibinfo {author}
  {\bibfnamefont {R.~N.}\ \bibnamefont {{Manchester}}}, \bibinfo {author}
  {\bibfnamefont {J.~R.}\ \bibnamefont {{Mattox}}}, \bibinfo {author}
  {\bibfnamefont {H.~A.}\ \bibnamefont {{Mayer-Hasselwander}}}, \bibinfo
  {author} {\bibfnamefont {P.~F.}\ \bibnamefont {{Michelson}}}, \bibinfo
  {author} {\bibfnamefont {C.}~\bibnamefont {{von Montigny}}}, \bibinfo
  {author} {\bibfnamefont {H.~I.}\ \bibnamefont {{Nel}}}, \bibinfo {author}
  {\bibfnamefont {D.~J.}\ \bibnamefont {{Nice}}}, \bibinfo {author}
  {\bibfnamefont {P.~L.}\ \bibnamefont {{Nolan}}}, \bibinfo {author}
  {\bibfnamefont {P.~V.}\ \bibnamefont {{Ramanamurthy}}}, \bibinfo {author}
  {\bibfnamefont {S.~L.}\ \bibnamefont {{Shemar}}}, \bibinfo {author}
  {\bibfnamefont {E.~J.}\ \bibnamefont {{Schneid}}}, \bibinfo {author}
  {\bibfnamefont {P.}~\bibnamefont {{Sreekumar}}}, \ and\ \bibinfo {author}
  {\bibfnamefont {J.~H.}\ \bibnamefont {{Taylor}}},\ }\bibfield  {title}
  {\enquote {\bibinfo {title} {{EGRET high-energy gamma-ray pulsar studies. 1:
  Young spin-powered pulsars}},}\ }\href {\doibase 10.1086/174895} {\bibfield
  {journal} {\bibinfo  {journal} {\apj}\ }\textbf {\bibinfo {volume} {436}},\
  \bibinfo {pages} {229--238} (\bibinfo {year} {1994})}\BibitemShut {NoStop}%
\bibitem [{\citenamefont {{Fierro}}\ \emph {et~al.}(1995)\citenamefont
  {{Fierro}}, \citenamefont {{Arzoumanian}}, \citenamefont {{Bailes}},
  \citenamefont {{Bell}}, \citenamefont {{Bertsch}}, \citenamefont {{Brazier}},
  \citenamefont {{Chiang}}, \citenamefont {{D'Amico}}, \citenamefont
  {{Dingus}}, \citenamefont {{Esposito}}, \citenamefont {{Fichtel}},
  \citenamefont {{Hartman}}, \citenamefont {{Hunter}}, \citenamefont
  {{Johnston}}, \citenamefont {{Kanbach}}, \citenamefont {{Kaspi}},
  \citenamefont {{Kniffen}}, \citenamefont {{Lin}}, \citenamefont {{Lyne}},
  \citenamefont {{Manchester}}, \citenamefont {{Mattox}}, \citenamefont
  {{Mayer-Hasselwander}}, \citenamefont {{Michelson}}, \citenamefont {{von
  Montigny}}, \citenamefont {{Nel}}, \citenamefont {{Nice}}, \citenamefont
  {{Nolan}}, \citenamefont {{Schneid}}, \citenamefont {{Shriver}},
  \citenamefont {{Sreekumar}}, \citenamefont {{Taylor}}, \citenamefont
  {{Thompson}},\ and\ \citenamefont {{Willis}}}]{Fierro1995}%
  \BibitemOpen
  \bibfield  {author} {\bibinfo {author} {\bibfnamefont {J.~M.}\ \bibnamefont
  {{Fierro}}}, \bibinfo {author} {\bibfnamefont {Z.}~\bibnamefont
  {{Arzoumanian}}}, \bibinfo {author} {\bibfnamefont {M.}~\bibnamefont
  {{Bailes}}}, \bibinfo {author} {\bibfnamefont {J.~F.}\ \bibnamefont
  {{Bell}}}, \bibinfo {author} {\bibfnamefont {D.~L.}\ \bibnamefont
  {{Bertsch}}}, \bibinfo {author} {\bibfnamefont {K.~T.~S.}\ \bibnamefont
  {{Brazier}}}, \bibinfo {author} {\bibfnamefont {J.}~\bibnamefont {{Chiang}}},
  \bibinfo {author} {\bibfnamefont {N.}~\bibnamefont {{D'Amico}}}, \bibinfo
  {author} {\bibfnamefont {B.~L.}\ \bibnamefont {{Dingus}}}, \bibinfo {author}
  {\bibfnamefont {J.~A.}\ \bibnamefont {{Esposito}}}, \bibinfo {author}
  {\bibfnamefont {C.~E.}\ \bibnamefont {{Fichtel}}}, \bibinfo {author}
  {\bibfnamefont {R.~C.}\ \bibnamefont {{Hartman}}}, \bibinfo {author}
  {\bibfnamefont {S.~D.}\ \bibnamefont {{Hunter}}}, \bibinfo {author}
  {\bibfnamefont {S.}~\bibnamefont {{Johnston}}}, \bibinfo {author}
  {\bibfnamefont {G.}~\bibnamefont {{Kanbach}}}, \bibinfo {author}
  {\bibfnamefont {V.~M.}\ \bibnamefont {{Kaspi}}}, \bibinfo {author}
  {\bibfnamefont {D.~A.}\ \bibnamefont {{Kniffen}}}, \bibinfo {author}
  {\bibfnamefont {Y.~C.}\ \bibnamefont {{Lin}}}, \bibinfo {author}
  {\bibfnamefont {A.~G.}\ \bibnamefont {{Lyne}}}, \bibinfo {author}
  {\bibfnamefont {R.~N.}\ \bibnamefont {{Manchester}}}, \bibinfo {author}
  {\bibfnamefont {J.~R.}\ \bibnamefont {{Mattox}}}, \bibinfo {author}
  {\bibfnamefont {H.~A.}\ \bibnamefont {{Mayer-Hasselwander}}}, \bibinfo
  {author} {\bibfnamefont {P.~F.}\ \bibnamefont {{Michelson}}}, \bibinfo
  {author} {\bibfnamefont {C.}~\bibnamefont {{von Montigny}}}, \bibinfo
  {author} {\bibfnamefont {H.~I.}\ \bibnamefont {{Nel}}}, \bibinfo {author}
  {\bibfnamefont {D.}~\bibnamefont {{Nice}}}, \bibinfo {author} {\bibfnamefont
  {P.~L.}\ \bibnamefont {{Nolan}}}, \bibinfo {author} {\bibfnamefont {E.~J.}\
  \bibnamefont {{Schneid}}}, \bibinfo {author} {\bibfnamefont {S.~K.}\
  \bibnamefont {{Shriver}}}, \bibinfo {author} {\bibfnamefont {P.}~\bibnamefont
  {{Sreekumar}}}, \bibinfo {author} {\bibfnamefont {J.~H.}\ \bibnamefont
  {{Taylor}}}, \bibinfo {author} {\bibfnamefont {D.~J.}\ \bibnamefont
  {{Thompson}}}, \ and\ \bibinfo {author} {\bibfnamefont {T.~D.}\ \bibnamefont
  {{Willis}}},\ }\bibfield  {title} {\enquote {\bibinfo {title} {{EGRET
  High-Energy gamma -Ray Pulsar Studies. II. Individual Millisecond
  Pulsars}},}\ }\href {\doibase 10.1086/175919} {\bibfield  {journal} {\bibinfo
   {journal} {\apj}\ }\textbf {\bibinfo {volume} {447}},\ \bibinfo {pages}
  {807} (\bibinfo {year} {1995})}\BibitemShut {NoStop}%
\bibitem [{\citenamefont {{Jungman}}\ \emph {et~al.}(1996)\citenamefont
  {{Jungman}}, \citenamefont {{Kamionkowski}},\ and\ \citenamefont
  {{Griest}}}]{Jungman1996}%
  \BibitemOpen
  \bibfield  {author} {\bibinfo {author} {\bibfnamefont {G.}~\bibnamefont
  {{Jungman}}}, \bibinfo {author} {\bibfnamefont {M.}~\bibnamefont
  {{Kamionkowski}}}, \ and\ \bibinfo {author} {\bibfnamefont {K.}~\bibnamefont
  {{Griest}}},\ }\bibfield  {title} {\enquote {\bibinfo {title}
  {{Supersymmetric dark matter}},}\ }\href {\doibase
  10.1016/0370-1573(95)00058-5} {\bibfield  {journal} {\bibinfo  {journal}
  {\physrep}\ }\textbf {\bibinfo {volume} {267}},\ \bibinfo {pages} {195--373}
  (\bibinfo {year} {1996})},\ \Eprint {http://arxiv.org/abs/hep-ph/9506380}
  {hep-ph/9506380} \BibitemShut {NoStop}%
\bibitem [{\citenamefont {{Chao}}\ \emph {et~al.}(2017)\citenamefont {{Chao}},
  \citenamefont {{Guo}}, \citenamefont {{Li}},\ and\ \citenamefont
  {{Shu}}}]{Chao2017}%
  \BibitemOpen
  \bibfield  {author} {\bibinfo {author} {\bibfnamefont {W.}~\bibnamefont
  {{Chao}}}, \bibinfo {author} {\bibfnamefont {H.-K.}\ \bibnamefont {{Guo}}},
  \bibinfo {author} {\bibfnamefont {H.-L.}\ \bibnamefont {{Li}}}, \ and\
  \bibinfo {author} {\bibfnamefont {J.}~\bibnamefont {{Shu}}},\ }\bibfield
  {title} {\enquote {\bibinfo {title} {{Electron Flavored Dark Matter}},}\
  }\href@noop {} {\bibfield  {journal} {\bibinfo  {journal} {ArXiv e-prints}\ }
  (\bibinfo {year} {2017})},\ \Eprint {http://arxiv.org/abs/1712.00037}
  {arXiv:1712.00037 [hep-ph]} \BibitemShut {NoStop}%
\bibitem [{\citenamefont {Ackermann}\ \emph {et~al.}(2011)\citenamefont
  {Ackermann} \emph {et~al.}}]{Ackermann:2011wa}%
  \BibitemOpen
  \bibfield  {author} {\bibinfo {author} {\bibfnamefont {M.}~\bibnamefont
  {Ackermann}} \emph {et~al.} (\bibinfo {collaboration} {Fermi-LAT}),\
  }\bibfield  {title} {\enquote {\bibinfo {title} {{Constraining Dark Matter
  Models from a Combined Analysis of Milky Way Satellites with the Fermi Large
  Area Telescope}},}\ }\href {\doibase 10.1103/PhysRevLett.107.241302}
  {\bibfield  {journal} {\bibinfo  {journal} {Phys. Rev. Lett.}\ }\textbf
  {\bibinfo {volume} {107}},\ \bibinfo {pages} {241302} (\bibinfo {year}
  {2011})},\ \Eprint {http://arxiv.org/abs/1108.3546} {arXiv:1108.3546
  [astro-ph.HE]} \BibitemShut {NoStop}%
\bibitem [{\citenamefont {Geringer-Sameth}\ and\ \citenamefont
  {Koushiappas}(2011)}]{GeringerSameth:2011iw}%
  \BibitemOpen
  \bibfield  {author} {\bibinfo {author} {\bibfnamefont {Alex}\ \bibnamefont
  {Geringer-Sameth}}\ and\ \bibinfo {author} {\bibfnamefont {Savvas~M.}\
  \bibnamefont {Koushiappas}},\ }\bibfield  {title} {\enquote {\bibinfo {title}
  {{Exclusion of canonical WIMPs by the joint analysis of Milky Way dwarfs with
  Fermi}},}\ }\href {\doibase 10.1103/PhysRevLett.107.241303} {\bibfield
  {journal} {\bibinfo  {journal} {Phys. Rev. Lett.}\ }\textbf {\bibinfo
  {volume} {107}},\ \bibinfo {pages} {241303} (\bibinfo {year} {2011})},\
  \Eprint {http://arxiv.org/abs/1108.2914} {arXiv:1108.2914 [astro-ph.CO]}
  \BibitemShut {NoStop}%
\bibitem [{\citenamefont {Tsai}\ \emph {et~al.}(2013)\citenamefont {Tsai},
  \citenamefont {Yuan},\ and\ \citenamefont {Huang}}]{Tsai:2012cs}%
  \BibitemOpen
  \bibfield  {author} {\bibinfo {author} {\bibfnamefont {Yue-Lin~Sming}\
  \bibnamefont {Tsai}}, \bibinfo {author} {\bibfnamefont {Qiang}\ \bibnamefont
  {Yuan}}, \ and\ \bibinfo {author} {\bibfnamefont {Xiaoyuan}\ \bibnamefont
  {Huang}},\ }\bibfield  {title} {\enquote {\bibinfo {title} {{A generic method
  to constrain the dark matter model parameters from Fermi observations of
  dwarf spheroids}},}\ }\href {\doibase 10.1088/1475-7516/2013/03/018}
  {\bibfield  {journal} {\bibinfo  {journal} {JCAP}\ }\textbf {\bibinfo
  {volume} {1303}},\ \bibinfo {pages} {018} (\bibinfo {year} {2013})},\ \Eprint
  {http://arxiv.org/abs/1212.3990} {arXiv:1212.3990 [astro-ph.HE]} \BibitemShut
  {NoStop}%
\bibitem [{\citenamefont {Ackermann}\ \emph {et~al.}(2015)\citenamefont
  {Ackermann} \emph {et~al.}}]{Ackermann:2015zua}%
  \BibitemOpen
  \bibfield  {author} {\bibinfo {author} {\bibfnamefont {M.}~\bibnamefont
  {Ackermann}} \emph {et~al.} (\bibinfo {collaboration} {Fermi-LAT}),\
  }\bibfield  {title} {\enquote {\bibinfo {title} {{Searching for Dark Matter
  Annihilation from Milky Way Dwarf Spheroidal Galaxies with Six Years of Fermi
  Large Area Telescope Data}},}\ }\href {\doibase
  10.1103/PhysRevLett.115.231301} {\bibfield  {journal} {\bibinfo  {journal}
  {Phys. Rev. Lett.}\ }\textbf {\bibinfo {volume} {115}},\ \bibinfo {pages}
  {231301} (\bibinfo {year} {2015})},\ \Eprint
  {http://arxiv.org/abs/1503.02641} {arXiv:1503.02641 [astro-ph.HE]}
  \BibitemShut {NoStop}%
\bibitem [{\citenamefont {Li}\ \emph {et~al.}(2016)\citenamefont {Li},
  \citenamefont {Liang}, \citenamefont {Duan}, \citenamefont {Shen},
  \citenamefont {Huang}, \citenamefont {Li}, \citenamefont {Fan}, \citenamefont
  {Liao}, \citenamefont {Feng},\ and\ \citenamefont {Chang}}]{Li:2015kag}%
  \BibitemOpen
  \bibfield  {author} {\bibinfo {author} {\bibfnamefont {Shang}\ \bibnamefont
  {Li}}, \bibinfo {author} {\bibfnamefont {Yun-Feng}\ \bibnamefont {Liang}},
  \bibinfo {author} {\bibfnamefont {Kai-Kai}\ \bibnamefont {Duan}}, \bibinfo
  {author} {\bibfnamefont {Zhao-Qiang}\ \bibnamefont {Shen}}, \bibinfo {author}
  {\bibfnamefont {Xiaoyuan}\ \bibnamefont {Huang}}, \bibinfo {author}
  {\bibfnamefont {Xiang}\ \bibnamefont {Li}}, \bibinfo {author} {\bibfnamefont
  {Yi-Zhong}\ \bibnamefont {Fan}}, \bibinfo {author} {\bibfnamefont {Neng-Hui}\
  \bibnamefont {Liao}}, \bibinfo {author} {\bibfnamefont {Lei}\ \bibnamefont
  {Feng}}, \ and\ \bibinfo {author} {\bibfnamefont {Jin}\ \bibnamefont
  {Chang}},\ }\bibfield  {title} {\enquote {\bibinfo {title} {{Search for
  gamma-ray emission from eight dwarf spheroidal galaxy candidates discovered
  in Year Two of Dark Energy Survey with Fermi-LAT data}},}\ }\href {\doibase
  10.1103/PhysRevD.93.043518} {\bibfield  {journal} {\bibinfo  {journal} {Phys.
  Rev.}\ }\textbf {\bibinfo {volume} {D93}},\ \bibinfo {pages} {043518}
  (\bibinfo {year} {2016})},\ \Eprint {http://arxiv.org/abs/1511.09252}
  {arXiv:1511.09252 [astro-ph.HE]} \BibitemShut {NoStop}%
\bibitem [{\citenamefont {Ade}\ \emph {et~al.}(2016)\citenamefont {Ade} \emph
  {et~al.}}]{Ade:2015xua}%
  \BibitemOpen
  \bibfield  {author} {\bibinfo {author} {\bibfnamefont {P.~A.~R.}\
  \bibnamefont {Ade}} \emph {et~al.} (\bibinfo {collaboration} {Planck}),\
  }\bibfield  {title} {\enquote {\bibinfo {title} {{Planck 2015 results. XIII.
  Cosmological parameters}},}\ }\href {\doibase 10.1051/0004-6361/201525830}
  {\bibfield  {journal} {\bibinfo  {journal} {Astron. Astrophys.}\ }\textbf
  {\bibinfo {volume} {594}},\ \bibinfo {pages} {A13} (\bibinfo {year}
  {2016})},\ \Eprint {http://arxiv.org/abs/1502.01589} {arXiv:1502.01589
  [astro-ph.CO]} \BibitemShut {NoStop}%
\bibitem [{\citenamefont {Feldman}\ \emph {et~al.}(2009)\citenamefont
  {Feldman}, \citenamefont {Liu},\ and\ \citenamefont {Nath}}]{Feldman:2008xs}%
  \BibitemOpen
  \bibfield  {author} {\bibinfo {author} {\bibfnamefont {Daniel}\ \bibnamefont
  {Feldman}}, \bibinfo {author} {\bibfnamefont {Zuowei}\ \bibnamefont {Liu}}, \
  and\ \bibinfo {author} {\bibfnamefont {Pran}\ \bibnamefont {Nath}},\
  }\bibfield  {title} {\enquote {\bibinfo {title} {{PAMELA Positron Excess as a
  Signal from the Hidden Sector}},}\ }\href {\doibase
  10.1103/PhysRevD.79.063509} {\bibfield  {journal} {\bibinfo  {journal} {Phys.
  Rev.}\ }\textbf {\bibinfo {volume} {D79}},\ \bibinfo {pages} {063509}
  (\bibinfo {year} {2009})},\ \Eprint {http://arxiv.org/abs/0810.5762}
  {arXiv:0810.5762 [hep-ph]} \BibitemShut {NoStop}%
\bibitem [{\citenamefont {Ibe}\ \emph {et~al.}(2009)\citenamefont {Ibe},
  \citenamefont {Murayama},\ and\ \citenamefont {Yanagida}}]{Ibe:2008ye}%
  \BibitemOpen
  \bibfield  {author} {\bibinfo {author} {\bibfnamefont {Masahiro}\
  \bibnamefont {Ibe}}, \bibinfo {author} {\bibfnamefont {Hitoshi}\ \bibnamefont
  {Murayama}}, \ and\ \bibinfo {author} {\bibfnamefont {T.~T.}\ \bibnamefont
  {Yanagida}},\ }\bibfield  {title} {\enquote {\bibinfo {title} {{Breit-Wigner
  Enhancement of Dark Matter Annihilation}},}\ }\href {\doibase
  10.1103/PhysRevD.79.095009} {\bibfield  {journal} {\bibinfo  {journal} {Phys.
  Rev.}\ }\textbf {\bibinfo {volume} {D79}},\ \bibinfo {pages} {095009}
  (\bibinfo {year} {2009})},\ \Eprint {http://arxiv.org/abs/0812.0072}
  {arXiv:0812.0072 [hep-ph]} \BibitemShut {NoStop}%
\bibitem [{\citenamefont {Guo}\ and\ \citenamefont {Wu}(2009)}]{Guo:2009aj}%
  \BibitemOpen
  \bibfield  {author} {\bibinfo {author} {\bibfnamefont {Wan-Lei}\ \bibnamefont
  {Guo}}\ and\ \bibinfo {author} {\bibfnamefont {Yue-Liang}\ \bibnamefont
  {Wu}},\ }\bibfield  {title} {\enquote {\bibinfo {title} {{Enhancement of Dark
  Matter Annihilation via Breit-Wigner Resonance}},}\ }\href {\doibase
  10.1103/PhysRevD.79.055012} {\bibfield  {journal} {\bibinfo  {journal} {Phys.
  Rev.}\ }\textbf {\bibinfo {volume} {D79}},\ \bibinfo {pages} {055012}
  (\bibinfo {year} {2009})},\ \Eprint {http://arxiv.org/abs/0901.1450}
  {arXiv:0901.1450 [hep-ph]} \BibitemShut {NoStop}%
\bibitem [{\citenamefont {Bi}\ \emph {et~al.}(2009)\citenamefont {Bi},
  \citenamefont {He},\ and\ \citenamefont {Yuan}}]{Bi:2009uj}%
  \BibitemOpen
  \bibfield  {author} {\bibinfo {author} {\bibfnamefont {Xiao-Jun}\
  \bibnamefont {Bi}}, \bibinfo {author} {\bibfnamefont {Xiao-Gang}\
  \bibnamefont {He}}, \ and\ \bibinfo {author} {\bibfnamefont {Qiang}\
  \bibnamefont {Yuan}},\ }\bibfield  {title} {\enquote {\bibinfo {title}
  {{Parameters in a class of leptophilic models from PAMELA, ATIC and
  FERMI}},}\ }\href {\doibase 10.1016/j.physletb.2009.06.009} {\bibfield
  {journal} {\bibinfo  {journal} {Phys. Lett.}\ }\textbf {\bibinfo {volume}
  {B678}},\ \bibinfo {pages} {168--173} (\bibinfo {year} {2009})},\ \Eprint
  {http://arxiv.org/abs/0903.0122} {arXiv:0903.0122 [hep-ph]} \BibitemShut
  {NoStop}%
\bibitem [{\citenamefont {Bi}\ \emph {et~al.}(2012)\citenamefont {Bi},
  \citenamefont {Yin},\ and\ \citenamefont {Yuan}}]{Bi:2011qm}%
  \BibitemOpen
  \bibfield  {author} {\bibinfo {author} {\bibfnamefont {Xiao-Jun}\
  \bibnamefont {Bi}}, \bibinfo {author} {\bibfnamefont {Peng-Fei}\ \bibnamefont
  {Yin}}, \ and\ \bibinfo {author} {\bibfnamefont {Qiang}\ \bibnamefont
  {Yuan}},\ }\bibfield  {title} {\enquote {\bibinfo {title} {{Breit-Wigner
  Enhancement Considering the Dark Matter Kinetic Decoupling}},}\ }\href
  {\doibase 10.1103/PhysRevD.85.043526} {\bibfield  {journal} {\bibinfo
  {journal} {Phys. Rev.}\ }\textbf {\bibinfo {volume} {D85}},\ \bibinfo {pages}
  {043526} (\bibinfo {year} {2012})},\ \Eprint {http://arxiv.org/abs/1106.6027}
  {arXiv:1106.6027 [hep-ph]} \BibitemShut {NoStop}%
\bibitem [{\citenamefont {Hisano}\ \emph {et~al.}(2011)\citenamefont {Hisano},
  \citenamefont {Kawasaki}, \citenamefont {Kohri}, \citenamefont {Moroi},
  \citenamefont {Nakayama},\ and\ \citenamefont {Sekiguchi}}]{Hisano:2011dc}%
  \BibitemOpen
  \bibfield  {author} {\bibinfo {author} {\bibfnamefont {Junji}\ \bibnamefont
  {Hisano}}, \bibinfo {author} {\bibfnamefont {Masahiro}\ \bibnamefont
  {Kawasaki}}, \bibinfo {author} {\bibfnamefont {Kazunori}\ \bibnamefont
  {Kohri}}, \bibinfo {author} {\bibfnamefont {Takeo}\ \bibnamefont {Moroi}},
  \bibinfo {author} {\bibfnamefont {Kazunori}\ \bibnamefont {Nakayama}}, \ and\
  \bibinfo {author} {\bibfnamefont {Toyokazu}\ \bibnamefont {Sekiguchi}},\
  }\bibfield  {title} {\enquote {\bibinfo {title} {{Cosmological constraints on
  dark matter models with velocity-dependent annihilation cross section}},}\
  }\href {\doibase 10.1103/PhysRevD.83.123511} {\bibfield  {journal} {\bibinfo
  {journal} {Phys. Rev.}\ }\textbf {\bibinfo {volume} {D83}},\ \bibinfo {pages}
  {123511} (\bibinfo {year} {2011})},\ \Eprint {http://arxiv.org/abs/1102.4658}
  {arXiv:1102.4658 [hep-ph]} \BibitemShut {NoStop}%
\bibitem [{\citenamefont {Bai}\ \emph {et~al.}(2017)\citenamefont {Bai},
  \citenamefont {Berger},\ and\ \citenamefont {Lu}}]{Bai:2017fav}%
  \BibitemOpen
  \bibfield  {author} {\bibinfo {author} {\bibfnamefont {Yang}\ \bibnamefont
  {Bai}}, \bibinfo {author} {\bibfnamefont {Joshua}\ \bibnamefont {Berger}}, \
  and\ \bibinfo {author} {\bibfnamefont {Sida}\ \bibnamefont {Lu}},\ }\bibfield
   {title} {\enquote {\bibinfo {title} {{Supersymmetric Resonant Dark Matter: a
  Thermal Model for the AMS-02 Positron Excess}},}\ }\href@noop {} {\
  (\bibinfo {year} {2017})},\ \Eprint {http://arxiv.org/abs/1706.09974}
  {arXiv:1706.09974 [hep-ph]} \BibitemShut {NoStop}%
\bibitem [{\citenamefont {Xiang}\ \emph {et~al.}(2017)\citenamefont {Xiang},
  \citenamefont {Bi}, \citenamefont {Lin},\ and\ \citenamefont
  {Yin}}]{Xiang:2017jou}%
  \BibitemOpen
  \bibfield  {author} {\bibinfo {author} {\bibfnamefont {Qian-Fei}\
  \bibnamefont {Xiang}}, \bibinfo {author} {\bibfnamefont {Xiao-Jun}\
  \bibnamefont {Bi}}, \bibinfo {author} {\bibfnamefont {Su-Jie}\ \bibnamefont
  {Lin}}, \ and\ \bibinfo {author} {\bibfnamefont {Peng-Fei}\ \bibnamefont
  {Yin}},\ }\bibfield  {title} {\enquote {\bibinfo {title} {{A dark matter
  model that reconciles tensions between the cosmic-ray $e^\pm$ excess and the
  gamma-ray and CMB constraints}},}\ }\href {\doibase
  10.1016/j.physletb.2017.09.003} {\bibfield  {journal} {\bibinfo  {journal}
  {Phys. Lett.}\ }\textbf {\bibinfo {volume} {B773}},\ \bibinfo {pages}
  {448--454} (\bibinfo {year} {2017})},\ \Eprint
  {http://arxiv.org/abs/1707.09313} {arXiv:1707.09313 [astro-ph.HE]}
  \BibitemShut {NoStop}%
\bibitem [{\citenamefont {Niu}\ \emph {et~al.}(2017)\citenamefont {Niu},
  \citenamefont {Li},\ and\ \citenamefont {Xu}}]{Niu:2017lts}%
  \BibitemOpen
  \bibfield  {author} {\bibinfo {author} {\bibfnamefont {Jia-Shu}\ \bibnamefont
  {Niu}}, \bibinfo {author} {\bibfnamefont {Tianjun}\ \bibnamefont {Li}}, \
  and\ \bibinfo {author} {\bibfnamefont {Fang-Zhou}\ \bibnamefont {Xu}},\
  }\bibfield  {title} {\enquote {\bibinfo {title} {{The Simple and Natural
  Interpretations of the DAMPE Cosmic Ray Electron/Positron Spectrum within Two
  Sigma Deviations}},}\ }\href@noop {} {\  (\bibinfo {year} {2017})},\ \Eprint
  {http://arxiv.org/abs/1712.09586} {arXiv:1712.09586 [hep-ph]} \BibitemShut
  {NoStop}%
\bibitem [{\citenamefont {{Clem}}\ \emph {et~al.}(1996)\citenamefont {{Clem}},
  \citenamefont {{Clements}}, \citenamefont {{Esposito}}, \citenamefont
  {{Evenson}}, \citenamefont {{Huber}}, \citenamefont {{L'Heureux}},
  \citenamefont {{Meyer}},\ and\ \citenamefont {{Constantin}}}]{Clem1996}%
  \BibitemOpen
  \bibfield  {author} {\bibinfo {author} {\bibfnamefont {J.~M.}\ \bibnamefont
  {{Clem}}}, \bibinfo {author} {\bibfnamefont {D.~P.}\ \bibnamefont
  {{Clements}}}, \bibinfo {author} {\bibfnamefont {J.}~\bibnamefont
  {{Esposito}}}, \bibinfo {author} {\bibfnamefont {P.}~\bibnamefont
  {{Evenson}}}, \bibinfo {author} {\bibfnamefont {D.}~\bibnamefont {{Huber}}},
  \bibinfo {author} {\bibfnamefont {J.}~\bibnamefont {{L'Heureux}}}, \bibinfo
  {author} {\bibfnamefont {P.}~\bibnamefont {{Meyer}}}, \ and\ \bibinfo
  {author} {\bibfnamefont {C.}~\bibnamefont {{Constantin}}},\ }\bibfield
  {title} {\enquote {\bibinfo {title} {{Solar Modulation of Cosmic
  Electrons}},}\ }\href {\doibase 10.1086/177340} {\bibfield  {journal}
  {\bibinfo  {journal} {\apj}\ }\textbf {\bibinfo {volume} {464}},\ \bibinfo
  {pages} {507} (\bibinfo {year} {1996})}\BibitemShut {NoStop}%
\bibitem [{\citenamefont {{Boella}}\ \emph {et~al.}(2001)\citenamefont
  {{Boella}}, \citenamefont {{Gervasi}}, \citenamefont {{Mariani}},
  \citenamefont {{Rancoita}},\ and\ \citenamefont {{Usoskin}}}]{Boella2001}%
  \BibitemOpen
  \bibfield  {author} {\bibinfo {author} {\bibfnamefont {G.}~\bibnamefont
  {{Boella}}}, \bibinfo {author} {\bibfnamefont {M.}~\bibnamefont {{Gervasi}}},
  \bibinfo {author} {\bibfnamefont {S.}~\bibnamefont {{Mariani}}}, \bibinfo
  {author} {\bibfnamefont {P.~G.}\ \bibnamefont {{Rancoita}}}, \ and\ \bibinfo
  {author} {\bibfnamefont {I.~G.}\ \bibnamefont {{Usoskin}}},\ }\bibfield
  {title} {\enquote {\bibinfo {title} {{Evidence for charge drift modulation at
  intermediate solar activity from the flux variation of protons and {$\alpha$}
  particles}},}\ }\href {\doibase 10.1029/2001JA900075} {\bibfield  {journal}
  {\bibinfo  {journal} {Journal of Geophysical Research (Space Physics)}\
  }\textbf {\bibinfo {volume} {106}},\ \bibinfo {pages} {29355--29362}
  (\bibinfo {year} {2001})}\BibitemShut {NoStop}%
\bibitem [{\citenamefont {{Usoskin}}\ \emph {et~al.}(2011)\citenamefont
  {{Usoskin}}, \citenamefont {{Bazilevskaya}},\ and\ \citenamefont
  {{Kovaltsov}}}]{Usoskin2011}%
  \BibitemOpen
  \bibfield  {author} {\bibinfo {author} {\bibfnamefont {I.~G.}\ \bibnamefont
  {{Usoskin}}}, \bibinfo {author} {\bibfnamefont {G.~A.}\ \bibnamefont
  {{Bazilevskaya}}}, \ and\ \bibinfo {author} {\bibfnamefont {G.~A.}\
  \bibnamefont {{Kovaltsov}}},\ }\bibfield  {title} {\enquote {\bibinfo {title}
  {{Solar modulation parameter for cosmic rays since 1936 reconstructed from
  ground-based neutron monitors and ionization chambers}},}\ }\href {\doibase
  10.1029/2010JA016105} {\bibfield  {journal} {\bibinfo  {journal} {Journal of
  Geophysical Research (Space Physics)}\ }\textbf {\bibinfo {volume} {116}},\
  \bibinfo {eid} {A02104} (\bibinfo {year} {2011})}\BibitemShut {NoStop}%
\bibitem [{\citenamefont {{Corti}}\ \emph {et~al.}(2016)\citenamefont
  {{Corti}}, \citenamefont {{Bindi}}, \citenamefont {{Consolandi}},\ and\
  \citenamefont {{Whitman}}}]{Corti2016}%
  \BibitemOpen
  \bibfield  {author} {\bibinfo {author} {\bibfnamefont {C.}~\bibnamefont
  {{Corti}}}, \bibinfo {author} {\bibfnamefont {V.}~\bibnamefont {{Bindi}}},
  \bibinfo {author} {\bibfnamefont {C.}~\bibnamefont {{Consolandi}}}, \ and\
  \bibinfo {author} {\bibfnamefont {K.}~\bibnamefont {{Whitman}}},\ }\bibfield
  {title} {\enquote {\bibinfo {title} {{Solar Modulation of the Local
  Interstellar Spectrum with Voyager 1, AMS-02, PAMELA, and BESS}},}\ }\href
  {\doibase 10.3847/0004-637X/829/1/8} {\bibfield  {journal} {\bibinfo
  {journal} {\apj}\ }\textbf {\bibinfo {volume} {829}},\ \bibinfo {eid} {8}
  (\bibinfo {year} {2016})},\ \Eprint {http://arxiv.org/abs/1511.08790}
  {arXiv:1511.08790 [astro-ph.HE]} \BibitemShut {NoStop}%
\bibitem [{\citenamefont {{Lin}}\ \emph {et~al.}(2017)\citenamefont {{Lin}},
  \citenamefont {{Bi}}, \citenamefont {{Feng}}, \citenamefont {{Yin}},\ and\
  \citenamefont {{Yu}}}]{Lin2016}%
  \BibitemOpen
  \bibfield  {author} {\bibinfo {author} {\bibfnamefont {S.-J.}\ \bibnamefont
  {{Lin}}}, \bibinfo {author} {\bibfnamefont {X.-J.}\ \bibnamefont {{Bi}}},
  \bibinfo {author} {\bibfnamefont {J.}~\bibnamefont {{Feng}}}, \bibinfo
  {author} {\bibfnamefont {P.-F.}\ \bibnamefont {{Yin}}}, \ and\ \bibinfo
  {author} {\bibfnamefont {Z.-H.}\ \bibnamefont {{Yu}}},\ }\bibfield  {title}
  {\enquote {\bibinfo {title} {{Systematic study on the cosmic ray antiproton
  flux}},}\ }\href {\doibase 10.1103/PhysRevD.96.123010} {\bibfield  {journal}
  {\bibinfo  {journal} {\prd}\ }\textbf {\bibinfo {volume} {96}},\ \bibinfo
  {eid} {123010} (\bibinfo {year} {2017})},\ \Eprint
  {http://arxiv.org/abs/1612.04001} {arXiv:1612.04001 [astro-ph.HE]}
  \BibitemShut {NoStop}%
\bibitem [{\citenamefont {Kamae}\ \emph {et~al.}(2005)\citenamefont {Kamae},
  \citenamefont {Abe},\ and\ \citenamefont {Koi}}]{Kamae:2004xx}%
  \BibitemOpen
  \bibfield  {author} {\bibinfo {author} {\bibfnamefont {Tuneyoshi}\
  \bibnamefont {Kamae}}, \bibinfo {author} {\bibfnamefont {Toshinori}\
  \bibnamefont {Abe}}, \ and\ \bibinfo {author} {\bibfnamefont {Tatsumi}\
  \bibnamefont {Koi}},\ }\bibfield  {title} {\enquote {\bibinfo {title}
  {{Diffractive interaction and scaling violation in p p ---> pi0 interaction
  and GeV excess in galactic diffuse gamma-ray spectrum of EGRET}},}\ }\href
  {\doibase 10.1086/426935} {\bibfield  {journal} {\bibinfo  {journal}
  {Astrophys. J.}\ }\textbf {\bibinfo {volume} {620}},\ \bibinfo {pages}
  {244--256} (\bibinfo {year} {2005})},\ \Eprint
  {http://arxiv.org/abs/astro-ph/0410617} {arXiv:astro-ph/0410617 [astro-ph]}
  \BibitemShut {NoStop}%
\bibitem [{\citenamefont {Kamae}\ \emph {et~al.}(2006)\citenamefont {Kamae},
  \citenamefont {Karlsson}, \citenamefont {Mizuno}, \citenamefont {Abe},\ and\
  \citenamefont {Koi}}]{Kamae:2006bf}%
  \BibitemOpen
  \bibfield  {author} {\bibinfo {author} {\bibfnamefont {Tuneyoshi}\
  \bibnamefont {Kamae}}, \bibinfo {author} {\bibfnamefont {Niklas}\
  \bibnamefont {Karlsson}}, \bibinfo {author} {\bibfnamefont {Tsunefumi}\
  \bibnamefont {Mizuno}}, \bibinfo {author} {\bibfnamefont {Toshinori}\
  \bibnamefont {Abe}}, \ and\ \bibinfo {author} {\bibfnamefont {Tatsumi}\
  \bibnamefont {Koi}},\ }\bibfield  {title} {\enquote {\bibinfo {title}
  {{Parameterization of Gamma, e+/- and Neutrino Spectra Produced by p-p
  Interaction in Astronomical Environment}},}\ }\href {\doibase 10.1086/513602,
  10.1086/505189} {\bibfield  {journal} {\bibinfo  {journal} {Astrophys. J.}\
  }\textbf {\bibinfo {volume} {647}},\ \bibinfo {pages} {692--708} (\bibinfo
  {year} {2006})},\ \bibinfo {note} {[Erratum: Astrophys. J.662,779(2007)]},\
  \Eprint {http://arxiv.org/abs/astro-ph/0605581} {arXiv:astro-ph/0605581
  [astro-ph]} \BibitemShut {NoStop}%
\bibitem [{\citenamefont {Evoli}\ \emph {et~al.}(2017)\citenamefont {Evoli},
  \citenamefont {Gaggero}, \citenamefont {Vittino}, \citenamefont {Di~Mauro},
  \citenamefont {Grasso},\ and\ \citenamefont {Mazziotta}}]{Evoli:2017vim}%
  \BibitemOpen
  \bibfield  {author} {\bibinfo {author} {\bibfnamefont {Carmelo}\ \bibnamefont
  {Evoli}}, \bibinfo {author} {\bibfnamefont {Daniele}\ \bibnamefont
  {Gaggero}}, \bibinfo {author} {\bibfnamefont {Andrea}\ \bibnamefont
  {Vittino}}, \bibinfo {author} {\bibfnamefont {Mattia}\ \bibnamefont
  {Di~Mauro}}, \bibinfo {author} {\bibfnamefont {Dario}\ \bibnamefont
  {Grasso}}, \ and\ \bibinfo {author} {\bibfnamefont {Mario~Nicola}\
  \bibnamefont {Mazziotta}},\ }\bibfield  {title} {\enquote {\bibinfo {title}
  {{Cosmic-ray propagation with DRAGON2: II. Nuclear interactions with the
  interstellar gas}},}\ }\href@noop {} {\  (\bibinfo {year} {2017})},\ \Eprint
  {http://arxiv.org/abs/1711.09616} {arXiv:1711.09616 [astro-ph.HE]}
  \BibitemShut {NoStop}%
\bibitem [{\citenamefont {{Jeannerot}}\ \emph {et~al.}(1999)\citenamefont
  {{Jeannerot}}, \citenamefont {{Zhang}},\ and\ \citenamefont
  {{Brandenberger}}}]{Jeannerot1999}%
  \BibitemOpen
  \bibfield  {author} {\bibinfo {author} {\bibfnamefont {R.}~\bibnamefont
  {{Jeannerot}}}, \bibinfo {author} {\bibfnamefont {X.}~\bibnamefont
  {{Zhang}}}, \ and\ \bibinfo {author} {\bibfnamefont {R.}~\bibnamefont
  {{Brandenberger}}},\ }\bibfield  {title} {\enquote {\bibinfo {title}
  {{Non-thermal production of neutralino cold dark matter from cosmic string
  decays}},}\ }\href {\doibase 10.1088/1126-6708/1999/12/003} {\bibfield
  {journal} {\bibinfo  {journal} {Journal of High Energy Physics}\ }\textbf
  {\bibinfo {volume} {12}},\ \bibinfo {eid} {003} (\bibinfo {year} {1999})},\
  \Eprint {http://arxiv.org/abs/hep-ph/9901357} {hep-ph/9901357} \BibitemShut
  {NoStop}%
\bibitem [{\citenamefont {{Lin}}\ \emph {et~al.}(2001)\citenamefont {{Lin}},
  \citenamefont {{Huang}}, \citenamefont {{Zhang}},\ and\ \citenamefont
  {{Brandenberger}}}]{Lin2001}%
  \BibitemOpen
  \bibfield  {author} {\bibinfo {author} {\bibfnamefont {W.~B.}\ \bibnamefont
  {{Lin}}}, \bibinfo {author} {\bibfnamefont {D.~H.}\ \bibnamefont {{Huang}}},
  \bibinfo {author} {\bibfnamefont {X.}~\bibnamefont {{Zhang}}}, \ and\
  \bibinfo {author} {\bibfnamefont {R.}~\bibnamefont {{Brandenberger}}},\
  }\bibfield  {title} {\enquote {\bibinfo {title} {{Nonthermal Production of
  Weakly Interacting Massive Particles and the Subgalactic Structure of the
  Universe}},}\ }\href {\doibase 10.1103/PhysRevLett.86.954} {\bibfield
  {journal} {\bibinfo  {journal} {Physical Review Letters}\ }\textbf {\bibinfo
  {volume} {86}},\ \bibinfo {pages} {954--957} (\bibinfo {year} {2001})},\
  \Eprint {http://arxiv.org/abs/astro-ph/0009003} {astro-ph/0009003}
  \BibitemShut {NoStop}%
\bibitem [{\citenamefont {{Yuan}}\ \emph {et~al.}(2012)\citenamefont {{Yuan}},
  \citenamefont {{Cao}}, \citenamefont {{Liu}}, \citenamefont {{Yin}},
  \citenamefont {{Gao}}, \citenamefont {{Bi}},\ and\ \citenamefont
  {{Zhang}}}]{Yuan2012}%
  \BibitemOpen
  \bibfield  {author} {\bibinfo {author} {\bibfnamefont {Q.}~\bibnamefont
  {{Yuan}}}, \bibinfo {author} {\bibfnamefont {Y.}~\bibnamefont {{Cao}}},
  \bibinfo {author} {\bibfnamefont {J.}~\bibnamefont {{Liu}}}, \bibinfo
  {author} {\bibfnamefont {P.-F.}\ \bibnamefont {{Yin}}}, \bibinfo {author}
  {\bibfnamefont {L.}~\bibnamefont {{Gao}}}, \bibinfo {author} {\bibfnamefont
  {X.-J.}\ \bibnamefont {{Bi}}}, \ and\ \bibinfo {author} {\bibfnamefont
  {X.}~\bibnamefont {{Zhang}}},\ }\bibfield  {title} {\enquote {\bibinfo
  {title} {{Gamma rays from warm WIMP dark matter annihilation}},}\ }\href
  {\doibase 10.1103/PhysRevD.86.103531} {\bibfield  {journal} {\bibinfo
  {journal} {\prd}\ }\textbf {\bibinfo {volume} {86}},\ \bibinfo {eid} {103531}
  (\bibinfo {year} {2012})},\ \Eprint {http://arxiv.org/abs/1203.5636}
  {arXiv:1203.5636 [astro-ph.HE]} \BibitemShut {NoStop}%
\bibitem [{\citenamefont {{Sommerfeld}}(1931)}]{Sommerfeld1931}%
  \BibitemOpen
  \bibfield  {author} {\bibinfo {author} {\bibfnamefont {A.}~\bibnamefont
  {{Sommerfeld}}},\ }\bibfield  {title} {\enquote {\bibinfo {title} {{{\"U}ber
  die Beugung und Bremsung der Elektronen}},}\ }\href {\doibase
  10.1002/andp.19314030302} {\bibfield  {journal} {\bibinfo  {journal} {Annalen
  der Physik}\ }\textbf {\bibinfo {volume} {403}},\ \bibinfo {pages} {257--330}
  (\bibinfo {year} {1931})}\BibitemShut {NoStop}%
\bibitem [{\citenamefont {{Hisano}}\ \emph {et~al.}(2005)\citenamefont
  {{Hisano}}, \citenamefont {{Matsumoto}}, \citenamefont {{Nojiri}},\ and\
  \citenamefont {{Saito}}}]{Hisano2005}%
  \BibitemOpen
  \bibfield  {author} {\bibinfo {author} {\bibfnamefont {J.}~\bibnamefont
  {{Hisano}}}, \bibinfo {author} {\bibfnamefont {S.}~\bibnamefont
  {{Matsumoto}}}, \bibinfo {author} {\bibfnamefont {M.~M.}\ \bibnamefont
  {{Nojiri}}}, \ and\ \bibinfo {author} {\bibfnamefont {O.}~\bibnamefont
  {{Saito}}},\ }\bibfield  {title} {\enquote {\bibinfo {title}
  {{Nonperturbative effect on dark matter annihilation and gamma ray signature
  from the galactic center}},}\ }\href {\doibase 10.1103/PhysRevD.71.063528}
  {\bibfield  {journal} {\bibinfo  {journal} {\prd}\ }\textbf {\bibinfo
  {volume} {71}},\ \bibinfo {eid} {063528} (\bibinfo {year} {2005})},\ \Eprint
  {http://arxiv.org/abs/hep-ph/0412403} {hep-ph/0412403} \BibitemShut {NoStop}%
\bibitem [{\citenamefont {{Arkani-Hamed}}\ \emph {et~al.}(2009)\citenamefont
  {{Arkani-Hamed}}, \citenamefont {{Finkbeiner}}, \citenamefont {{Slatyer}},\
  and\ \citenamefont {{Weiner}}}]{Arkani-Hamed2009}%
  \BibitemOpen
  \bibfield  {author} {\bibinfo {author} {\bibfnamefont {N.}~\bibnamefont
  {{Arkani-Hamed}}}, \bibinfo {author} {\bibfnamefont {D.~P.}\ \bibnamefont
  {{Finkbeiner}}}, \bibinfo {author} {\bibfnamefont {T.~R.}\ \bibnamefont
  {{Slatyer}}}, \ and\ \bibinfo {author} {\bibfnamefont {N.}~\bibnamefont
  {{Weiner}}},\ }\bibfield  {title} {\enquote {\bibinfo {title} {{A theory of
  dark matter}},}\ }\href {\doibase 10.1103/PhysRevD.79.015014} {\bibfield
  {journal} {\bibinfo  {journal} {\prd}\ }\textbf {\bibinfo {volume} {79}},\
  \bibinfo {eid} {015014} (\bibinfo {year} {2009})},\ \Eprint
  {http://arxiv.org/abs/0810.0713} {arXiv:0810.0713 [hep-ph]} \BibitemShut
  {NoStop}%
\bibitem [{\citenamefont {Griest}\ and\ \citenamefont
  {Seckel}(1991)}]{Griest1990}%
  \BibitemOpen
  \bibfield  {author} {\bibinfo {author} {\bibfnamefont {Kim}\ \bibnamefont
  {Griest}}\ and\ \bibinfo {author} {\bibfnamefont {David}\ \bibnamefont
  {Seckel}},\ }\bibfield  {title} {\enquote {\bibinfo {title} {{Three
  exceptions in the calculation of relic abundances}},}\ }\href {\doibase
  10.1103/PhysRevD.43.3191} {\bibfield  {journal} {\bibinfo  {journal} {Phys.
  Rev.}\ }\textbf {\bibinfo {volume} {D43}},\ \bibinfo {pages} {3191--3203}
  (\bibinfo {year} {1991})}\BibitemShut {NoStop}%
\bibitem [{\citenamefont {Gondolo}\ and\ \citenamefont
  {Gelmini}(1991)}]{Gondolo1990}%
  \BibitemOpen
  \bibfield  {author} {\bibinfo {author} {\bibfnamefont {Paolo}\ \bibnamefont
  {Gondolo}}\ and\ \bibinfo {author} {\bibfnamefont {Graciela}\ \bibnamefont
  {Gelmini}},\ }\bibfield  {title} {\enquote {\bibinfo {title} {{Cosmic
  abundances of stable particles: Improved analysis}},}\ }\href {\doibase
  10.1016/0550-3213(91)90438-4} {\bibfield  {journal} {\bibinfo  {journal}
  {Nucl. Phys.}\ }\textbf {\bibinfo {volume} {B360}},\ \bibinfo {pages}
  {145--179} (\bibinfo {year} {1991})}\BibitemShut {NoStop}%
\bibitem [{\citenamefont {{Jin}}\ \emph {et~al.}(2017)\citenamefont {{Jin}},
  \citenamefont {{Yue}}, \citenamefont {{Zhang}},\ and\ \citenamefont
  {{Chen}}}]{Jin2017_dampe}%
  \BibitemOpen
  \bibfield  {author} {\bibinfo {author} {\bibfnamefont {H.-B.}\ \bibnamefont
  {{Jin}}}, \bibinfo {author} {\bibfnamefont {B.}~\bibnamefont {{Yue}}},
  \bibinfo {author} {\bibfnamefont {X.}~\bibnamefont {{Zhang}}}, \ and\
  \bibinfo {author} {\bibfnamefont {X.}~\bibnamefont {{Chen}}},\ }\bibfield
  {title} {\enquote {\bibinfo {title} {{Cosmic ray $e^{+} e^{-}$ spectrum
  excess and peak feature observed by the DAMPE experiment from dark
  matter}},}\ }\href@noop {} {\bibfield  {journal} {\bibinfo  {journal} {ArXiv
  e-prints}\ } (\bibinfo {year} {2017})},\ \Eprint
  {http://arxiv.org/abs/1712.00362} {arXiv:1712.00362 [astro-ph.HE]}
  \BibitemShut {NoStop}%
\bibitem [{\citenamefont {{Huang}}\ \emph {et~al.}(2017)\citenamefont
  {{Huang}}, \citenamefont {{Wu}}, \citenamefont {{Zhang}},\ and\ \citenamefont
  {{Zhou}}}]{Huang2017}%
  \BibitemOpen
  \bibfield  {author} {\bibinfo {author} {\bibfnamefont {X.-J.}\ \bibnamefont
  {{Huang}}}, \bibinfo {author} {\bibfnamefont {Y.-L.}\ \bibnamefont {{Wu}}},
  \bibinfo {author} {\bibfnamefont {W.-H.}\ \bibnamefont {{Zhang}}}, \ and\
  \bibinfo {author} {\bibfnamefont {Y.-F.}\ \bibnamefont {{Zhou}}},\ }\bibfield
   {title} {\enquote {\bibinfo {title} {{Origins of sharp cosmic-ray electron
  structures and the DAMPE excess}},}\ }\href@noop {} {\bibfield  {journal}
  {\bibinfo  {journal} {ArXiv e-prints}\ } (\bibinfo {year} {2017})},\ \Eprint
  {http://arxiv.org/abs/1712.00005} {arXiv:1712.00005 [astro-ph.HE]}
  \BibitemShut {NoStop}%
\bibitem [{\citenamefont {{Yang}}\ \emph {et~al.}(2017)\citenamefont {{Yang}},
  \citenamefont {{Su}},\ and\ \citenamefont {{Zhao}}}]{Yang2017}%
  \BibitemOpen
  \bibfield  {author} {\bibinfo {author} {\bibfnamefont {F.}~\bibnamefont
  {{Yang}}}, \bibinfo {author} {\bibfnamefont {M.}~\bibnamefont {{Su}}}, \ and\
  \bibinfo {author} {\bibfnamefont {Y.}~\bibnamefont {{Zhao}}},\ }\bibfield
  {title} {\enquote {\bibinfo {title} {{Dark Matter Annihilation from Nearby
  Ultra-compact Micro Halos to Explain the Tentative Excess at \~{}1.4 TeV in
  DAMPE data}},}\ }\href@noop {} {\bibfield  {journal} {\bibinfo  {journal}
  {ArXiv e-prints}\ } (\bibinfo {year} {2017})},\ \Eprint
  {http://arxiv.org/abs/1712.01724} {arXiv:1712.01724 [astro-ph.HE]}
  \BibitemShut {NoStop}%
\bibitem [{\citenamefont {{Ge}}\ and\ \citenamefont {{He}}(2017)}]{Ge2017}%
  \BibitemOpen
  \bibfield  {author} {\bibinfo {author} {\bibfnamefont {S.-F.}\ \bibnamefont
  {{Ge}}}\ and\ \bibinfo {author} {\bibfnamefont {H.-J.}\ \bibnamefont
  {{He}}},\ }\bibfield  {title} {\enquote {\bibinfo {title} {{Flavor Structure
  of the Cosmic-Ray Electron/Positron Excesses at DAMPE}},}\ }\href@noop {}
  {\bibfield  {journal} {\bibinfo  {journal} {ArXiv e-prints}\ } (\bibinfo
  {year} {2017})},\ \Eprint {http://arxiv.org/abs/1712.02744} {arXiv:1712.02744
  [astro-ph.HE]} \BibitemShut {NoStop}%
\bibitem [{\citenamefont {{Cholis}}\ \emph {et~al.}(2017)\citenamefont
  {{Cholis}}, \citenamefont {{Karwal}},\ and\ \citenamefont
  {{Kamionkowski}}}]{Cholis2017}%
  \BibitemOpen
  \bibfield  {author} {\bibinfo {author} {\bibfnamefont {I.}~\bibnamefont
  {{Cholis}}}, \bibinfo {author} {\bibfnamefont {T.}~\bibnamefont {{Karwal}}},
  \ and\ \bibinfo {author} {\bibfnamefont {M.}~\bibnamefont {{Kamionkowski}}},\
  }\bibfield  {title} {\enquote {\bibinfo {title} {{Features in the Spectrum of
  Cosmic-Ray Positrons from Pulsars}},}\ }\href@noop {} {\bibfield  {journal}
  {\bibinfo  {journal} {ArXiv e-prints}\ } (\bibinfo {year} {2017})},\ \Eprint
  {http://arxiv.org/abs/1712.00011} {arXiv:1712.00011 [astro-ph.HE]}
  \BibitemShut {NoStop}%
\bibitem [{\citenamefont {{Niu}}\ \emph {et~al.}(2017)\citenamefont {{Niu}},
  \citenamefont {{Li}}, \citenamefont {{Zong}}, \citenamefont {{Xue}},\ and\
  \citenamefont {{Wang}}}]{Niu2017_DAV}%
  \BibitemOpen
  \bibfield  {author} {\bibinfo {author} {\bibfnamefont {J.-S.}\ \bibnamefont
  {{Niu}}}, \bibinfo {author} {\bibfnamefont {T.}~\bibnamefont {{Li}}},
  \bibinfo {author} {\bibfnamefont {W.}~\bibnamefont {{Zong}}}, \bibinfo
  {author} {\bibfnamefont {H.-F.}\ \bibnamefont {{Xue}}}, \ and\ \bibinfo
  {author} {\bibfnamefont {Y.}~\bibnamefont {{Wang}}},\ }\bibfield  {title}
  {\enquote {\bibinfo {title} {{Probing the Dark Matter-Electron Interactions
  via Hydrogen-Atmosphere Pulsating White Dwarfs}},}\ }\href@noop {} {\bibfield
   {journal} {\bibinfo  {journal} {ArXiv e-prints}\ } (\bibinfo {year}
  {2017})},\ \Eprint {http://arxiv.org/abs/1709.08804} {arXiv:1709.08804
  [astro-ph.HE]} \BibitemShut {NoStop}%
\bibitem [{\citenamefont {{Maurin}}\ \emph {et~al.}(2014)\citenamefont
  {{Maurin}}, \citenamefont {{Melot}},\ and\ \citenamefont
  {{Taillet}}}]{Maurin2014}%
  \BibitemOpen
  \bibfield  {author} {\bibinfo {author} {\bibfnamefont {D.}~\bibnamefont
  {{Maurin}}}, \bibinfo {author} {\bibfnamefont {F.}~\bibnamefont {{Melot}}}, \
  and\ \bibinfo {author} {\bibfnamefont {R.}~\bibnamefont {{Taillet}}},\
  }\bibfield  {title} {\enquote {\bibinfo {title} {{A database of charged
  cosmic rays}},}\ }\href {\doibase 10.1051/0004-6361/201321344} {\bibfield
  {journal} {\bibinfo  {journal} {\aap}\ }\textbf {\bibinfo {volume} {569}},\
  \bibinfo {eid} {A32} (\bibinfo {year} {2014})},\ \Eprint
  {http://arxiv.org/abs/1302.5525} {arXiv:1302.5525 [astro-ph.HE]} \BibitemShut
  {NoStop}%
\bibitem [{\citenamefont {Foreman-Mackey}\ \emph {et~al.}(2016)\citenamefont
  {Foreman-Mackey}, \citenamefont {Vousden}, \citenamefont {Price-Whelan},
  \citenamefont {Pitkin}, \citenamefont {Zabalza}, \citenamefont {Ryan},
  \citenamefont {Emily}, \citenamefont {Smith}, \citenamefont {Ashton},
  \citenamefont {Cruz}, \citenamefont {Kerzendorf}, \citenamefont {Caswell},
  \citenamefont {Hoyer}, \citenamefont {Barbary}, \citenamefont {Czekala},
  \citenamefont {Hogg},\ and\ \citenamefont {Brewer}}]{corner}%
  \BibitemOpen
  \bibfield  {author} {\bibinfo {author} {\bibfnamefont {Dan}\ \bibnamefont
  {Foreman-Mackey}}, \bibinfo {author} {\bibfnamefont {Will}\ \bibnamefont
  {Vousden}}, \bibinfo {author} {\bibfnamefont {Adrian}\ \bibnamefont
  {Price-Whelan}}, \bibinfo {author} {\bibfnamefont {Matt}\ \bibnamefont
  {Pitkin}}, \bibinfo {author} {\bibfnamefont {Victor}\ \bibnamefont
  {Zabalza}}, \bibinfo {author} {\bibfnamefont {Geoffrey}\ \bibnamefont
  {Ryan}}, \bibinfo {author} {\bibnamefont {Emily}}, \bibinfo {author}
  {\bibfnamefont {Michael}\ \bibnamefont {Smith}}, \bibinfo {author}
  {\bibfnamefont {Gregory}\ \bibnamefont {Ashton}}, \bibinfo {author}
  {\bibfnamefont {Kelle}\ \bibnamefont {Cruz}}, \bibinfo {author}
  {\bibfnamefont {Wolfgang}\ \bibnamefont {Kerzendorf}}, \bibinfo {author}
  {\bibfnamefont {Thomas~A}\ \bibnamefont {Caswell}}, \bibinfo {author}
  {\bibfnamefont {Stephan}\ \bibnamefont {Hoyer}}, \bibinfo {author}
  {\bibfnamefont {Kyle}\ \bibnamefont {Barbary}}, \bibinfo {author}
  {\bibfnamefont {Ian}\ \bibnamefont {Czekala}}, \bibinfo {author}
  {\bibfnamefont {David~W.}\ \bibnamefont {Hogg}}, \ and\ \bibinfo {author}
  {\bibfnamefont {Brendon~J.}\ \bibnamefont {Brewer}},\ }\href {\doibase
  10.5281/zenodo.45906} {\enquote {\bibinfo {title} {corner.py: corner.py
  v1.0.2},}\ } (\bibinfo {year} {2016})\BibitemShut {NoStop}%
\end{thebibliography}

%

\end{document}